\documentstyle[12pt]{report}
\input psfig.tex
  \def\singlespacing{\parskip 5 pt plus 1 pt \baselineskip 13 pt
      \lineskip 7 pt \normallineskip 7 pt}
  \def\doublespacing{\parskip 5 pt plus 1 pt \baselineskip 25 pt
      \lineskip 13 pt \normallineskip 13 pt}
\begin{document}
\def\slash{\not{}{\mskip-3.mu}}
\def\ra{\rightarrow}
\def\lra{\leftrightarrow}
\def\bea{\begin{eqnarray}}
\def\ena{\end{eqnarray}}
\def\beq{\begin{equation}}
\def\enq{\end{equation}}
\def\cw{\cos \theta_W}
\def\sw{\sin \theta_W}
\def\tw{\tan \theta_W}
\def\eg{${\it e.g.}$}
\def\ie{${\it i.e.}$}
\def\etc{${\it etc}$}
\def\kln{\kappa_{L}^{NC}}
\def\krn{\kappa_{R}^{NC}}
\def\klc{\kappa_{L}^{CC}}
\def\krc{\kappa_{R}^{CC}}
\def\ttz{{\mbox {\,$t$-${t}$-$Z$}\,}}
\def\bbz{{\mbox {\,$b$-${b}$-$Z$}\,}}
\def\tta{{\mbox {\,$t$-${t}$-$A$}\,}}
\def\bba{{\mbox {\,$b$-${b}$-$A$}\,}}
\def\tbw{{\mbox {\,$t$-${b}$-$W$}\,}}
\def\tbW{{\mbox {\,$t$-${b}$-$W$}\,}}
\def\tltlz{{\mbox {\,$t_L$-$\overline{t_L}$-$Z$}\,}}
\def\blblz{{\mbox {\,$b_L$-$\overline{b_L}$-$Z$}\,}}
\def\brbrz{{\mbox {\,$b_R$-$\overline{b_R}$-$Z$}\,}}
\def\tlblw{{\mbox {\,$t_L$-$\overline{b_L}$-$W$}\,}}
\def\beq{\begin{equation}}
\def\enq{\end{equation}}
\def\pbarp{ \bar{{\rm p}} {\rm p} }
\def\pp{ {\rm p} {\rm p} }
\def\ipb{ {\rm pb}^{-1} }
\def\ifb{ {\rm fb}^{-1} }
\def\stds{\strut\displaystyle}
\def\SST{\scriptscriptstyle}
\def\TT{\textstyle}
\def\ra{\rightarrow}
\def\cro{\cropen{12pt}}
\def\mf{m_f}
\def\mb{m_b}
\def\mt{m_t}
\def\MW2{M^2_W}
\def\MZ{M_Z}
\def\Cw{C_w}
\def\Sw{S_w}
\def\mHn{m_{\SST H^{\SST 0}} }
\def\mh0{m_{\SST h^{\SST 0}} }
\def\mHp{m_{\SST H^{\pm}} }
\def\mA0{m_{\SST A^{\SST 0}} }
\def\mH{m_{\SST H} }
\def\mHs{m^2_{\SST H} }
\def\qq{q_1 \bar{q}_2}
\def\jj{j_1 j_2}
\def\ee{e^+ e^-}
\def\ptW{P_{\SST T}^{\SST W} }
\def\DR{\Delta R}
\def\fL{f_{\SST L}}
\def\yW{y_{\SST W}}
\def\qs{\theta^\ast}
\def\MWW{M_{\SST WW}}
\def\EWQCD{(1.1)~}
\def\eg{${\it e.g.}$}
\def\ie{${\it i.e.}$}
\def\etc{${\it etc}$}
\def\etal{${\it et al.}$}
\def\hatt{ \hat {\rm T} }
\def\ETslash{\not{\hbox{\kern-4pt $E_T$}}}
\def\mynot#1{\not{}{\mskip-3.5mu}#1}
\def\sss{\scriptscriptstyle}
\def\rtS{\sqrt{S}}
\def\ra{\rightarrow}
\def\d{{\rm d}}
\def\M {{\cal M}}   
\def\qgtb{q' g \ra q t \bar b}
\def\Wgtb{q' g (W^+ g) \ra q t \bar b}
\def\ubdt{q' b \ra q t}
\def\udbt{q' \bar q \ra W^* \ra t \bar b}
\def\Wbt{W^+ b \ra t}
\def\ggtt{q \bar q, \, g g \ra t \bar t}
\def\ttb{t \bar t}
\def\Wt{W t}
\def\gbtW{g b \ra W^- t}
\def\width{\Gamma(t \ra b W^+)}
\def\tevs{Di-TeV}
\def\flong{f_{\rm Long}}
\def\ltap{\;\centeron{\raise.35ex\hbox{$<$}}{\lower.65ex\hbox{$\sim$}}\;}
\def\gtap{\;\centeron{\raise.35ex\hbox{$>$}}{\lower.65ex\hbox{$\sim$}}\;}
\def\gsim{\mathrel{\gtap}}
\def\lsim{\mathrel{\ltap}}
\def\del{\partial }
\def\D0{D\O~}

\def\beq{\begin{equation}}
\def\enq{\end{equation}}
\def\bea{\begin{eqnarray}}
\def\ena{\end{eqnarray}}
\def\bec{\begin{center}}
\def\enc{\end{center}}
\def\eg{${\it e.g.}$}
\def\ie{${\it i.e.}$}
\def\etc{${\it etc}$}
\def\tbw{{\mbox {\,$t$-${b}$-$W$}\,}}
\def\tbW{{\mbox {\,$t$-${b}$-$W$}\,}}
\def\klc{{\kappa_{L}}^{CC}}
\def\krc{{\kappa_{R}}^{CC}}
\def\ra{\rightarrow}
\def\st{{\sin\theta}}
\def\ct{{\cos\theta}}
\def\sp{{\sin\phi}}
\def\cp{{\cos\phi}}
\def\M {{\cal M}}
\def\veps{{\varepsilon}}
\def\slash{\not{}{\mskip-3.mu}}
\newcommand{\myabs}[1]{{| {\vec{#1}} \, |}}
\newcommand{\br}[1]{{\langle {#1} |}}
\newcommand{\kt}[1]{{| {#1} \rangle}}
\newcommand{\bra}[2]{{\langle \, {\hat{#1}} \, {#2} |}}
\newcommand{\ket}[2]{{| \, {\hat{#1}} \, {#2} \rangle}}
\newcommand{\bk}[2]{{\langle \, {#1} | \, {#2} \rangle}}
\newcommand{\braket}[4]{{\langle \, {\hat{#1}} \, {#2} | \, {\hat{#3}} \, {#4} \rangle}}
\newcommand{\bsk}[5]{{\langle \, {\hat{#1}} \, {#2} | \slash {#3} \,
| \, {\hat{#4}} \, {#5} \rangle}}
\newcommand{\bssk}[6]{{\langle \, {\hat{#1}} \, {#2} | \slash {#3} \slash {#4} \, 
| \, {\hat{#5}} \, {#6} \rangle}}
\newcommand{\bsssk}[7]{{\langle \, {\hat{#1}} \, {#2} | \slash {#3} \slash {#4}\slash {#5} \, | \, {\hat{#6}} \, {#7} \rangle}}
\newcommand{\vect}[2]{\pmatrix{{#1}\cr{#2}\cr}}
\newcommand{\mat}[4]{\pmatrix{{#1}&{#2}\cr{#3}&{#4}\cr}}
\def\sone{\mat{0}{1}{1}{0}}
\def\stwo{\mat{0}{-i}{i}{0}}
\def\sthr{\mat{1}{0}{0}{-1}}
\def\gamz{\mat{0}{1}{1}{0}}
\def\gamj{\mat{0}{-\sigma_j}{\sigma_j}{0}}
\def\gam5{\mat{1}{0}{0}{-1}}
\def\Ppm{{1 \over 2} (1 {\pm} \gamma^5)}
\def\Pp{\mat{1}{0}{0}{0}}
\def\Pm{\mat{0}{0}{0}{1}}

\def\klc{{\kappa_{L}}^{CC}}
\def\krc{{\kappa_{R}}^{CC}}
\def\ra{\rightarrow}
\def\st{{\sin\theta}}
\def\ct{{\cos\theta}}
\def\sp{{\sin\phi}}
\def\cp{{\cos\phi}}
\def\M {{\cal M}}
\def\slash{\not{}{\mskip-3.mu}}
\def\gamz{\mat{0}{1}{1}{0}}
\def\gamj{\mat{0}{-\sigma_j}{\sigma_j}{0}}
\def\gam5{\mat{1}{0}{0}{-1}}
\def\Ppm{{1 \over 2} (1 {\pm} \gamma^5)}
\def\Pp{\mat{1}{0}{0}{0}}
\def\Pm{\mat{0}{0}{0}{1}}
\def\rts{\sqrt{s}}
\def\sh{\hat{s}}
\def\rsh{\sqrt{\hat{s}}}

\def\phip{e^{i\phi}}
\def\phipsq{e^{2i\phi}}
\def\phim{e^{-i\phi}}
\def\phimsq{e^{-2i\phi}}
\def\vtwo{\sqrt{2}\, }
\def\vmt{\sqrt{m_t}\, }
\def\vmttwo{\sqrt{2m_t}\, }
\def\vbp{\sqrt{E_b+p}\, }
\def\vbm{\sqrt{E_b-p}\, }
\def\epplus{(E_b+p)}
\def\epminus{(E_b-p)}
\def\ra{\rightarrow}
\def\Wtb{W-t-b}
\def\eg{{\it e.g.}}
\def\etc{{\it etc.}}
\def\Lag{L}
\def\MWW{M_{WW}}
\def\ie{{\it i.e.}}
\def\tow{{m_t \over M_W}}
\def\fol{f_1^L}
\def\for{f_1^R}
\def\cpyftl{f_2^L}
\def\cpyftr{f_2^R}
\def\ctt{\cos{\theta \over 2}}
\def\stt{\sin{\theta \over 2}}
\def\rt{\sqrt{2}}
\def\klc{{\kappa_{L}}^{CC}}
\def\krc{{\kappa_{R}}^{CC}}
\def\ra{\rightarrow}
\def\st{{\sin\theta}}
\def\ct{{\cos\theta}}
\def\sp{{\sin\phi}}
\def\cp{{\cos\phi}}
\def\M {{\cal M}}
\def\flong{f_{\rm Long}}
\def\veps{{\varepsilon}}
\def\slash{\not{}{\mskip-3.mu}}
\def\gamz{\mat{0}{1}{1}{0}}
\def\gamj{\mat{0}{-\sigma_j}{\sigma_j}{0}}
\def\gam5{\mat{1}{0}{0}{-1}}
\def\Ppm{{1 \over 2} (1 {\pm} \gamma^5)}
\def\Pp{\mat{1}{0}{0}{0}}
\def\Pm{\mat{0}{0}{0}{1}}

\def\klc{{\kappa_{L}}^{CC}}
\def\krc{{\kappa_{R}}^{CC}}
\def\ra{\rightarrow}
\def\st{{\sin\theta}}
\def\ct{{\cos\theta}}
\def\sp{{\sin\phi}}
\def\cp{{\cos\phi}}
\def\M {{\cal M}}
\def\ugtb{q' g \ra q t \bar b}
\def\gutb{g q' \ra q t \bar b}
\def\ubdt{q' b \ra q t}
\def\budt{b q' \ra q t}
\def\veps{{\varepsilon}}
\def\slash{\not{}{\mskip-3.mu}}

\def\ra{\rightarrow}
\def\st{{\sin\theta}}
\def\ct{{\cos\theta}}
\def\sp{{\sin\phi}}
\def\cp{{\cos\phi}}
\def\M {{\cal M}}

\pagestyle{plain}
\pagenumbering{roman}
\doublespacing
 
%
%
\pagestyle{empty}

\begin{flushright}
MSUHEP--050727
\end{flushright}
\begin{center}
{\large PHYSICS OF SINGLE-TOP QUARK PRODUCTION AT HADRON COLLIDERS}
\end{center}
\vspace{.5in}
\begin{center}
By

Douglas Olaf Carlson\\
\vspace{.35in}
A DISSERTATION\\
\vspace{.35in}
Submitted to\\
Michigan State University \\
in partial fulfillment of the requirements \\
for the Degree of \\
\vspace{.35in}
DOCTOR OF PHILOSOPHY \\
\vspace{.6in}
Department of Physics and Astronomy
\end{center}
\begin{center}
1995
\end{center}



\newpage
\pagestyle{empty}
\begin{center}
ABSTRACT \\ [15pt]

{PHYSICS OF SINGLE-TOP QUARK PRODUCTION AT HADRON COLLIDERS} \\ [30pt]

By \\ [12pt]

Douglas Olaf Carlson \\ [30pt]

\end{center}
\def\tevs{Di-TeV}

We discuss the physics of single-top quark production and decay
at hadron colliders, such as the Tevatron, the \tevs~and the LHC.
Our study includes how to measure the mass and the width of the
top quark produced from a {\hbox{single-$t$}} or a
single-$\bar t$ process.
We also show how to probe new physics by studying the couplings
of $t$-${b}$-$W$ and show what can be improved from measuring
the production rate of single-top events. We also discuss how to probe 
CP properties of the top quark by measuring the single-top production
rate.  Finally, we present a Monte Carlo study on the detection of 
single-top events in hadron collisions.

\newpage


\pagestyle{plain}
\pagenumbering{roman}
\vspace{4in}
\begin{center}
For Dawn.
\end{center}
\setcounter{page}{3}
\newpage

\begin{center} {\large ACKNOWLEDGEMENTS}  \end{center}

\bigskip

I would like to express my deepest gratitude to the people who have contributed
to me and to this work:

To my thesis advisor, C.--P. Yuan, for his calm guidance over the past 
three years;

To my mentor, Chip Brock, for sparking my interest in this field and
introducing me to C.--P.;

To the members of my Thesis Committee:  Chip Brock, Wu--Ki Tung,
S. D. Mahanti and Horace Smith, for their carefull reading of the 
manuscript;

To my colleagues in the High Energy Theory group:  Glenn Ladinsky
and Pankaj Agrawal for many illuminating discussions; and my graduate
student colleagues:  Ehab Malkawi, Mike Wiest, Csaba Balasz, Liang--Hung
Lai and Xiaoning Wong for our lunch gatherings;

To my wife, Dawn, for her love and support;

To my parents, Jerry and Justine, for everything.

\newpage


\tableofcontents
\clearpage
\singlespacing

\addcontentsline{toc}{chapter}{LIST OF TABLES}
\listoftables
\clearpage

\addcontentsline{toc}{chapter}{LIST OF FIGURES}
\listoffigures

\singlespacing
   \pagestyle{plain}
   \pagenumbering{arabic}
   \setcounter{page}{1}
   \makeatletter
   \def\@evenfoot{}
   \def\@evenhead{\hfil\thepage\hfil}
   \def\@oddhead{\@evenhead}
   \def\@oddfoot{\@evenfoot}
   \makeatother

\chapter{Introduction to the Standard Model}

The Standard Model (SM) of elementary particle physics
\cite{Griffiths,HalzenMartin,MandlShaw,Renton} is a 
Yang--Mills gauge field theory with symmetry 
\beq
{\rm SU(3)}_{C}\times {\rm SU(2)}_{L}\times {\rm U(1)}_{Y}. 
\label{sm}
\enq
It has been
very successful in explaining and predicting experimental data.
The ${\rm SU(3)}_{C}$ sector governs the strong force of the SM 
and is known as quantum
chromodynamics (QCD). The ${\rm SU(2)}_{L}\times {\rm U(1)}_{Y}$ sector
unifies the electromagnetic and weak forces, collectively known
as the electroweak force.
The unification of the electromagnetic and weak forces
is accomplished in the SM via the mixing of the neutral
${\rm SU(2)}_{L}$ gauge boson and the hypercharge gauge boson of 
${\rm U(1)}_{Y}$.
Masses are introduced in a gauge invariant way through spontaneous 
symmetry breaking which gives rise to the as yet undiscovered Higgs
boson ($H$). This process of spontaneous
symmetry breaking is known as the Higgs mechanism.

The standard model does not incorporate the gravitational force.
So far, no definitive quantum field theory of gravity
exists. Gravity affects all massive particles, however,
gravitational interactions with elementary particles are
too weak and can be ignored.

In the standard model there are three
generations of leptons and quarks as listed in Table~\ref{fm}. 
Associated with each force is one or several gauge bosons as
listed in Table~\ref{bm}. For reference we list the masses
of each particle as found in Reference~\cite{databook}.
So far, the only particle for which a 
discovery is lacking is the scalar Higgs boson. 

In what follows, we briefly describe the particle spectrum, 
particle interactions and the 18 independent parameters which 
constitute the standard model.

\begin{table}
\caption{Lepton and Quark Masses}
\label{fm}
\begin{center}
\vspace{.5cm}
\begin{tabular}{lccc}
  Particle          &  Symbol    & Mass (GeV) &\\ \hline \hline \\
  Electron neutrino & $\nu_e$    & 0          &\\
  Electron 	    & $e$        & 0.00051    & First\\
  Up quark 	    & $u$	 & 0.002 to 0.008 & Generation\\
  Down quark        & $d$        & 0.005 to 0.015 &\\ \\
  Muon neutrino     & $\nu_\mu$  & 0          &\\
  Muon 	            & $\mu$      & 0.106      & Second\\
  Charm quark 	    & $c$	 & 1.3 to 1.7 & Generation\\
  Strange quark     & $s$        & 0.1 to 0.3 &\\ \\
  Tau neutrino      & $\nu_\tau$ & 0          &\\
  Tau 	            & $\tau$     & 1.78       & Third\\
  Top quark 	    & $t$	 & 174        & Generation\\
  Bottom quark      & $b$        & 4.7 to 5.3 & \\ \\ \hline\hline
\end{tabular}
\end{center}
\end{table}

\begin{table}
\caption{Boson Masses}
\label{bm}
\begin{center}
\vspace{.5cm}
\begin{tabular}{lccl}
  Particle &  Symbol   & Mass (GeV)       & \\ \hline \hline \\
  Photon   & $\gamma$  & 0                & Electromagnetic Force\\
  W Boson  & $W^{\pm}$ & 80.22            & Charged Weak Force\\
  Z Boson  & $Z^0$     & 91.187           & Neutral Weak Force\\
  Gluon    & $G$       & 0                & Strong Force\\ \\
  Higgs    & $H$       & $60 < m_H < 800$ & Spontaneous Symmetry Breaking
\\ \\ \hline\hline
\end{tabular}
\end{center}
\end{table}

\section{ The Matter Spectrum of Spin-$1\over2$ Fermions }
\begin{table}
\caption{Quantum numbers of the fermion spectrum}
\label{qn}
\begin{center}
\vspace{.5cm}
\begin{tabular}{lcccc}
Chirality   & $Q$ &$T^3_{\rm{W}}$& $Y$ & $C$\\ \hline \hline \\
${\nu_e}_L$ &  0  &  1/2 & -1   & 0        \\
$e_L$       & -1  & -1/2 & -1   & 0        \\ \\
${u}_L$     & 2/3 &  1/2 & 1/3  & $r,g,b$  \\
$d_L$       &-1/3 & -1/2 & 1/3  & $r,g,b$  \\ \\
$e_R$       & -1  &   0  & -2   & 0        \\ \\
$u_R$       & 2/3 &   0  & 4/3  & $r,g,b$  \\ \\
$d_R$       &-1/3 &   0  &-2/3  & $r,g,b$  \\ \\ \hline\hline
\end{tabular}
\end{center}
\end{table}
The matter spectrum consists of twelve fermions
which are organized in Table~\ref{fm} 
according to the symmetry structure of Equation~(\ref{sm}).
Each successive generation is a more massive copy of the previous
generation, so only the quantum numbers of the first generation
are shown in Table~\ref{qn}.
Under the ${\rm SU(2)}_{L}$ sector of the SM,
left-handed fermions transform as weak isospin ($T_{\rm{W}}$) doublets, 
\beq
\ell_L = {\pmatrix{{\nu_e}\cr{e}\cr}}_L, \quad
q_L = {\pmatrix{{u}\cr{d}\cr}}_L, 
\label{doublet}
\enq
whereas right-handed fermions transform as singlets,
\beq(e)_R,\quad (u)_R,\quad (d)_R.
\enq
Since neutrinos are massless Dirac fermions in the SM, 
there are no right-handed neutrinos.
Once the third component of Weak isospin 
$T^3_{\rm{W}}$ is assigned, the values of hypercharge
$Y$ can be determined to
cancel the chiral anomalies~\cite{donoghue}. 
With these quantum numbers in place, the charge quantization is determined by
\beq
Q = T^3_{\rm{W}} + {Y\over 2}.
\enq

Only quarks transform under the ${\rm SU(3)}_{C}$ sector of the SM. Each quark
flavor, (\ie, $q = u, d, s, c, b, t$), carries $red(r), green(g)$ or $blue(b)$
color charge and therefore transforms as a triplet:
\beq
\Psi_q = {\pmatrix{{q_r}\cr{q_g}\cr{q_b}\cr}}. 
\enq

\section{ Force Mediators as Spin-1 Gauge Bosons }
Associated with the group structure of the SM are twelve generators
and each generator is associated with a gauge boson. Therefore,
${\rm U(1)}_{Y}$ has one generator, the neutral hypercharge gauge boson
$B_\mu$. ${\rm SU(2)}_{L}$ has three ($2 \times 2 - 1 = 3$) generators,
two of which are charged ${\rm SU(2)}_{L}$ gauge bosons $W_\mu^{\pm}$ and
one neutral ${\rm SU(2)}_{L}$ gauge boson $W_\mu^3$. Finally, ${\rm SU(3)}_{C}$
has eight ($3 \times 3 - 1 = 8$) generators and therefore, eight gluons,
$G_\mu^a, a=1,2,\ldots,8$.

As stated earlier, electroweak unification is accomplished by mixing
the $B_\mu$ and $W_\mu^3$ gauge bosons.  Formally this is accomplished
via the following rotation,
\beq
\ \pmatrix{{Z^0_\mu}\cr {A_\mu}\cr} = \ \pmatrix {{\cw} & {-\sw}\cr
{\sw} & {\cw}\cr}\ \pmatrix {{W_\mu^3}\cr {B_\mu}\cr},
\label{mix}
\enq
where $\theta_W$, called the weak mixing angle, 
is chosen such that $A_\mu$ only couples with
charged particles. $A_\mu$ is then identified as the photon field
in quantum electrodynamics (QED)
and additionally, a neutral weak force, $Z^0_\mu$ , is obtained.

At this point in the theory, all fermions and
gauge bosons are massless to preserve
gauge invariance.  In the next two sections we describe how the force mediators
interact with fermions and amongst themselves.  The Higgs mechanism
is also introduced to incorporate mass in a gauge invariant way.

\section{ Gauge Invariant Interactions of Fermion and Gauge Boson Fields }
We begin with the lagrangian for a massless free fermion field $\Psi$,
\bea
{\cal{L}}_{FK} = \overline{\Psi} \, i\!\!\! \slash{\partial}\,\Psi .
\label{lfk}
\ena
Equation~(\ref{lfk}) is the kinetic term for fermion fields.\footnote{
$\slash{\partial} = \partial_{\mu} \gamma^{\mu}$ where $\partial_{\mu}$ is the
Lorentz invariant space-time derivative and $\gamma^{\mu}$ are the
Dirac matrices. }
To make the lagrangian gauge invariant, we introduce the gauge covariant
derivative
\bea
\partial_{\mu} \ra D_{\mu} = \partial_{\mu} - i g_1 {Y\over 2} B_{\mu} -
i g_2 {\tau^j\over 2} W^j_{\mu} - i g_3 {\lambda^a\over 2} G^a_{\mu}, 
\label{cov}
\ena
where $j = 1,2,3$ and $a = 1,2,\ldots,8$.
The $\tau^j$'s are the Pauli matrices defined in Appendix~A 
and $T^j_W = {\tau^j\over 2}$ is the Weak isospin. 
The ($3\times 3$) matrices (generators)
$\lambda^a, a=1,2,\ldots,8$, are the ${\rm SU(3)}_{C}$ matrices.
As a result, we obtain gauge interaction terms in the lagrangian.

The $B_{\mu}$ term acts on all fields with different $Y$'s, including
leptons and quarks in Table~\ref{qn} (same for the other two generations)
and the Higgs doublet field $\Phi$
discussed in Section~1.5 below.  The $W^j_{\mu}$ term acts only on 
the ${\rm SU(2)}_{L}$ doublets with non-zero $T^3_{\rm W}$ and 
the field $\Phi$. 
In the
process of obtaining electroweak unification the coupling constants,
$g_1$ and $g_2$ become related through the weak mixing angle $\theta_W$
via 
\bea
{g_2\over g_1} = \tw,
\ena
where $g_1 \sw = e$.
The value of $e$ is related to the fine structure constant $\alpha$
by $\alpha = e^2/{4 \pi}$.

For electroweak interactions, the colors of the quarks have to be the same
since ${\rm SU(2)}_{L}\times {\rm U(1)}_{Y}$ does not act on color space. 
Also, the $A_\mu$ and $Z^0_\mu$ fields do not induce quark or lepton flavor
changing at the Born level, (\ie\, no $u\lra c\lra t$, $d\lra s\lra b$,
${\nu_e}\lra {\nu_{\mu}}\lra {\nu_{\tau}}$ 
or $e\lra \mu\lra \tau$ transitions occur).
However, charged current transitions via $W_\mu^{\pm}$ do occur and 
for quarks, are not restricted by generation (\ie\, 
$u, c, t\lra d, s, b$ transitions are allowed). Lepton flavor changes are 
restricted to generation due to massless neutrinos.  Finally,
the SM does not allow for direct lepton-quark transitions.

The $G^a_{\mu}$ term in Equation~(\ref{cov}) only acts on colored
fermions, \ie\, quarks. The coupling strength $g_3 = g_S$ is 
universal for all colored quarks. Analogous to the fine structure constant
$\alpha$ in electromagnetic theory is the strong coupling constant
$\alpha_S = g_S^2/{4 \pi}$ in QCD. Here again, no flavor changes occur.

\section{ Gauge Field Self Interactions }
To complete the lagrangian for massless fermions and gauge bosons and their
interactions, we must introduce the kinetic term for gauge bosons,
\bea
{\cal{L}}_{GK} = -{1\over 4} {\cal{B}}_{\mu \nu} {\cal{B}}^{\mu \nu}
       -{1\over 4} {\cal{W}}^i_{\mu \nu} {\cal{W}}^{i \mu \nu}
       -{1\over 4} {\cal{G}}^a_{\mu \nu} {\cal{G}}^{a \mu \nu},
\ena
where
\beq
{\cal{B}}_{\mu \nu}=\partial _\mu B_\nu -\partial _\nu B_\mu,
\enq
\beq
{\cal{W}}^i_{\mu \nu}=\partial _\mu W^i_\nu -\partial _\nu W^i_\mu
+ g_2 \epsilon ^{ijk} W^j_\mu W^k_\nu,
\enq
\beq
{\cal{G}}^a_{\mu \nu }=\partial _\mu G^a_\nu -\partial _\nu G^a_\mu
+ g_S f^{abc} G^b_\mu G^c_\nu.
\enq
The Lie group structure constants $\epsilon ^{ijk}$ and $f^{abc}$
are defined through the following anti-commutation relations,
\bea
{\left[{\tau^i\over 2}\, ,{\tau^j\over 2} \right]} = 
i\, \epsilon^{ijk}\, {\tau^k\over 2},\quad i,j,k = 1,2,3
\ena
and
\bea
{\left[{\lambda^a\over 2}\, ,{\lambda^b\over 2} \right]} = 
i\, f_{abc}\, {\lambda^c\over 2},\quad a,b,c = 1,2,\ldots,8
\ena
for ${\rm SU(2)}$ and ${\rm SU(3)}$, respectively.

We note that the pure Yang-Mills terms, 
$-{1\over 4} {\cal{W}}^i_{\mu \nu} {\cal{W}}^{i \mu \nu}$
and 
$-{1\over 4} {\cal{G}}^a_{\mu \nu} {\cal{G}}^{a \mu \nu}$ 
contain factors
that are trilinear and quadrilinear in $W^i_{\mu}$ and 
$G^a_{\mu}$.
These Yang-Mills terms expand out partially as
\bea
\cdots -g_2\, \epsilon^{ijk}\, (\partial_{\mu}{W^i_\nu})W^{j \mu}
W^{k \nu} - {g_2^2\over 4}\, \epsilon^{ijk}\, \epsilon^{ilm}\, 
{W^j_\mu}{W^k_\nu} W^{l \mu} W^{m \nu}
\ena
and
\bea
\cdots -g_S\, f^{abc}\, (\partial_{\mu}{G^a_\nu}) G^{b \mu}
G^{c \nu} - {g_S^2\over 4}\, f^{abc}\, f^{ade}\, {G^b_\mu}{G^c_\nu}
G^{d \mu} G^{e \nu}
\ena
respectively, 
and  correspond to self-couplings of non-abelian gauge fields.
This is fundamentally different than in the abelian case where, in
QED, photons do not directly couple with photons. It also accounts for
the short distance interaction of the strong force, despite
the fact that the gluon is massless. As can be seen in Chapter~2,
the triple gluon interaction contributes to $t \bar{t}$ production
via $gg \ra t \bar{t}$ at hadron colliders.

\section{ The Higgs Mechanism }
The goal of the Higgs mechanism is to introduce mass to the particles in the
SM in a gauge invariant way.  We begin by defining a complex doublet scalar
field $\Phi$ composed of four real scalar fields $H, \phi^0, \phi^1$ and
$\phi^2$ where
\bea
\Phi = {1\over \sqrt{2}} {\pmatrix{{v + H + i\phi^0}\cr
				  {i\phi^1 - \phi^2}\cr}}
     =                   {\pmatrix{{{v + H + i\phi^0}\over \sqrt{2}}\cr
				   {i\phi^-}\cr}}.
\ena
The quantum numbers of the Higgs field are as shown in Table~\ref{Hqn}.
\begin{table}
\caption{Quantum numbers of the Higgs doublet}
\label{Hqn}
\begin{center}
\vspace{.5cm}
\begin{tabular}{ccccc}
                         & $Q$ &$T^3_{\rm{W}}$& $Y$ & $C$\\ \hline \hline \\
${{v + H + i\phi^0}\over \sqrt{2}}$ & 0 & 1/2 &-1 & 0 \\ \\
$i \phi^-$                          &-1 &-1/2 &-1 & 0 \\ \\ \hline\hline
\end{tabular}
\end{center}
\end{table}

The fields $\phi^0$ and $\phi^{\pm} = (\phi^1 \mp i\phi^2)/\sqrt{2}$ are the
unphysical 
would-be Goldstone bosons associated with spontaneous symmetry breaking.
They give rise to the masses of the gauge bosons $W^\pm$ and $Z^0$.
One physical field thus remains, which is the Higgs field $H$.  The constant
$v\simeq 246$ GeV is the scale characterizing the symmetry breaking scale
and is called the vacuum expectation value of $\Phi$,  where
\beq
\langle \Phi \rangle_{0} \equiv \langle 0 | \Phi | 0 \rangle = 
\vect{{v\over\sqrt{2}}}{0}.
\enq

The lagrangian for the Higgs sector is
\beq
{\cal L}_{\Phi}={({D_\mu} \Phi )^{\dagger}}({D^\mu} \Phi )
-{\lambda \over 2}({\Phi^{\dagger}} \Phi )^2 - 
\mu ^2({\Phi^{\dagger}} \Phi ),
\enq
with
\beq
{D_\mu} \Phi = \left[{\partial_{\mu} - i g_1 {Y\over 2} B_{\mu} -
i g_2 {\tau^j\over 2} W^j_{\mu}}\right] \Phi
\enq 
as in Equation~(\ref{cov}) without the gluon interaction.
 
If $\mu < 0$ and $\lambda > 0$ then the minimum of the potential
energy occurs at 
\beq
v = \sqrt{{-2\mu}\over \lambda}.
\enq
The Goldstone bosons $\phi^{\pm}$ and $\phi^0$
are ``eaten'' by the vector bosons $W^{\pm}$ and $Z^0$, respectively,
where 
\bea
M_W = {1\over2} g_2 v\:\: {\rm and}\:\: M_Z = {M_W\over\cw}.
\ena
Therefore, $W^{\pm}$ and $Z^0$ have three polarization states;
two transverse and one longitudinal. The massless photon and
gluon have only the two transverse polarization states.  The Higgs mass
is given by
\beq
m_H = v \sqrt{\lambda}.
\enq

To introduce fermion mass in a gauge invariant way one introduces the
Higgs mechanism through Yukawa coupling interactions. For the first
generation
\bea
\lefteqn{{\cal L}_{Yukawa} =
{{\sqrt{2} m_u}\over v}\,(\bar{u}_L\,\bar{d}_L)\,\Phi\,u_R +
{{\sqrt{2} m_d}\over v}\,(\bar{u}_L\,\bar{d}_L)\,(-i\tau_2\Phi^*)\,d_R} 
\nonumber \\ 
& &+ {{\sqrt{2} m_e}\over v}\,(\bar{\nu}_L\,\bar{e}_L)\,(-i\tau_2\Phi^*)\,e_R +
\rm{hermitian\,\,conjugate}
\ena
where $m_u$, $m_d$ and $m_e$ are up quark, down quark and electron masses,
respectively (neutrinos are massless) and
\bea
-i\tau_2\Phi^* = \mat{0}{-1}{1}{0}\,\Phi^* = 
\vect{i\phi^+}{{{v + H + i\phi^0}\over \sqrt{2}}}.
\ena

As mentioned in Section~1.3, neutral currents coupled to $\gamma$, $Z^0$
and $G$ do not change flavor, although $G$ changes color, but charged
currents coupled to $W^{\pm}$ do change flavor.  For leptons, the flavor
change does not exceed generational bounds, due to the massless neutrino.
However, there is a chance that an up quark, for instance, can change to
a down quark, a strange quark or even a bottom quark.  This is called
quark mixing, which is due to the weak eigenstates 
(indicated by the subscript ``Weak'') of quarks being 
different than the mass eigenstates (indicated by the subscript ``Mass''). 

By convention, the three charge 2/3 quarks $u$, $c$ and $t$ are unmixed:
\bea
\pmatrix{{u}\cr{c}\cr{t}\cr}_{\rm Weak} = 
\pmatrix{{u}\cr{c}\cr{t}\cr}_{\rm Mass}.
\ena
All the mixing is therefore expressed in terms of a ($3\times 3$) 
unitary matrix $V$ operating on the charge (-1/3) quarks $d$, $s$ and $b$:
\bea \lefteqn{
\pmatrix{{d}\cr{s}\cr{b}\cr}_{\rm Weak} \equiv 
\pmatrix{{V_{ud}}&{V_{us}}&{V_{ub}}\cr{V_{cd}}&{V_{cs}}&{V_{cb}}\cr
{V_{td}}&{V_{ts}}&{V_{tb}}\cr}\,
\pmatrix{{d}\cr{s}\cr{b}\cr}_{\rm Mass}} \nonumber \\ \nonumber \\ & & \approx
\pmatrix{{1-{\lambda^2}/2}&{\lambda}&{A{\lambda^3}(\rho-i\eta)}\cr
        {-\lambda}&{1-{\lambda^2}/2}&{A{\lambda^2}}\cr
        {A{\lambda^3}(1-\rho-i\eta)}&{-A{\lambda^2}}&{1}\cr}\, 
\pmatrix{{d}\cr{s}\cr{b}\cr}_{\rm Mass}.
\ena
The matrix $V$ is known as the Cabibbo--Kobayashi--Maskawa matrix (CKM), which
consists of three mixing angles and one phase. The second parameterization
is due to Wolfenstein~\cite{wolfy}, where $\lambda \simeq 0.22$, 
$A \simeq 1$, $\eta \simeq 0.5$ and $-0.4 \leq \rho \leq 0.2$.
$CP$ violation, which is the violation of combined charge conjugation
$C$ and the parity transformation $P$, is characterized by the 
$CP$ violating phase in $\rho - i\eta$.

\section{ Review }
In this introduction to the SM, we have outlined how the SM is constructed
based on gauge invariance and the Higgs mechanism.  Although the SM
has been successful in describing experimental data, there are 18
independent parameters which must be determined experimentally.
These parameters are:
\begin{itemize} 
\item nine fermion masses: $m_e, m_\mu, m_\tau, m_u, m_d, m_c, m_s, m_t,
m_b$ (neutrinos are massless),
\item four CKM parameters: $\lambda, A, \eta, \rho$,
\item four electroweak parameters: $e, \theta_W, M_W, m_H$,
\item the strong coupling constant: $\alpha_S$.
\end{itemize}

\chapter{ Introduction to the Top Quark }

If the ${\rm SU}(2)$ structure of the Standard Model (SM) holds, 
the top quark ($t$) has to exist as
the weak isospin partner of the bottom quark ($b$) \cite{kane}.
If the coupling of \ttz~ is as predicted in the SM, then
from  LEP and SLAC experiments
the mass of the top quark ($m_t$) has to be larger
than half of the $Z$-boson mass ($\sim 45$\,GeV)
 independent of how the top quark decays. 
If the coupling of \tbw~ is as predicted in the SM, then
from the measurement of the total width of the
$W$--boson, by measuring the ratio of the event rates of 
$\pbarp \ra W ( \ra \ell \nu)$ to $\pbarp \ra Z ( \ra \ell^+ \ell^-)$
\footnote{A proton is denoted by $\rm p$ and an anti-proton by $\bar{\rm p}$},
the mass of the top quark ($m_t$) has to be larger
than 62 GeV independent 
of how the top quark decays \cite{wwidth}. 
From examining the radiative corrections to low energy observables, 
such as the $\rho$ parameter\footnote{ 
$\rho= \frac{M^2_W}{M^2_Z\,{\cos}^2 \theta_W}$, where $M_W$ (or $M_Z$) is
the mass of $W^\pm$ (or $Z$) boson.
$\theta_W$ is the weak mixing angle.
$\rho$ has been measured to the accuracy of about 0.1\%.
}  
which is proportional to $m_t^2$ at the one 
loop level \cite{tinirho}, $m_t$ has to be less than about 200\,GeV.  
Based upon analysis of a broad range of Electroweak data, 
the mass of the SM top quark is expected to be in the 
vicinity of 150 to 200 GeV \cite{mele,alta,lep94,sld94}.
Independently, from the direct search at the Tevatron, 
the top quark has been discovered and found to have mass of 
$m_t = 176 \pm 8 \, {\rm (stat.)} \pm 10 {\rm (sys.)}$\,GeV 
from CDF data \cite{CDFft}, and
$m_t = 199^{+29}_{-21}\, {\rm (stat.)} \pm 22 {\rm (sys.)}$\,GeV 
from \D0 data \cite{D0ft}.

For a heavy top quark, $m_t$ is of the order of the 
electroweak symmetry breaking 
scale $v={(\sqrt{2}G_F)}^{-1/2}=246$\,GeV. In fact, recall the
Yukawa coupling interaction, this time for third generation quarks,
\bea
{\cal L}_{Yukawa} =
{{\sqrt{2} m_t}\over v}\,(\bar{t}_L\,\bar{b}_L)\,\Phi\,t_R +
{{\sqrt{2} m_b}\over v}\,(\bar{t}_L\,\bar{b}_L)\,(-i\tau_2\Phi^*)\,b_R.
\ena
We see that 
\bea
{{\sqrt{2} m_t}\over v} \sim 1,
\ena 
for $m_t = 175$ GeV.
Because the generation of fermion mass can be closely related to the 
electroweak symmetry breaking \cite{pczh,sekh}, 
effects from new physics should be more apparent in the 
top quark sector than any other light sector of the electroweak theory.  
Thus, the top quark system may be used to probe the symmetry breaking sector.
A few examples were discussed in Ref.~\cite{ehab} to illustrate that 
different models of electroweak symmetry breaking mechanism will induce 
different interactions among the top quark 
and the $W$-- and $Z$--bosons. Therefore, hopefully through studying 
the top quark system one may
eventually learn about the symmetry breaking sector of the electroweak
theory. 

The most important consequence of a heavy top quark
is that to a good approximation it decays as a free quark because its
lifetime is short and it does not have time to
bind with light quarks before it decays \cite{decay}.  
Furthermore, because the heavy top quark
has the weak two-body decay $t \ra b W^+$, it will
analyze its own polarization.
Thus we can use the polarization
properties of the top quark
as additional observables to test the SM and 
to probe new physics.
In the SM, the heavy top quark produced from the usual 
QCD process, at the Born level, is unpolarized. 
However, top quarks will have longitudinal
polarization if weak effects are present in their production \cite{qqttew}.
For instance, the top quark produced from the $W$--gluon 
fusion process is left-hand polarized. 
With a large number of top quark events, it
will be possible to test the polarization effects of top quarks.

\begin{figure}
\centerline{\hbox{\psfig{figure=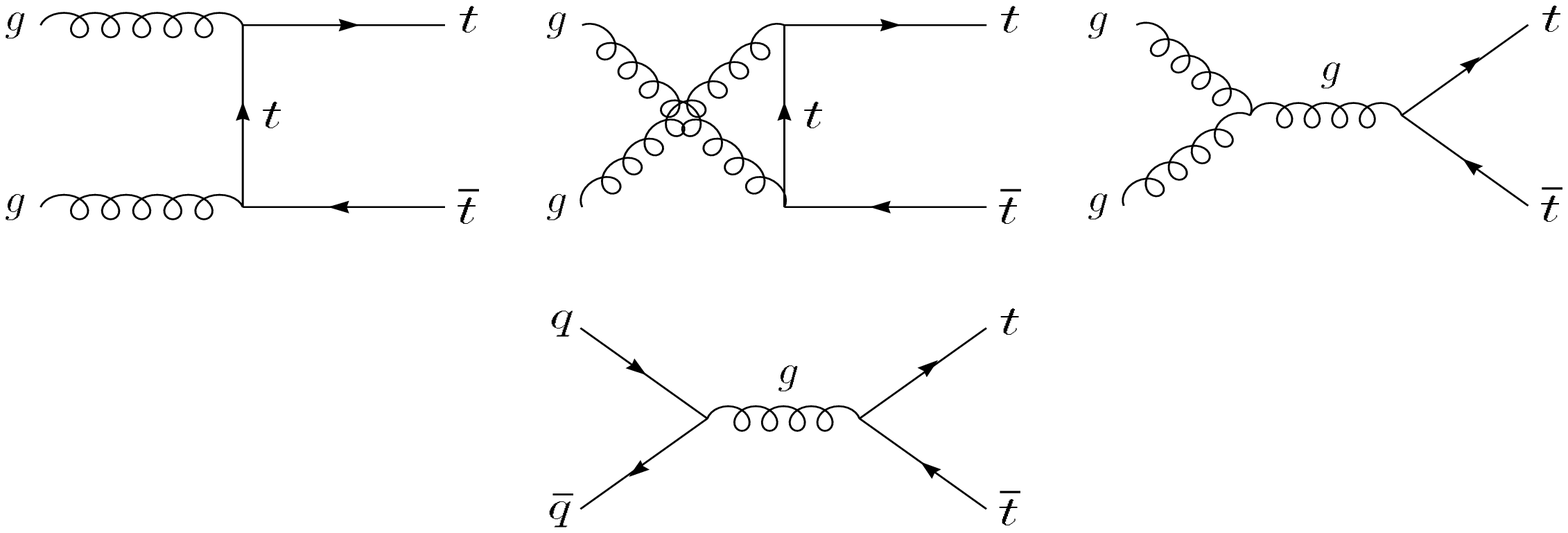,height=2.in}}}
\caption{ Diagrams contributing to the QCD production of $\ggtt$ }
\label{ttbAll}
\end{figure}
How to detect a SM top quark pair produced via the QCD processes 
$\ggtt$, as shown in Figure~\ref{ttbAll},
has been extensively studied in the literature \cite{argon}.
In this paper we will concentrate on how to detect and study 
the top quark produced from the single-top quark processes
$\Wgtb$, $\ubdt$, $\gbtW$, and $\udbt$.
For the single-top production we will only consider the decay mode of 
$t \ra b W^+ \ra b \ell^+ \nu$, with $\ell^+=e^+ \,{\rm or}\,\nu^+$.
(The branching ratio for this decay mode is
${\rm Br}=\frac{2}{9}$.) 

The rest of this paper is organized as follows.
In Chapter~3 we  discuss the production rates of top quarks at hadron
colliders. Following that, we will discuss in Chapters~4 and~5, 
respectively, how to measure the mass and
the width of the top quark.
In Chapter~6 we discuss what we have learned about 
the couplings of the top quark to the weak gauge bosons and show what 
can be improved from measuring the production rate of single-top quark events.
We will also discuss in Chapter~7 the potential of the Tevatron
as a $\pbarp$ collider
to probe CP properties of the top quark by simply measuring the  
single-top quark production rate.  
Finally, in Chapter~8 we present a Monte Carlo study on the detection of 
a single-top quark event in hadron collisions.
Various unique features of the kinematics of the
 single-top quark signal 
will be discussed. Chapter~9 contains our conclusions.
Based upon the results of the FNAL CDF and \D0, the mass of 
the SM top quark $m_t$ is not likely to be lighter than,
say, 140\,GeV. 
Throughout this paper we will use $\mt = 180$\,GeV (or $140$\,GeV) 
as an example of a heavy (or a light) top quark for our studies.

\chapter{ The Single-Top Production Mechanism}

In this chapter we consider the production rate of 
a single-top quark 
at the Tevatron, the \tevs~(the upgraded Tevatron)
and the LHC (Large Hadron Collider) colliders.
In referring to single-top production, unless stated otherwise, 
we will concentrate only on the positive charge mode 
({\it i.e.}, only including single-$t$, but not single-$\bar t$).
The colliders we consider are the Tevatron 
(a $\pbarp$ collider) with the Main Injector at
$\rtS = 2\,$TeV, the Di-TeV (a $\pbarp$ collider) at $4\,$TeV 
and the LHC (a $\pp$ collider) at $\rtS = 14\,$TeV 
with an integrated luminosity of
1\,$\ifb$, 10\,$\ifb$, and 100\,$\ifb$, respectively.\footnote{
In reality, the integrated luminosity can be higher than the 
ones used here. For instance, with a couple of years of running 
a $2\,$TeV Tevatron can accumulate, say, 10\,$\ifb$ luminosity.
Similarly, it is not out of question to have a $4\,$TeV Di-TeV 
to deliver an integrated luminosity of about 100\,$\ifb$.}

\begin{figure}
\centerline{\hbox{
\psfig{figure=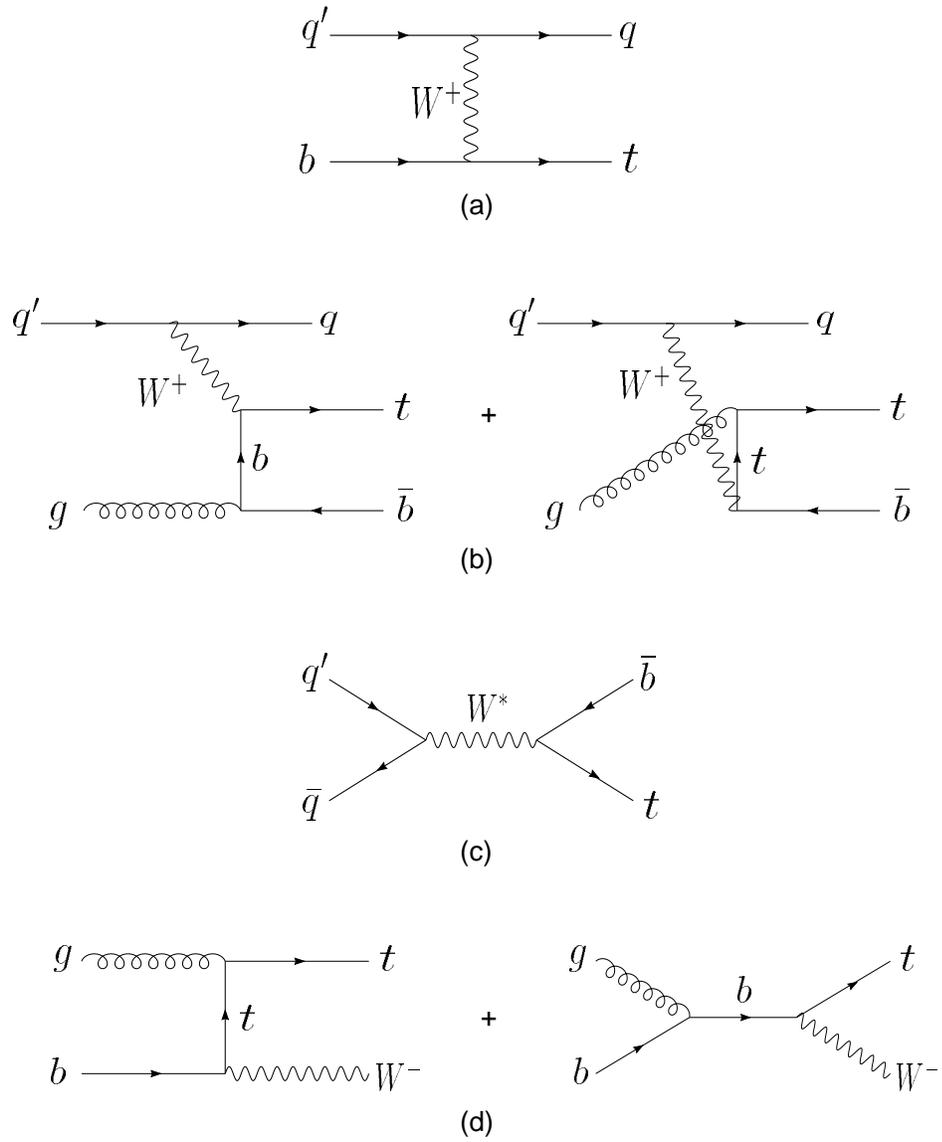,height=6in}}}
\caption{ Diagrams for various single-top quark processes.}
\label{newdiag}
\end{figure}

A single-top quark signal can be produced from either the $W$--gluon 
fusion process $\Wgtb$ (or $\ubdt$) \cite{sally,wgtb}, 
the Drell-Yan type process $\udbt$ (also known as
``$W^*$'' production) \cite{cortese}, 
or $\Wt$ production via $\gbtW$ \cite{galwt}.
The corresponding Feynman diagrams for these processes are shown in
Figure~\ref{newdiag}.

\begin{figure}[p]
\centerline{\hbox{
\psfig{figure=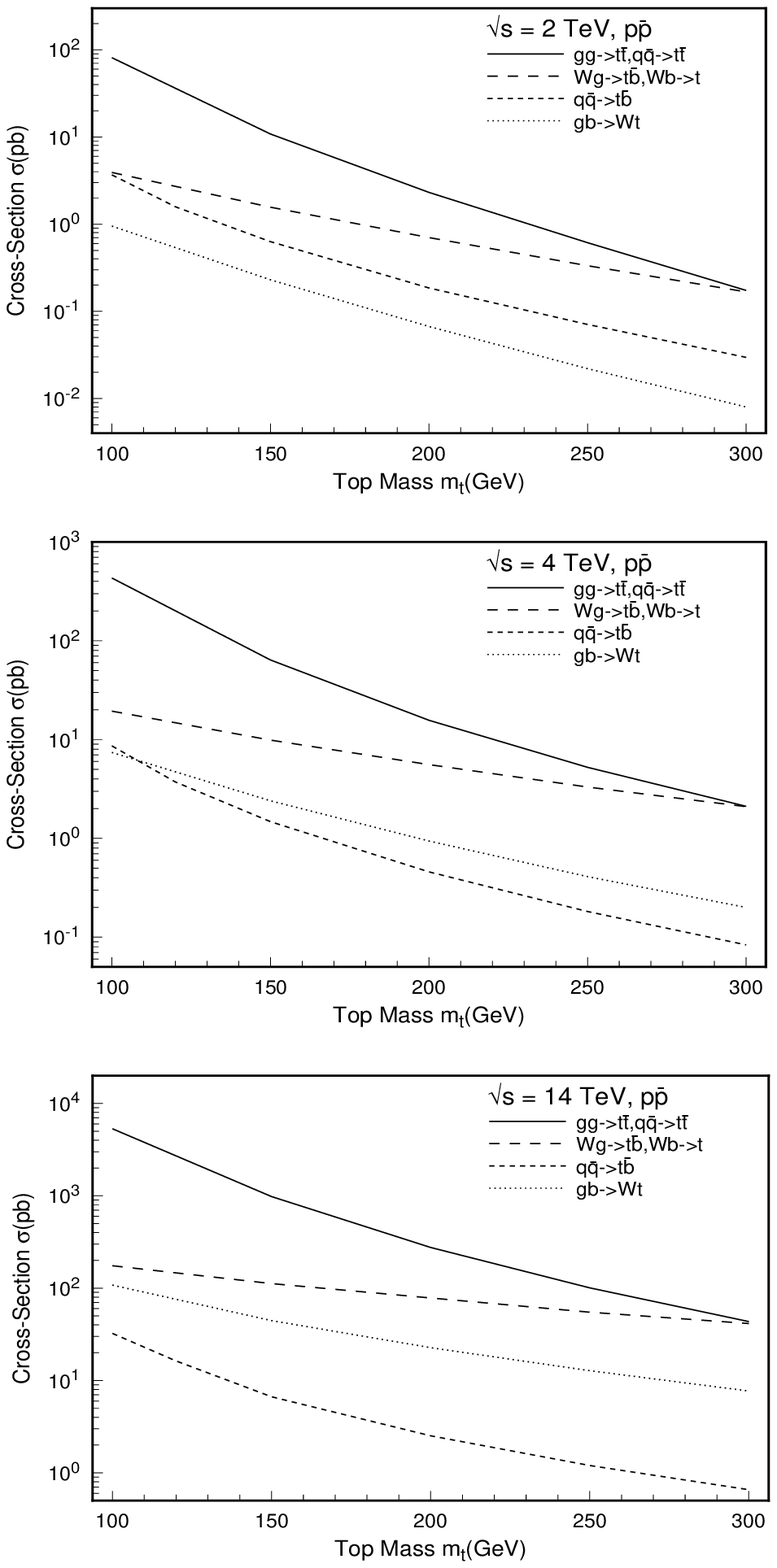,height=8in}}}
\caption{ Rate in [pb] for $\ggtt$, $\Wgtb$, $\udbt$ and $\gbtW$ at 
various energies of $\pbarp$ colliders.}
\label{fig1a}
\end{figure}

\begin{figure}[p]
\centerline{\hbox{
\psfig{figure=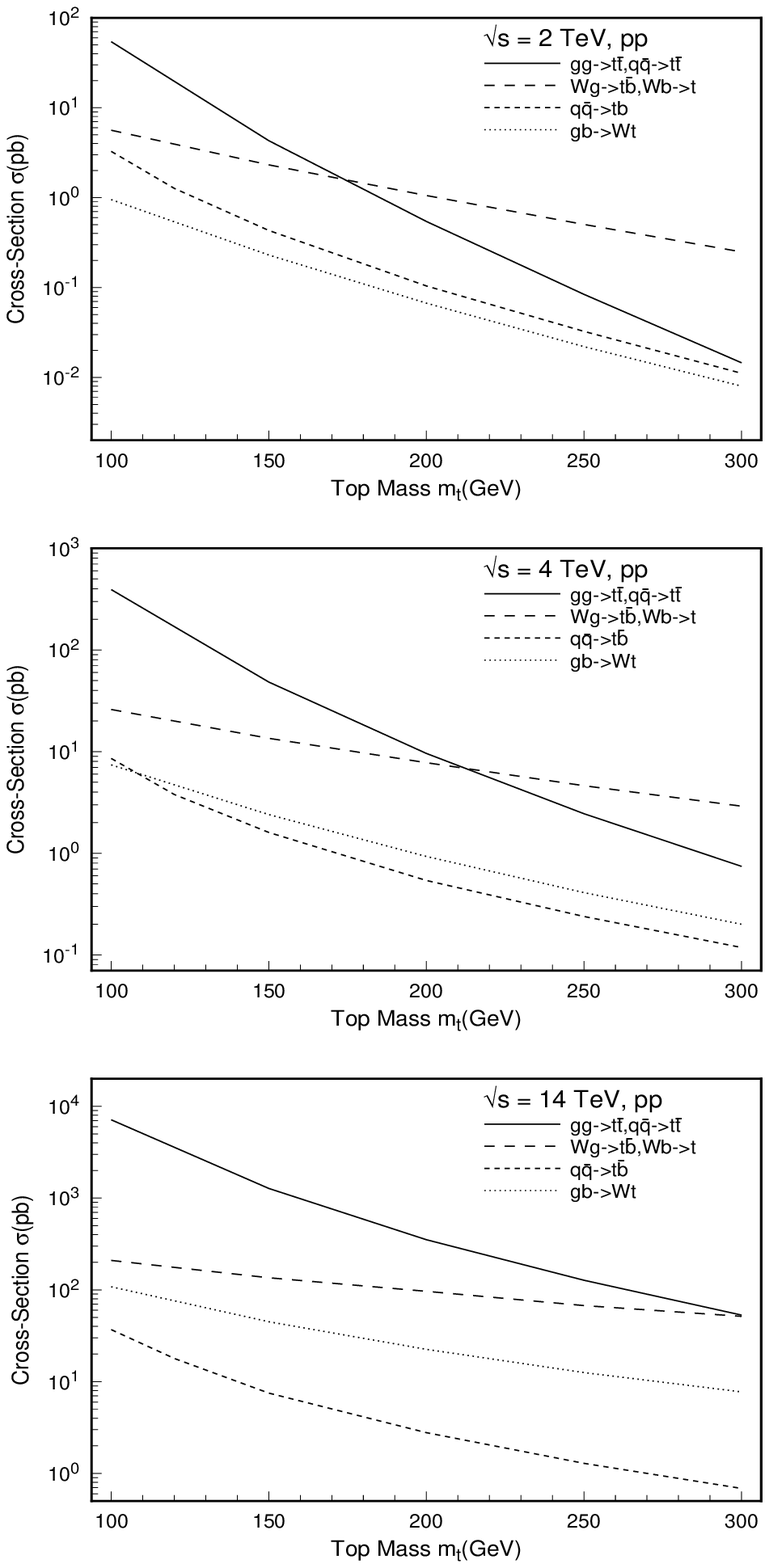,height=8in}}}
\caption{ Rate in [pb] for $\ggtt$, $\Wgtb$, $\udbt$ and $\gbtW$ at 
various energies of $\pp$ colliders.}
\label{fig1b}
\end{figure}

In Figures~\ref{fig1a} and~\ref{fig1b} 
we show the total cross sections of these processes
for the Tevatron, the \tevs~and the LHC energies referred to above.
For reference we include plots of the cross sections of top quarks
as a function of $m_t$ in both the $\pbarp$ collisions, 
shown in Figure~\ref{fig1a}, and $\pp$ collisions, shown in
Figure~\ref{fig1b}.  
The parton distribution function (PDF) used in our calculation is 
the leading order set CTEQ2L \cite{pdf}.
We note that taking the $\Lambda_{\rm QCD}$
value given in CTEQ2L PDF we obtain $\alpha_s(M_Z)=0.127$
which is about 15\% larger than 
the value of 0.110 in CTEQ2M PDF \cite{pdf}.
We found that if we rescale the $t \bar t$ production 
rates obtained from CTEQ2L PDF with born level amplitudes by
the ratio of $\alpha_s^2(Q, \Lambda_{QCD})$ from CTEQ2M 
and that from CTEQ2L,
which yields 0.7 for $Q=M_Z$, then our total rates are 
in good agreement with those obtained using 
NLO PDF and NLO amplitudes \cite{qcdtt}, see, for
example, Reference~\cite{smith}.
Hereafter we shall use the scaled results for our rates.
The constituent cross sections are all calculated at tree 
level for simplicity 
to study the kinematics of the top quark and its decay products. 

To include the production rates for
both single-$t$ and  single-$\bar t$ events
at  $\pbarp$ colliders,  
a factor of 2 should be multiplied to the single-$t$ rates shown in 
Figures~\ref{fig1a} and~\ref{fig1b}
because the parton luminosity for single-$\bar t$ 
production is the same as that for single-$t$.  
Similarly, at $\pp$ colliders the rates
should be multiplied by about $1.5$
for the center-of-mass energy ($\rtS$) of the collider
up to approximately $4$ TeV, 
but almost a factor of two at higher energies (say, $\rtS \geq 8$ TeV
up to about 14 TeV)
because the relevant parton luminosities for
producing a single-$t$ and a single-$\bar t$ event in $\pp$ 
collisions are different.
As shown in Figures~\ref{fig1a} and~\ref{fig1b}
the total rate for single-top production is about the 
same at $\pbarp$ and $\pp$ colliders for 
 $\rtS \geq 8$\,TeV  because the relevant 
valence and sea quark parton distributions
are about equal for $100 \,{\rm GeV} \, < m_t < 300$\,GeV.
For smaller $\rtS$, up to about $4$ TeV, a $\pbarp$ collider 
is preferred over
a $\pp$ collider for heavy top quark production because of
its larger parton luminosities.  
Similarly, for $t \bar t$ pair productions at small $\sqrt{S}$,
 the quark initiated process $ q \bar q \ra t \bar t$
is more important than the gluon fusion process
$gg \ra t \bar t$.
At $\rtS$ from $8$ to $14$ TeV the $\ttb$ rate 
is about the same in $\pbarp$ and $\pp$ collisions
because the $gg \ra t \bar t$ subprocess becomes dominant.

\vspace{4mm}
\begin{table}
\caption{ Rates of the above processes 
for $\mt = 180 (140)$\,GeV. (Branching ratios are not included here.)
For $\protect\rtS = 2\,$TeV and $4\,$TeV we include rates for a $\pbarp$
machine. At $\protect\rtS = 14$ TeV the rates are for a $\pp$ machine.
For the single-top rates we only include single-$t$ production.}
\label{trates}
\begin{center}
\begin{tabular}{|l||l|l|l|l|r|}              \hline
            &   \multicolumn{4}{c|}{Cross Section (pb)}  \\ \hline
$\rtS$(TeV) & $\ggtt$   & $\qgtb$ (or $\ubdt$)& $\udbt$  & $\gbtW$  \\ \hline
2           & 4.5(16)   & 1(2)     & 0.3(0.8) & 0.1(0.3) \\ \hline
4           & 26(88)    & 7(11)    & 0.8(2.1) & 1.3(2.9) \\ \hline 
14          & 430(1300) & 100(140) & 4.6(11)  & 3.6(8.8) \\ \hline 
\end{tabular}
\end{center}
\vspace{4mm}
\end{table}
For later reference in this paper, 
we show the rates of the above processes in Table~\ref{trates}
for $\mt = 180 (140)$\,GeV. (Branching ratios are not included here.)
For $\rtS = 2$ and 4 TeV we include only the rates for a $\pbarp$ 
machine, whereas at $\rtS = 14$ TeV the rates are for a $\pp$ machine.
Again, for the single-top rates we only include $t$ production.

Both in Figures~\ref{fig1a} and~\ref{fig1b} and Table~\ref{trates},
we have given the cross section of single-top quarks produced from either
the $\Wgtb$ or $\ubdt$ processes. From now on, we will refer to 
this production rate as the rate of the $W$--gluon fusion process.
The single-top quark produced from the $W$--gluon fusion process 
involves a very important and not 
yet well-developed technique of handling the kinematics of 
a {\it heavy} $b$ parton inside a hadron. 
Thus the kinematics of the top quark produced from this process can
not be accurately calculated yet.
However, the total event rate for 
single-top quark production via this process
can be estimated using the method proposed in Reference~\cite{wuki}.
The total rate for the $W$--gluon fusion process
involves the ${\cal O}(\alpha^2)$
($2 \ra 2$) process $\ubdt$
plus the ${\cal O}(\alpha^2 \alpha_s)$ ($2 \ra 3$)
process $q' g (W^+ g) \ra q t \bar b $ 
(where the gluon splits to $b \bar b$)
minus the {\it splitting} piece $g \ra b \bar b \,\otimes\, \ubdt$
in which $b \bar b$ are nearly collinear.
\begin{figure}
\centerline{\hbox{\psfig{figure=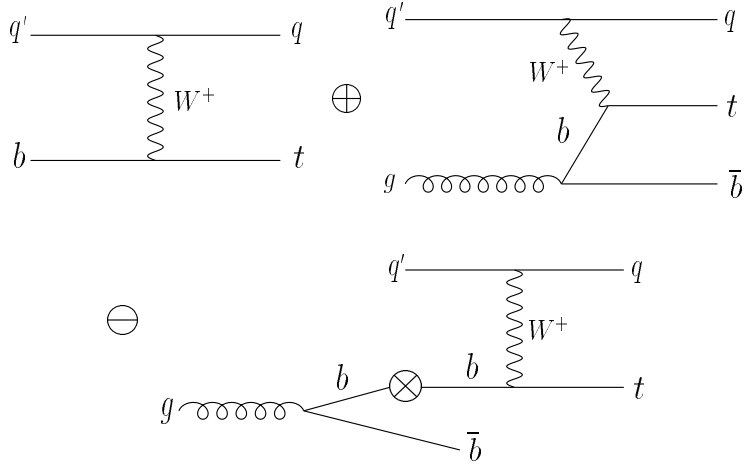,height=2.5in}}}
\caption{ Feynman diagrams illustrating the subtraction procedure
for calculating the total rate for $W$--gluon fusion:
$\ubdt\,\oplus\, q' g (W^+ g) \ra q t \bar b \,
\ominus\, (g \ra b \bar b \,\otimes\, \ubdt)$. }
\label{feynm}
\end{figure}
These processes are shown diagrammatically in Figure~\ref{feynm}.
The helicity amplitudes and the cross sections 
for these processes are given in Appendices A and B respectively.

The splitting piece is subtracted to avoid double counting the regime in
which the $b$ propagator in the ($2 \ra 3$) process is close 
to on-shell.\footnote{
The total rate of the $(2 \ra 3)$ process is extensively 
discussed in the Appendix C.}
The procedure is to resum the large logarithm 
$\alpha_s \ln (m_t^2/m_b^2)$ in the $W$--gluon fusion process
to all orders in $\alpha_s$ and include part of the higher order
${\cal O}(\alpha^2 \alpha_s)$ corrections to  
its production rate.
($m_b$ is the mass of the bottom quark.)
We note that to obtain the complete ${\cal O}(\alpha^2 \alpha_s)$ 
corrections beyond just the leading log contributions
one should also include 
virtual corrections to the ($2 \ra 2$) process, but
we shall ignore these non-leading contributions in this work.
Using the prescription described as above  
we found that the total rate of the $W$--gluon fusion process
is about $25\%$ less as compared to the ($2 \ra 2$) event rate for 
$\mt = 180\,(140)$ GeV regardless of 
the energy or the  type ({\it i.e.}, $\pp$ or $\pbarp$)
of the machine.
\begin{figure}[p]
\centerline{\hbox{
\psfig{figure=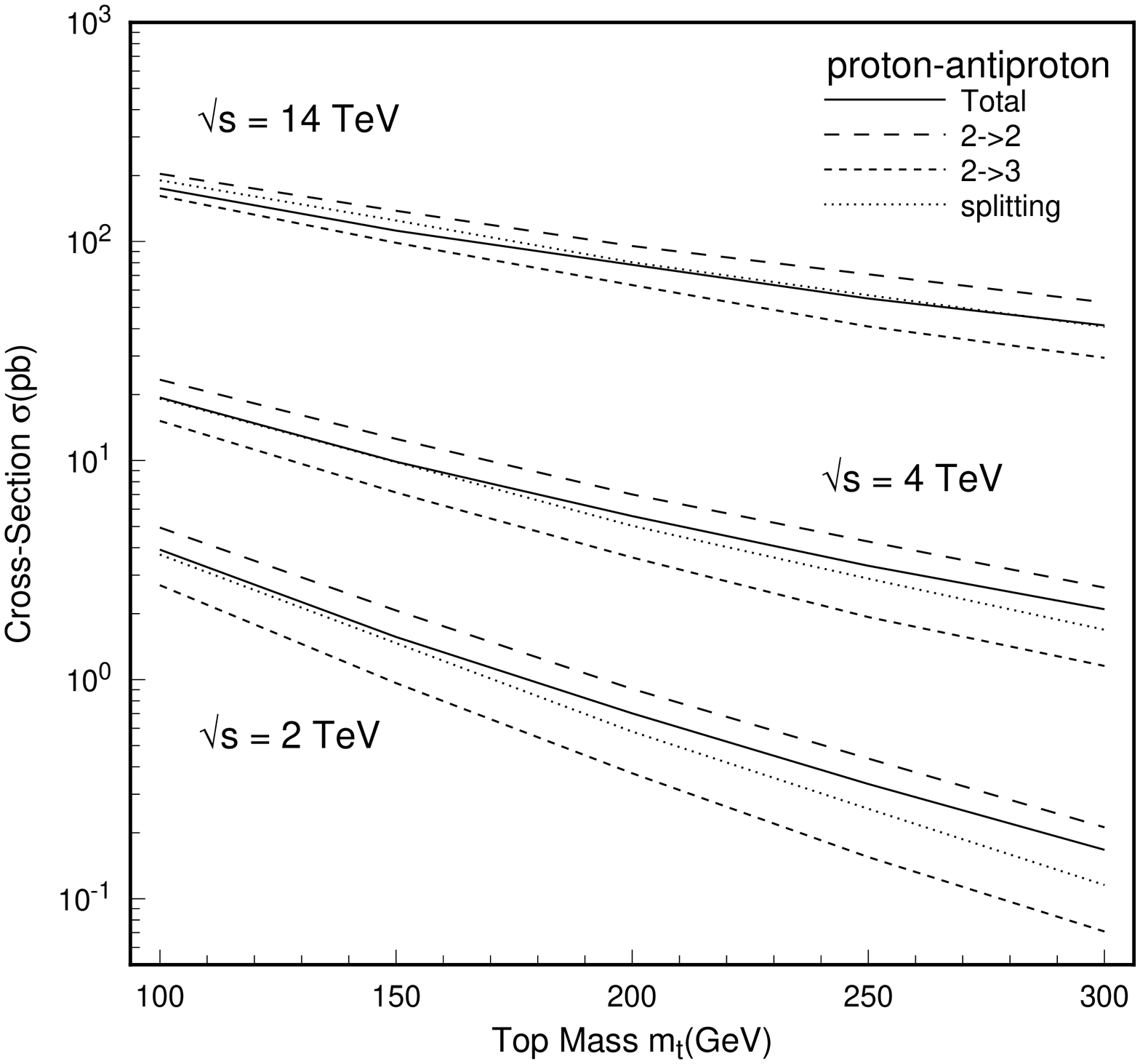,height=4.5in}}}
\caption{ Rate in [pb] for single-$t$ production: $\ubdt$ 
$(2 \ra 2)$, $\qgtb$ $(2 \ra 3)$ and 
the {\it splitting} piece $g \ra b \bar b \,\otimes\, \ubdt$ 
in which $b \bar b$ are collinear.  The rates are for 
$\pbarp$ colliders. }
\label{fig2a}
\end{figure}
\begin{figure}[p]
\centerline{\hbox{
\psfig{figure=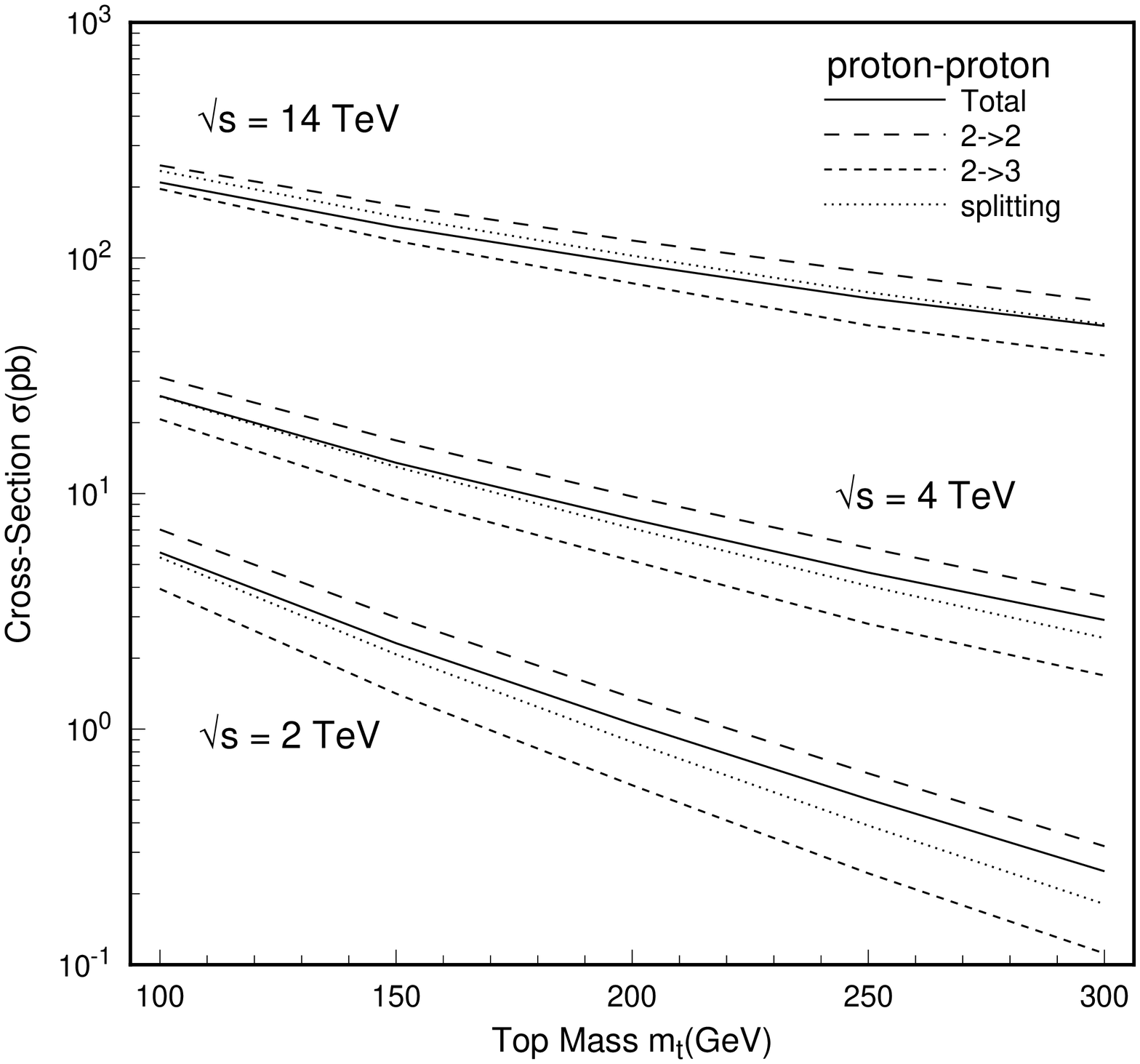,height=4.5in}}}
\caption{ Rate in [pb] for single-$t$ production: $\ubdt$ 
$(2 \ra 2)$, $\qgtb$ $(2 \ra 3)$ and 
the {\it splitting} piece $g \ra b \bar b \,\otimes\, \ubdt$ 
in which $b \bar b$ are collinear.  The rates are for 
$\pp$ colliders. }
\label{fig2b}
\end{figure}
In Figures~\ref{fig2a} and~\ref{fig2b}
we show the total rate of $W$--gluon fusion  versus $\mt$ with scale
$Q = \mt$ as well as a breakdown of the 
contributing processes at the Tevatron, the \tevs~and the LHC.

\begin{figure}[p]
\centerline{\hbox{
\psfig{figure=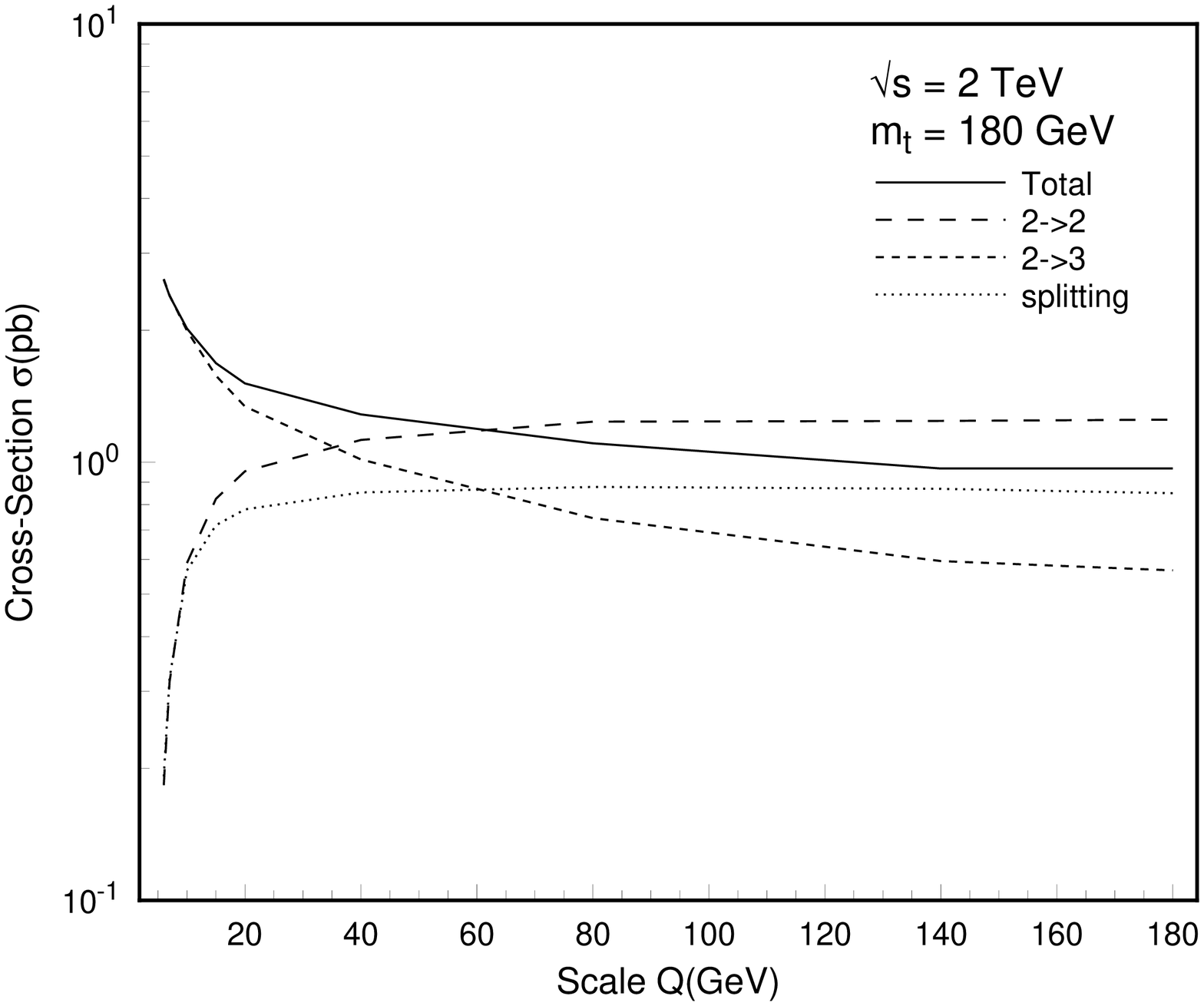,height=4.5in}}}
\caption{ Rate of $W$--gluon fusion process versus scale $Q$ 
for $m_t = 180\,$GeV and $\protect\rtS=2$\,TeV. }
\label{scale}
\end{figure}

To estimate the uncertainty in the production rate 
due to the choice of the scale $Q$ in evaluating 
the strong coupling constant $\alpha_s$ and the 
parton distributions, we show
in Figure~\ref{scale} the scale dependence
of the $W$--gluon fusion rate.  
As shown in the figure, although the individual rate from 
either ($2 \ra 2$), ($2 \ra 3$) or the splitting piece
is relatively sensitive to the choice of the scale,
the total rate as defined by $(2 \ra 2)\, + \, (2 \ra 3) \, - \, 
({\rm splitting\,piece})$ only varies by about 30\% 
for $M_W/2 < Q < 2 m_t$ at the Tevatron. At the \tevs~and the LHC, 
it varies by about 30\% and 10\%, respectively.
Based upon the results shown in Figure~\ref{scale}, we argue 
that $Q < M_W/2$ probably is not a good choice as the relevant 
scale for the production of the top 
quark from the $W$--gluon fusion
process because the total rate rapidly increases
by about a factor of 2 in the low $Q$ regime. 
In view of the prescription adopted in calculating the total rate,
the only relevant scales are the top quark mass $m_t$ and the 
virtuality of the $W$-line in the scattering amplitudes.
Since the typical transverse momentum
of the quark ($q$), which comes from the initial quark ($q'$) 
after emitting the $W$-line, is about half of the $W$--boson mass,
the typical virtuality of the $W$-line is about 
$M_W/2 \simeq 40$\,GeV. 
$m_b \simeq 5$\,GeV is thus not an appropriate scale 
to be used in calculating the $W$--gluon fusion rate using 
our prescription. 
We note that in the ($2 \ra 2$) process the
$b$ quark distribution effectively contains sums to order 
$[\alpha_s \ln(Q/{\mb})]^n$ from $n$-fold collinear
gluon emission, whereas the subtraction term (namely, the splitting piece)
contains only first order
in $\alpha_s \ln({\it Q}/{\mb})$.  Therefore, as 
${\it Q} \ra {\mb}$ the 
($2 \ra 2$) process picks up only the leading order in
$\alpha_s \ln({\it Q}/{\mb})$ and so gets largely 
cancelled in calculating the total rate.
Consequently, as shown in Figure~\ref{scale}, the total rate is about the
same as the ($2 \ra 3$) rate for $Q \ra m_b$.
We also note that at $Q \simeq  {M_W}/2$, 
the ($2 \ra 2$) and ($2 \ra 3$) processes have about the same rate.
As $Q$ increases the ($2 \ra 2$) rate gradually increases 
while the ($2 \ra 3$) rate decreases such that the total rate is 
not sensitive to the scale $Q$.
It is  easy to see also that the total rates calculated via this prescription 
will not be sensitive to the choice of PDF although each individual 
piece can have different results from different PDF's, based upon the
factorization of the QCD theory \cite{wuki}.

Another single-top quark production mechanism is the Drell-Yan type process 
$\udbt$. As shown in Figures~\ref{fig1a} and~\ref{fig1b}, 
for top quarks with mass
on the order of 180\,GeV the rate for $W^*$ production is about one fifth
that of $W$--gluon fusion at $\sqrt{S}=2$\,TeV. The $W^*$ process becomes
much less important for a heavier top quark.
This is because at higher invariant masses $\hat s$ 
(for producing a heavier top quark) of the $t \bar b$ system, 
$W^*$ production suffers the usual $1/{\hat s}$ suppression 
in the constituent cross section. However, in the $W$--gluon fusion
process the constituent cross section does not fall off as $1/{\hat s}$
but flattens out asymptotically to $1/M_W^2$. 
(The analytical results of these amplitudes
are given in Appendix A for reference.)
For colliders with higher energies, 
therefore with large range of $\hat s$,
the $W^*$ production mechanism for
heavy top quarks becomes much less important. However, the kinematics of the
top quarks produced from this process are different from those in the 
$W$--gluon fusion events.
Moreover, possible new physics may introduce a high mass state 
(say, particle $V$) to couple strongly with the $t \bar b$ system such that 
the production rate from $q' \bar q \ra W^* \ra V \ra t \bar b$ 
can largely deviate from the SM $W^*$ rate.\footnote{
This is similar to the speculations made in Reference~\cite{hill}
for having some high mass resonance in $t \bar t$ production.} 
We will however not discuss it in detail here because
its rate is highly model dependent.

The $W$--gluon fusion process becomes more important for a heavier top quark.
Why?
Effectively,  the $W$--gluon fusion process
can be viewed as the scattering of a longitudinal
$W$--boson ($W_L$) with gluon to produce a top quark and a bottom
anti-quark ($W^+_L g \ra t \bar b$) after applying the effective-$W$ 
approximation \cite{dawson}. For large $\hat s$ this scattering 
process is equivalent to ($\phi^+ g \ra t \bar b$) where $\phi^+$
is the corresponding Goldstone boson of the gauge boson $W^+$ due
to the Goldstone Equivalence Theorem \cite{equiv,equivrad}.
Since the coupling of 
$t$-$b$-$\phi$ is proportional to the mass of the top quark, the 
constituent cross section of the $W$--gluon fusion process
grows like $m_t^2/M_W^2$ when $m_t$ increases.
This explains why the $W$--gluon fusion rate only decreases 
slightly as the mass of the top quark increases even
though both the parton luminosity and the available phase space
decrease for a heavier top quark.
In contrast, the $t \bar t$ pair production rate from the QCD processes
decreases more rapidly as $m_t$ increases because the constituent
cross section of $\ggtt$ goes as $1/{\hat s}$ and the phase space 
for producing a $t \bar t$ pair is smaller than that for producing 
a single-$t$. Therefore, the $W$--gluon fusion process becomes 
more important for the production of a heavy top quark.

Before closing this chapter, we note that
the Effective-$W$ approximation has been the essential tool 
used in studying the strongly interacting longitudinal $W$ system to probe the
symmetry breaking sector at the supercolliders such as the LHC \cite{wwww}.
By studying single-top production from the 
$W$--gluon fusion process at the Tevatron,
one can learn about the validity of the 
Effective-$W$ approximation prior 
to the supercolliders.

\chapter{Measuring the Top Quark Mass}
                                         
By the year 2000, we expect results from the Tevatron (with 1\,$\ifb$) 
and results from LEP-200, giving an error of about $50$\,MeV on $M_W$.
Due to Veltman's screening theorem, 
the low energy data are not sensitive to the mass of the Higgs boson
\cite{veltman}. 
For a heavy Higgs boson, the low energy data
can at most depend on $m_H$ logarithmically
up to the one loop level. 
Therefore, within the SM one needs to also know the mass of the 
top quark to within about $5$\,GeV to start getting
useful information on $m_H$ with an uncertainty less than
a few hundred GeV. This can be done by studying  radiative 
corrections to the low energy
data which include LEP, SLC, and neutrino experiments
\cite{mele,alta,lep94,sld94}.
(Of course, $m_H$ will be measured to better 
precision if it is detected from direct production at colliders.)

How accurate can the mass of the top quark be measured at hadron colliders?
At hadron colliders, $m_t$ can be measured in the $\ttb$ events
by several methods \cite{argon,cdfd0}.
\begin{figure}
\centerline{\hbox{\psfig{figure=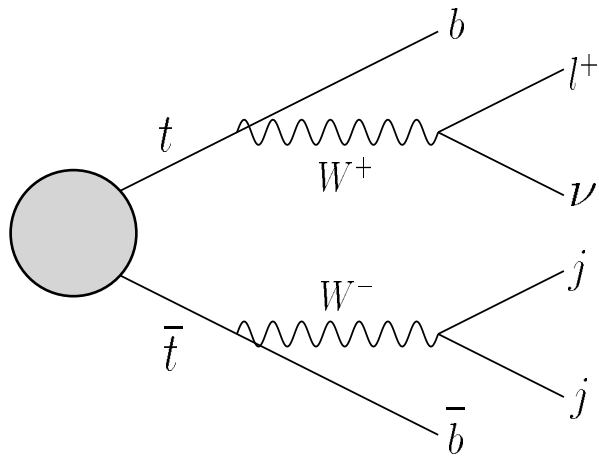,height=2.5in}}}
\caption{ The lepton+jet decay mode of $\ttb$ production. }
\label{decay2}
\end{figure}
The first method is to use the lepton+jet decay mode of
the $\ttb$ pair, as shown in Figure~\ref{decay2}.
This is done by reconstructing the invariant mass of the three jets in 
the opposite hemisphere
from the isolated lepton $\ell$ ($= e \,{\rm or}\, \mu$) 
in $t \ra bW (\ra \ell \nu)$,
and requiring that two of the three jets reconstruct to a 
$W$ and the third be tagged as a $b$-jet.
\begin{figure}
\centerline{\hbox{\psfig{figure=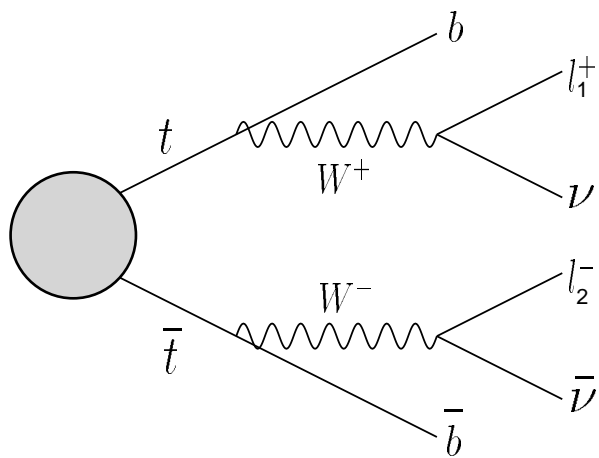,height=2.5in}}}
\caption{ The di-lepton decay mode of $\ttb$ production. }
\label{decay1}
\end{figure}
The second method is to use the di-lepton decay mode of the $\ttb$
pair, as shown in Figure~\ref{decay1}. Here it is required that
both $W$'s decay leptonically.
In addition, one of the $b$'s must decay semileptonically
to measure the mass distribution of the non-isolated 
lepton $\ell_{b}$ (from $b$ decay) and one of the two 
isolated leptons ($\ell_1$ and $\ell_2$ from $W^\pm$ decay) 
which is closer to $\ell_{b}$.
The third method is to measure the cross section of the 
di-lepton decay mode of the $\ttb$ pair. 
At the LHC, there will be about $10^8$ $\ttb$ pairs produced in one 
year of running for $\mt < 200$\,GeV.
With such a large number of events, the ATLAS and CMS collaborations
concluded that $m_t$ can be measured with a precision 
of $\leq 5$\,GeV using the first method described above
and with about a factor of 2 improvement using the second method 
\cite{atlas,cms}.
A similar conclusion was also drawn by the CDF and the \D0 collaborations 
for the Tevatron with Main Injector after the upgrade of their 
detectors \cite{chip}.  This is remarkable given that
the $\ttb$ cross section at the Tevatron is smaller 
by about two orders of magnitude as compared with that at the
LHC, as shown in Figures~\ref{fig1a} and~\ref{fig1b}.
 
Next, we would like to discuss how to measure the mass of the top 
quark in the $W$--gluon fusion process. 
Since $m_t$ has been measured by the FNAL CDF and \D0 groups
in the $\ttb$ events \cite{CDFft,D0ft}, 
why do we care?
To check whether it is a SM top quark, we should verify its production rate
predicted by the SM for other production processes
such as the single-top quark process.

Suppose the coupling of \tbw~is not of the SM nature, then
we would find that the single-top quark production rate of 
the $W$--gluon fusion process is different from the SM prediction
because its production rate is directly proportional to the square
of this coupling. (We will discuss more on this point in Chapter~6.)
Hence, without knowing the nature of the \tbw~interactions
one can not use the production rates of the single-top 
quark events to measure $m_t$.
Alternatively, we propose two methods to measure $m_t$ in the single-top
quark events. We will refer to them as the fourth and the fifth method.
The fourth method is a slight variation of the second method.
Instead of measuring the invariant mass of the leptons, we propose to 
directly measure the invariant mass ($m_{b\ell}$) of the $\ell$ and $b$
in $t \ra b W (\ra \ell \nu)$.  
We expect that the efficiency of $b$ tagging 
using the displaced vertex is higher  for detecting a heavier top quark,
and the $b$ jet energy measurement 
is better for $b$ having  larger transverse momentum from
a heavy top quark decay. 
Thus $m_{b\ell}$ can be used to measure
the mass of a SM top quark. 
The details of our Monte Carlo study are given in Chapter~8
for a single-top quark event.
In the $\ttb$ event there are two $b$'s,
therefore this method may not work as well as in the single-top 
event which only contains one $b$.  
\begin{figure}[p]
\centerline{\hbox{\psfig{figure=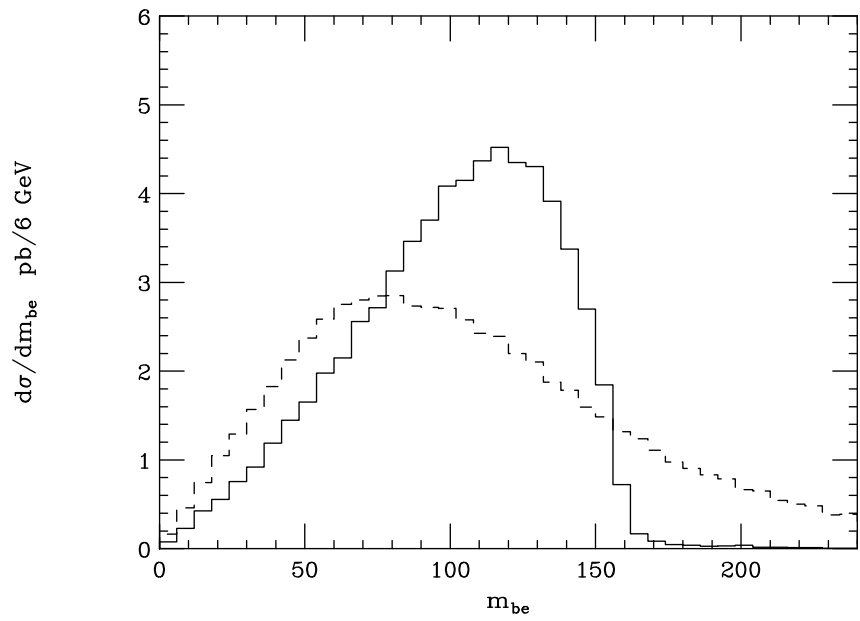,height=4.5in}}}
\caption{ Distributions of $m_{b\ell}$ (solid) and 
$m_{{\bar b}\ell}$ (dash) in $t \bar t$ events for a 180 GeV top quark. }
\label{figttb}
\end{figure}
However it is not entirely impossible to use this method because, 
as shown in Figure~\ref{figttb},
the sum of the invariant mass distributions of 
$b\ell$ and $\bar{b} \ell$ for a 180 GeV top quark
still show a bump near the region that the distribution of 
$m_{b\ell}$ peaks.
(With a larger sample of $t \bar t$ events one might be able to afford
using the electric charge of the soft-lepton from $b$-decay to 
separate $b$ from $\bar b$ on an event-by-event basis at the cost of 
the small branching ratio of $b \ra \mu + X$, of about 10\%.) 
We will explain in more detail how to use $\flong$ 
(the fraction of longitudinal $W$--bosons from top quark decay), 
derived from the distribution of
$m_{b\ell}$, to measure $m_t$ in Chapter~6.

The fifth method is to reconstruct the invariant mass of the top quark
in the $t \ra b W (\ra \ell \nu)$ decay mode by 
measuring  the missing transverse momentum 
and  choosing a two-fold solution 
of the longitudinal momentum of the neutrino from the mass constraint of
the $W$ boson.
In Chapter~8 we conclude that it 
is possible to measure $m_t$ using 
either of these last two methods to
a precision of 5\,GeV at the Tevatron 
($\sqrt{S}=2\,$TeV) with 1\,$\ifb$ integrated luminosity.
We also find that after applying all the kinematical 
cuts to suppress the dominant
background  $W+ b \bar b$,
at most $10\%$ of $W^*$ events contribute to
single-top production for a 180 (140)\,GeV top quark.
The SM $W^*$ production rate is already much smaller than 
the $W$--gluon fusion rate for a heavier top quark, 
therefore the contribution from the $W^*$ is not important in our
study although we do include its small effects in our analysis 
as described in Chapter~8.

\chapter{Measuring the Top Quark Width}

As shown in Reference~\cite{steve} the intrinsic width of the top quark 
can not be measured at a high energy hadron collider such as the LHC 
through the usual QCD processes.\footnote{
In Reference~\cite{steve}, the effects of QCD radiation in top 
quark decay (at one loop level) to the measurement of $m_t$
in $t \bar t$ events produced in hadron collisions was studied. 
It was concluded that 
the peak position of the $m_t$ distribution remains about the same as
the tree level result, but the shape is different.
It was also found that the $m_{b\ell}$ distribution is not
sensitive to QCD radiation in top decay.
} For instance,
the intrinsic width of a 150 GeV Standard Model top quark is 
about 1 GeV, and the full width at half maximum of the reconstructed 
top quark invariant mass (from $t \ra b W (\ra \,jets)$ decay mode)
is about $10$ GeV after including the detector resolution
effects by smearing the final state parton momenta. 
Here, the ratio of the measured width and the intrinsic width for a 150 GeV 
top quark is about a factor of 10. For a heavier top quark, 
this ratio may be slightly improved because the jet energy can be better 
measured. (The detector resolution $\Delta E/E$ for a QCD jet with 
energy $E$ is proportional to $1/\sqrt{E}$.)
A similar conclusion was
also given from a hadron level analysis presented in the SDC Technical
Design Report which concluded that reconstructing the top quark invariant
mass gave a width of 9 GeV for a 150 GeV top quark \cite{sdc}.
Is there a way to measure the top quark width $\width$, say, within
a factor of 2 or better, at hadron colliders?
Yes, it can in principle be measured in single-top events.

The width $\width$ can be measured by counting the production rate of top
quarks from the $W$--$b$ fusion process which is {\it equivalent} 
to the $W$--gluon fusion process by a proper treatment of the bottom 
quark and the $W$ boson as partons inside the hadron.
The $W$--boson which interacts with the $b$-quark to produce the top
quark can be treated as an on-shell boson 
in the leading log approximation \cite{dawson,effw}.
The result is that even 
under the approximations considered, 
a factor of 2 uncertainty in the production
rate for this process gives a 
factor of 2 uncertainty in the measurement of $\width$.
This is already  much better than what can be measured
from the invariant mass distribution of the jets from
the decay of top quarks in the $t \bar t$ events produced via 
the usual QCD processes.
More precisely, as argued in Chapter~3, the production rate of
single-top events at the Tevatron can probably be known within about 30\%,
thus it implies $\width$ can be measured to 
about the same accuracy.\footnote{
Strictly speaking, from the production rate of single-top events,
one measures the sum of all the possible partial decay widths, such as
$\Gamma(t \ra b W^+) +\Gamma(t \ra s W^+) +\Gamma(t \ra d W^+) + \cdots$,
therefore, this measurement is really measuring the width of
$\Gamma(t \ra XW^+)$ where $X$ can be more than one particle state 
as long as it originates from the partons inside the proton (or anti-proton).
In the SM, $\Gamma(t \ra b W^+)$ is about equal to 
the total width of the top quark.} 
Therefore, this is an extremely important measurement
because it directly tests the couplings of~\tbw. 

$W$--gluon fusion can also tell us about the CKM matrix element
$|V_{tb}|$. Assuming only three generations of quarks, the constraints
from low energy data together with unitarity of the CKM matrix 
require $|V_{tb}|$ to be in 0.9988 to 0.9995 at the 90\% confidence 
level~\cite{databook}. 
As noted in Reference~\cite{databook} 
the low energy data do not preclude there being
more than three generations of quarks (assuming the same 
interactions as described by the SM). Moreover, the entries deduced 
from unitarity might be altered when the CKM matrix is expanded to 
accommodate more generations. When there are more than three generations
the allowed ranges (at 90\% CL) of the matrix element
$|V_{tb}|$ can be anywhere  between 0 and 0.9995~\cite{databook}.
Since  $|V_{tb}|$  is directly involved in 
single-top production via $W$--gluon fusion,
any deviation from SM value in $|V_{tb}|$ will 
either enhance or suppress the production rate of single-top events.
It can therefore be measured by simply counting the single-top
event rates. For instance, if the single-top production rate is
measured to within 30\%, then $|V_{tb}|$ is determined to within 15\%.

In conclusion, after the top quark is found, 
the branching ratio of $t \ra b W^+(\ra \ell^+\nu)$ can be measured 
from the ratio of $(2\ell+\,jets)$ and $(1\ell+\,jets)$
 rates in $t \bar t$ events.
The measured single-top quark event rate is equal to the single-top
production rate multiplied by the branching ratio of
$t \ra b W^+(\ra \ell^+\nu)$ for the  $(1 \ell+\,jets)$ mode and
the same $t$-$b$-$W$ couplings appearing in the decay of $t$
in this process appear also in the production of $t$.
Thus, a model independent measurement of the decay 
width~$\width$ can  be made by simply counting the production rate 
of $t$ in the $W$--gluon fusion process.  
Should the top quark width be found to be different from the SM expectations, 
we would then have to look for 
non-standard decay modes of the top quark.
We note that it is important to measure at least one
partial width (say, $\width$) precisely in order to discriminate between
different models of new physics, if any. In the SM, the 
partial width $\width$
is about the same as the total width of the top quark at the tree level
because of the small CKM matrix element 
$|V_{ts}|$, thus measuring the 
single-top quark production rate measures the 
lifetime of the top quark.

\chapter{Top Quark Couplings to the $W$ Gauge Boson }

It is equally important to ask what kind of interactions the~\tbw 
vertex might involve \cite{toppol}.
For instance, one should examine the form factors of~\tbw 
which result from  higher order corrections due to
SM strong and/or electroweak interactions.  
It is even more interesting to examine these form factors to test the
plausibility of having  {\it nonuniversal} gauge couplings
of~\tbw due to some dynamical symmetry breaking scenario \cite{pecc,ehab}.

The QCD~\cite{qcd} and the electroweak~\cite{elecw} corrections to the
decay process $t \ra b W^+$ in the SM have
been done in the literature. 
The most general operators for this coupling are
described by the interaction lagrangian
\begin{eqnarray}
L&=&\ {g\over \sqrt{2}}\left[ W^-_\mu\bar{b}\gamma^\mu
                (f_1^L P_-+f_1^R P_+)t -{1\over M_W}
\del_\nu W^-_\mu\bar{b}\sigma^{\mu\nu}(f_2^LP_-+f_2^RP_+)t \right] \nonumber \\ 
 & &+ {g\over \sqrt{2}}\left[W^+_\mu\bar{t}\gamma^\mu
      ({f_1^L}^* P_-+{f_1^R}^* P_+)b -{1\over M_W}
      \del_\nu W^+_\mu\bar{t}\sigma^{\mu\nu}
      ({f_2^R}^*P_-+{f_2^L}^*P_+)b \right] \, , \nonumber \\
 & & \qquad \,\,  
\label{eqlag}
\end{eqnarray}
where $P_\pm ={1\over 2}(1\pm \gamma_5)$,
$i\sigma^{\mu\nu}=-{1\over 2}[\gamma^\mu,\gamma^\nu]$ and
the superscript $*$ denotes the complex conjugate.
In general, the form factors $f_{1}^{L,R}$ and $f_{2}^{L,R}$ can be complex.
Note that in Equation~(\ref{eqlag}), if there is a relative phase between
$f^L_1$ and $f^R_2$ or between $f^R_1$ and $f^L_2$, then CP is violated.
For instance, in the limit of $m_b=0$ ,
a CP-violating observable will have a coefficient proportional to
${\rm Im}(f^L_1 {f^R_2}^*)$ for a left-handed bottom quark, 
and ${\rm Im}(f^R_1 {f^L_2}^*)$ for a right-handed 
bottom quark~\cite{toppol}.
(We will discuss  more on CP violation in Chapter~7.)
If the $W$--boson can be off--shell then there are additional form factors
such as
\beq
\del^\mu W^-_\mu \bar{b}(f_3^LP_-+f_3^RP_+)t
+\del^\mu W^+_\mu \bar{t}({f_3^R}^* P_- + {f_3^L}^* P_+)b \, ,
\label{eqlag2}
\enq
which vanish for an on--shell $W$--boson or when the off--shell $W$--boson
couples to massless on--shell fermions.
Here, we only consider on--shell $W$--bosons for
 $m_t> M_W + m_b$.
At tree level in the SM the form factors are
$f_1^L = 1$ and $f_1^R = f_2^L = f_2^R = 0$.
These form factors will in general affect the
experimental observables related to the top quark, such as 
the fraction of longitudinal $W$'s produced in top quark decays.
 
The fraction ($\flong$) of longitudinally polarized $W$--bosons, produced
in the rest frame of the decaying top
quark, strongly depends on the form factors 
$f_1^{L,R}$ and $f_2^{L,R}$, as shown in Appendix~C.
Hence, $\flong$ is a useful observable for 
measuring these form factors.
The definition of $\flong$ is simply the 
ratio of the number of longitudinally
polarized $W$--bosons produced with respect to the total number of
$W$--bosons produced in top quark decays:
\beq
\flong={\Gamma(\lambda_W=0)
\over{\Gamma(\lambda_W=0)+\Gamma(\lambda_W=-)+\Gamma(\lambda_W=+)}}.
\enq
We use $\Gamma(\lambda_W)$ to refer to the decay rate for a
top quark to decay into a $W$--boson with polarization $\lambda_W$.
($\lambda_W=-, +, 0$ denotes a left-handed, right-handed,
and longitudinal $W$--boson.)
Clearly, the polarization of the $W$--boson depends on the form factors
$f_1$ and $f_2$.\footnote{
The fraction of longitudinal $W$'s in top quark decays 
contributed from the form factor $f_1^R$ is the same as 
that from  $f_1^L$ \cite{toppol}.}
 Therefore, one can measure the polarization of the
$W$--boson to measure these form factors.
As shown in Appendix C, the polarization of the 
$W$--boson can be determined by the 
angular distribution of the lepton, say, $e^+$ in the rest frame of $W^+$ in
the decay mode $t \ra b W^+ (\ra e^+ \nu)$. 
However, the reconstruction of the $W$--boson rest frame (to measure 
its polarization) could be a non-trivial matter due to the missing 
longitudinal momentum ($P_{\SST Z}$) (with a two-fold ambiguity)
of the neutrino ($\nu$) from $W$ decay.
Fortunately, as shown in Equation~(\ref{mbe1}), 
one can determine the polarization
of the $W$--boson without reconstructing its rest frame by using the
Lorentz-invariant observable $m_{be}$, the invariant mass of 
$b$ and $e$ from $t$ decay.

The polar angle $\theta^*_{e^+}$
distribution of the $e^+$ in the rest frame
of the $W^+$ boson, whose z-axis is defined to be the moving direction of
the $W^+$ boson in the rest frame of the top quark, can be written in terms of
$m_{be}$ through the following derivation:
\begin{eqnarray}
\cos \theta^*_{e^+} &=& {{ E_e E_b - p_e \cdot p_b }\over
{|\vec{\bf p}_e| |\vec{\bf p}_b| }}       \nonumber \\
&\simeq & 1-{p_e \cdot p_b \over E_e E_b}
= 1-{2 m_{be}^2 \over m_t^2 - M_W^2}.  
\label{mbe1}
\end{eqnarray}
The energies $E_e$ and $E_b$ are evaluated in the rest frame of
the $W^+$ boson from the top quark decay and are given by
\begin{eqnarray}
E_e &=& {M_W^2+m_e^2-m_\nu^2 \over 2 M_W}, \qquad |\vec{\bf p}_e|=
\sqrt{E_e^2-m_e^2}, \nonumber \\ 
E_b &=& {m_t^2-M_W^2-m_b^2 \over 2 M_W}, \qquad |\vec{\bf p}_b|=
\sqrt{E_b^2-m_b^2}. 
\label{mbe2}
\end{eqnarray}
$m_e$ ($m_\nu$) denotes the mass of $e^+$ ($\nu_e$) for the
sake of bookkeeping.
The first line in Equation~(\ref{mbe1}) is exact when using Equation~(\ref{mbe2}),
while the second line of Equation~(\ref{mbe1}) holds in the limit of $m_b=0$.
It is now trivial to find $\flong$ by first calculating the 
$\cos\theta^*_{e+}$ distribution
then fitting it according to the decay amplitudes 
of the $W$--boson from top quark decay, as given in the Appendix~C.
In what follows we will show how to use the distribution of
$m_{be}$ to measure the mass of the top quark and 
its couplings to the $W$--boson. 

In Reference~\cite{ehab}, we considered the effective couplings
\beq 
 W-t_{L}-b_{L}:\,\, \frac{g}{2\sqrt{2}}\frac{ 1 + \kappa^{CC}_{L}}{2}
 \gamma_{\mu}(1-\gamma_{5})\, 
\enq
and
\beq 
 W-t_{R}-b_{R}:\,\, \frac{g}{2\sqrt{2}}\frac{ \kappa^{CC}_{R}}{2}
 \gamma_{\mu}(1+\gamma_{5})\, 
\enq
derived from an electroweak chiral lagrangian with the symmetry 
${\rm SU(2)}_{L}\times {\rm U(1)}_Y$ broken down to ${\rm U(1)}_{\rm EM}$.
(Here, $\kappa_L^{CC}=f_1^L - 1$, and $\kappa_R^{CC}=f_1^R$.)
At the Tevatron and the LHC, heavy top quarks are 
predominantly produced from the QCD process
$gg, q \bar q \ra t \bar t$ and the $W$--gluon fusion process
$qg (Wg) \ra t \bar{b}, \bar{t} b$.  
In the former process, one can probe $\klc$ and $\krc$ from the decay of the 
top quark to a bottom quark and a $W$--boson. In the latter process, 
these non-standard couplings can also 
be measured by simply counting the production 
rates of signal events with a single $t$ or $\bar t$.
Let us discuss $\klc$ and $\krc$ in more detail as follows.

\section{From the Decay of Top Quarks}

To probe $\klc$ and $\krc$ from the decay of the 
top quark to a bottom quark and a $W$--boson, one needs to measure the 
polarization of the $W$--boson, which can be measured from 
the distribution of the invariant mass $m_{b\ell}$.
 For a massless $b$, the $W$--boson from top
quark decay can only be either longitudinally or left-hand polarized for 
a left-hand charged current ($\krc=0$). For a right-hand 
charged current ($\klc=-1$) the $W$--boson can only be either longitudinally 
or right-hand polarized. 
(Note that the handedness of the $W$--boson is reversed for a massless
$\bar b$ from $\bar t$ decays.)
\begin{figure}
\centerline{\hbox{\psfig{figure=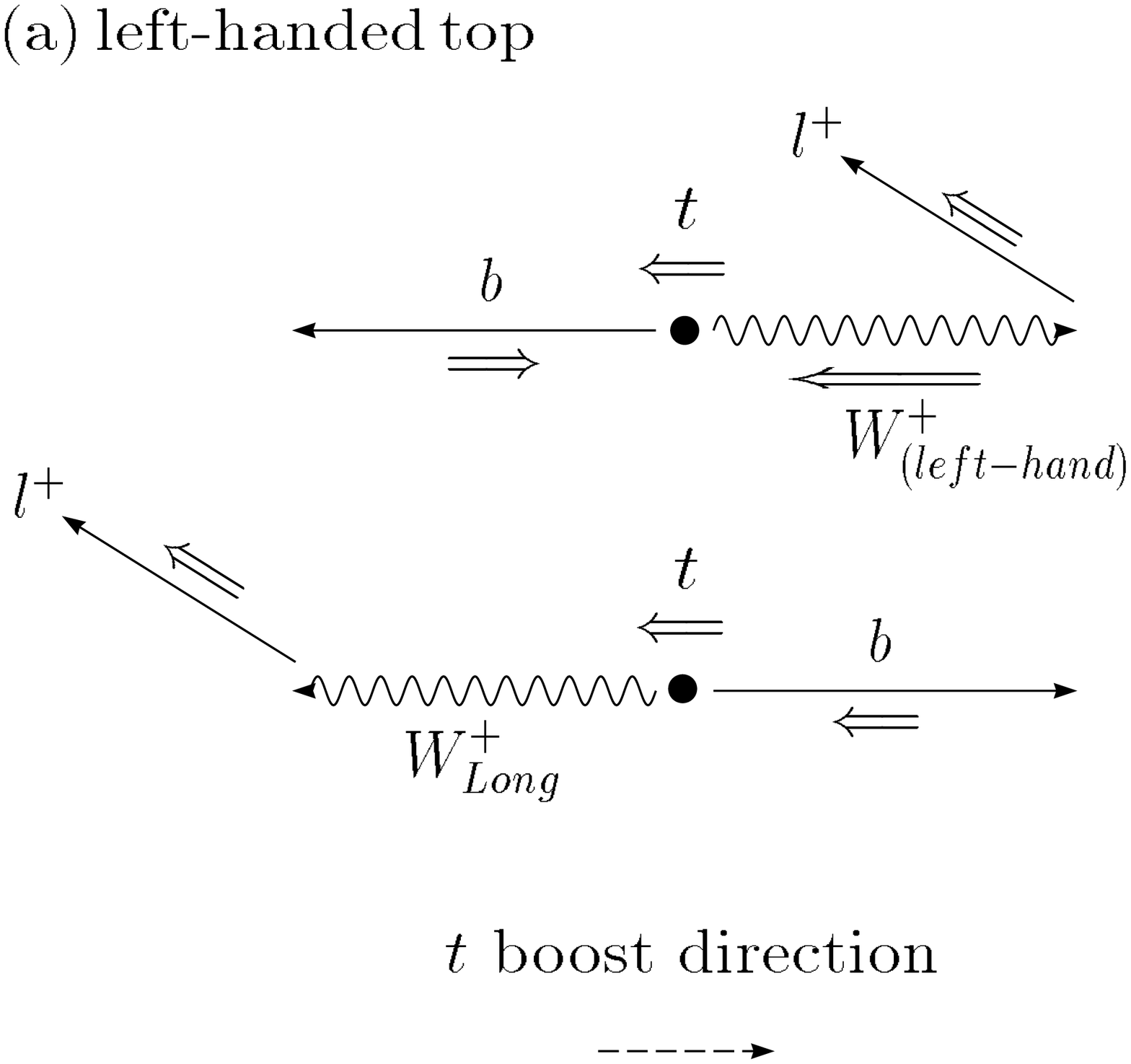,height=2.5in}
	\hspace{1.in}\psfig{figure=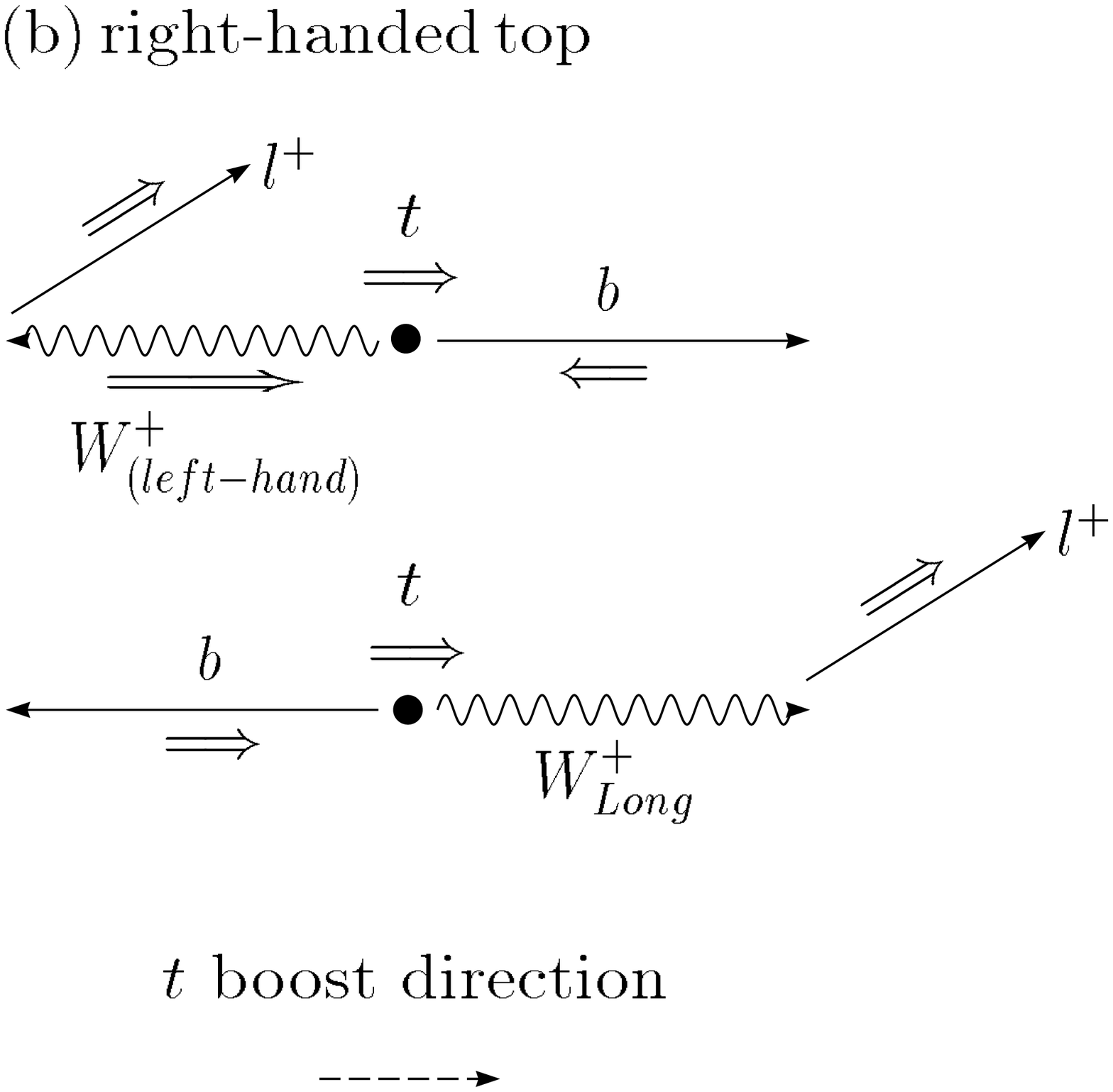,height=2.5in}}}
\caption{ For a left-handed $\tbw$ vertex. }
\label{left}
\end{figure}
\begin{figure}
\centerline{\hbox{\psfig{figure=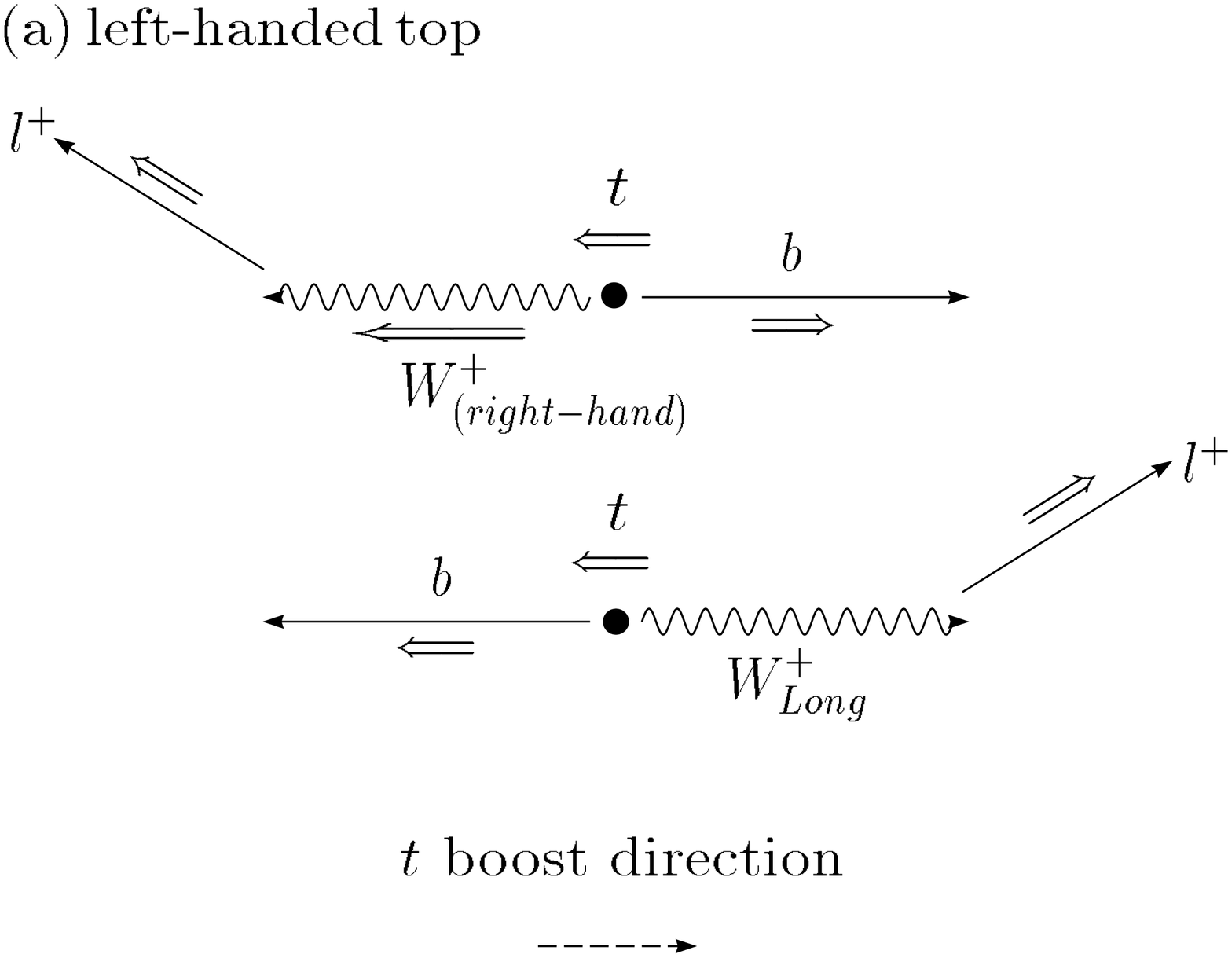,height=2.5in}
		  \psfig{figure=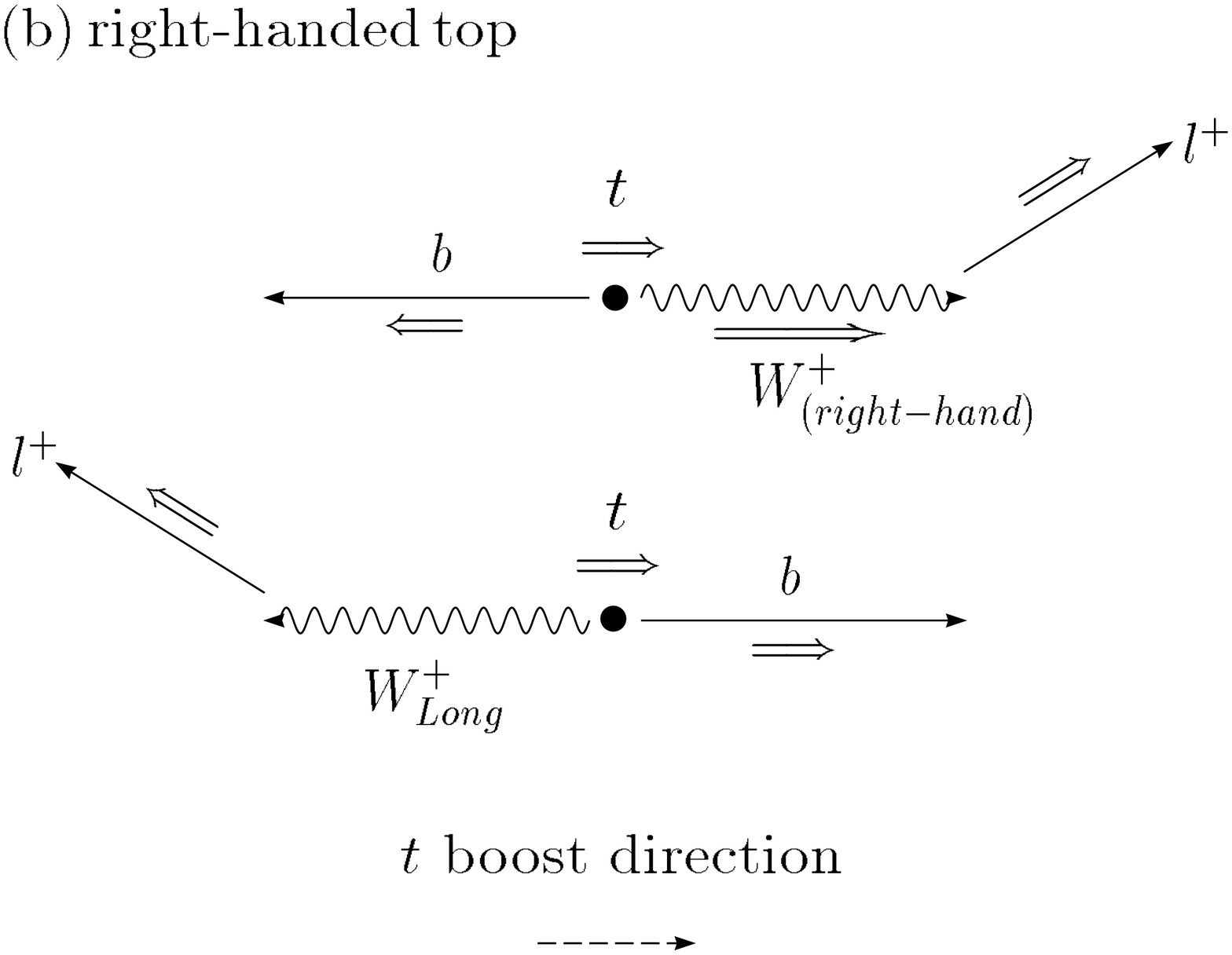,height=2.5in}}}
\caption{ For a right-handed $\tbw$ vertex. }
\label{right}
\end{figure}
This is a consequence of helicity conservation, as diagrammatically
shown in Figures~\ref{left} and~\ref{right} for a polarized 
top quark. In these figures we show the preferred 
moving direction of the lepton from a polarized $W$--boson in the rest 
frame of a polarized top quark for 
either a left-handed or a right-handed $\tbw$ vertex. 
As indicated in these figures, the invariant mass
$m_{b \ell}$ depends on the polarization of the $W$--boson from the decay
of a polarized top quark. 
\begin{figure}[p]
\centerline{\hbox{\psfig{figure=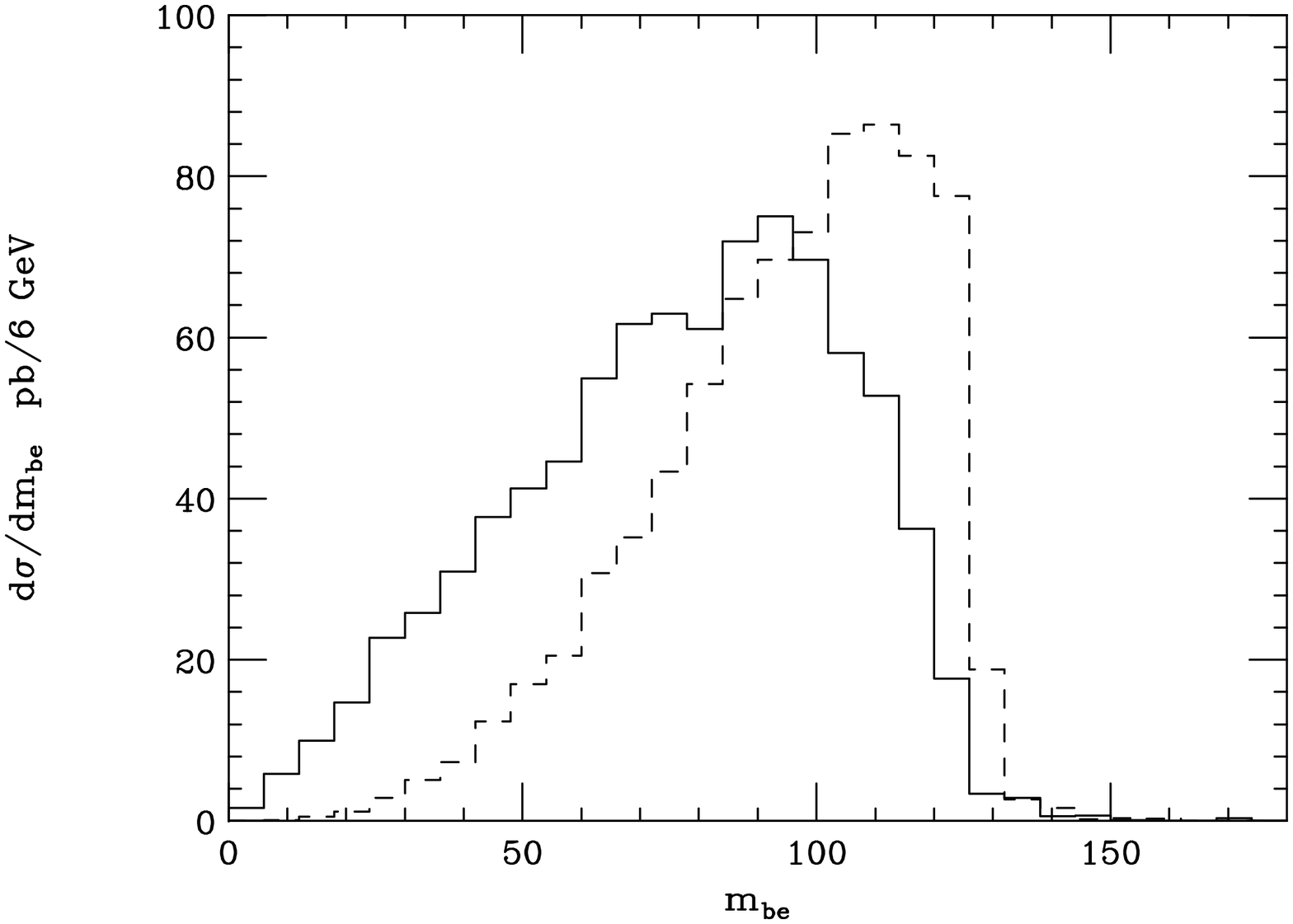,height=4.5in}}}
\caption{ $m_{b{\ell}}$ distribution for SM top quark 
(solid) and for pure right-handed~$\tbW$ coupling of 
$tbW$(dash).}
\label{mbe}
\end{figure}
\begin{figure}[p]
\centerline{\hbox{\psfig{figure=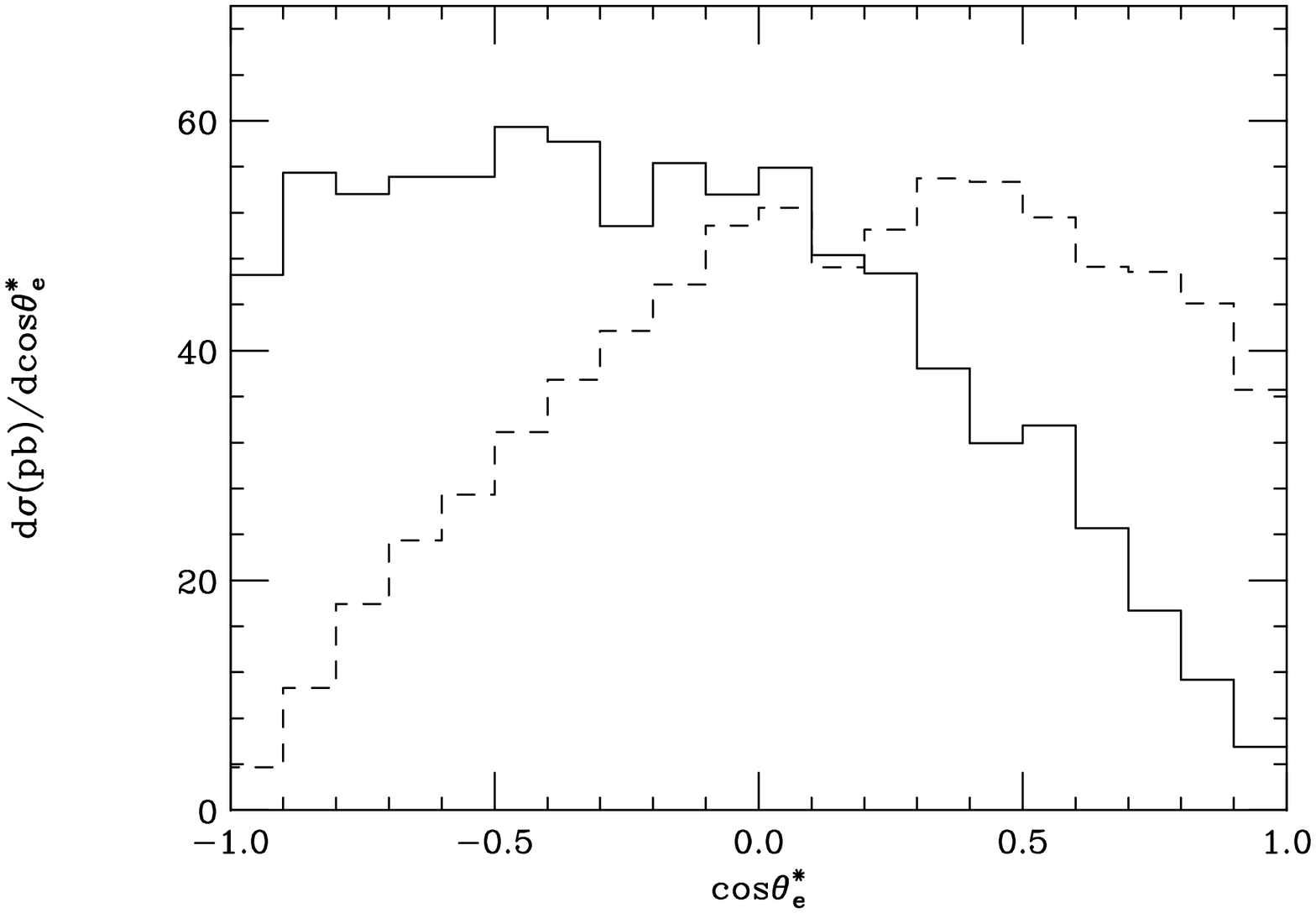,height=4.5in}}}
\caption{ $\cos \theta^*_{\ell}$ distribution for SM top quark
(solid) and for pure right-hand~$\tbW$ coupling of
$tbW$(dash).}
\label{thesta}
\end{figure}
Also, $m_{b \ell}$ is preferentially larger for a pure 
right-handed $\tbw$ vertex than a pure left-handed one.
This is clearly shown in Figure~\ref{mbe}, in which 
the peak of the $m_{b{\ell}}$ 
distribution is shifted to the right and the distribution falls 
off sharply at the upper mass limit for a pure right-handed $\tbw$ vertex.
In terms of  $\cos \theta^*_{\ell}$, their difference is shown in
 Figure~\ref{thesta}. 
However, in both cases the fraction ($\flong$) of longitudinal $W$'s 
from top quark decay is enhanced by ${m_t}^2/{2{M_W}^2}$ as compared 
to the fraction of transversely polarized $W$'s \cite{toppol}, namely,
\beq
\flong = { {m_t^2 \over 2 M_W^2 } \over { 1 + {m_t^2 \over 2 M_W^2 } } } \,.
\enq
 Therefore, for a heavier 
top quark, it is more difficult to untangle the $\klc$ and $\krc$ 
contributions. 
On the other hand, because of the very same reason, the mass of a
heavy top quark can be accurately measured from $\flong$ (discussed below)
irrespective of the nature of the $\tbw$ couplings
(either left-handed or right-handed). 

The QCD production rate of $t \bar t$ is 
obviously independent of $\klc$ and $\krc$. (Here 
we assume the electroweak production rate of $q\bar{q}\ra A,Z \ra t\bar t$
remains small as in the SM.)
Let us estimate how well the couplings  $\klc$ and $\krc$
can be measured at the Tevatron, the \tevs, and the LHC.
First, we need to know the production rates of the top 
quark pairs from the QCD processes.
As shown in Table~\ref{trates},
 the QCD production rate of $gg,q\bar{q}\ra t \bar t$ 
for a 180\,GeV top quark
is about 4.5\,pb, 26\,pb and 430\,pb at the Tevatron, the \tevs, and the LHC,
 respectively. 
For simplicity, let's consider the $\ell^\pm \, + \geq 3\, {\rm jet}$ 
decay mode whose 
branching ratio is ${\rm Br}=2 {\frac{2}{9}} {\frac{6}{9}} = \frac{8}{27}$,
where the charged lepton $\ell^\pm$ can be either $e^\pm$ or $\mu^\pm$. 
We assume the experimental detection 
efficiency ($\epsilon$), which includes 
both the kinematic acceptance and the efficiency of $b$-tagging,
to be 15\% for the signal event \cite{CDF}. 
Let's further assume that there is
no ambiguity in picking up the right $b$ ($\bar b$)
to combine with the charged lepton $\ell^+$ ($\ell^-$)
to reconstruct $t$ (or $\bar t$), then in total there are 
$4.5\,{\rm pb}\,\times \, 10^3\,{\rm pb}^{-1}\,
\times \,{\frac{8}{27}}\,\times \,0.15=200$
reconstructed $t \bar t$ events to be used in measuring 
$\klc$ and $\krc$ at $\sqrt{S}= 2$\,TeV. 
The same calculation at
the \tevs~and the LHC yields 1100 and 19000 reconstructed 
$t \bar t$ events, respectively.
Given the number of reconstructed top quark events,
one can fit the $m_{b\ell}$ distribution to measure
$\klc$ and $\krc$. 
For example we have done a study for the Tevatron. 
Let us assume the effects of 
new physics only modify the SM results ($f_1^L=1$ and $f_1^R=0$
at Born level) slightly and the form factors
$f_2^{L,R}$ are as small as expected from the usual dimensional
analysis~\cite{geor2,dimens}.\footnote{
The coefficients of the form factors $f_2^{L,R}$, assumed to be 
induced through loop effects,  will be a factor of 
${1 \over 16 \pi^2}$ smaller than that of the form factors 
$f_1^{L,R}$.}
We summarize our results on the accuracy of measuring 
$f_1^{L,R}$ for various luminosities 
in Table~\ref{formfac} \cite{danf1}.
(Only statistical errors are included at the 95\% confidence level.)
\vspace{4mm}
\begin{table}
\caption{
Results on the accuracy of measuring 
$f_1^{L,R}$ for various luminosities.
(Only statistical errors are included at the 95\% confidence level.)
}
\label{formfac}
\begin{center}
\begin{tabular}{|l|l|l|l|l|r|}                                    \hline
Integrated  & Number of            & &            &            \\ 
Luminosity  & reconstructed        & ${\Delta f_1^L}\over f_1^L$ 
& $\Delta f_1^R$ & ${\Delta m_t} \over m_t$                    \\
$\ifb$      & $t \bar t $ events.   & &            &            \\ \hline 
1           & 200       & $8\%  $    & $\pm 0.5 $ & $4\%  $    \\ \hline
3           & 600       & $4\%  $    & $\pm 0.3 $ & $2\%  $    \\ \hline
10          & 2000      & $2\%  $    & $\pm 0.2 $ & $1\%  $    \\ \hline
\end{tabular}
\end{center}
\vspace{4mm}
\end{table}

In the same table ({\it i.e.},Table~\ref{formfac}) 
we also show our estimate on how well the mass of the top
quark $m_t$ can be measured from $\flong$. 
By definition of $\flong$, for a SM top quark 
({\it i.e.}, $f_1^L=1$ and $f_1^R=0$), the distribution of  
  $\cos \theta^*_{\ell}$ has the functional form
\beq
F(\cos \theta^*_{\ell}) \sim 
(1 - \flong) \left( { 1- \cos \theta^*_{\ell} \over 2 } \right)^2 
+ \flong \left( { \sin \theta^*_{\ell} \over \sqrt{2} } \right)^2 \, .
\enq
Therefore, $\flong$ can be calculated by fitting with the distribution 
of $\cos \theta^*_{\ell}$, or equivalently with
the distribution of $m_{b\ell}$. 
We prefer to measure $\klc$ and $\krc$ using the distributions of
$m_{b\ell}$ than of $\cos \theta^*_{\ell}$ because 
the former can be directly calculated from the 
measured momenta of $b$ and $\ell$. However,
to convert from the distributions of $m_{b\ell}$ to
$\cos \theta^*_{\ell}$, as given in Equation~(\ref{mbe1}), the effects
from the width of the $W$--boson and the top quark might slightly 
distort the distribution of $\cos \theta^*_{\ell}$.
(Notice that in the full calculation of the scattering amplitudes
the widths of the $W$--boson and the top quark have to be included 
in the Breit-Wigner form to generate a finite event rate.)

However, in reality, the momenta of the bottom quark and the 
charged lepton will be smeared by
detector effects and another problem in this analysis is 
the identification of the right $b$ to reconstruct $t$.
There are three possible strategies to improve the efficiency of identifying
the right $b$.  One is to demand a large invariant mass of the $t \bar t$
system so that $t$ is boosted and its decay products are collimated.
Namely, the right $b$ will be moving closer to the lepton from $t$ decay.
This can be easily enforced by demanding leptons with a larger transverse 
momentum.
Another is to identify the soft (non-isolated) lepton from $\bar b$ decay
(with a branching ratio ${\rm Br}(\bar b \ra \mu^{+} X) \sim 10\%$).  
The other is to statistically determine the electric charge of the 
$b$-jet (or $\bar b$-jet) to be $1/3$ (or $-1/3$) \cite{lepjet}.
All of these methods may further reduce the reconstructed signal rate by
an order of magnitude. How will these affect our conclusion on 
the determination of the non-universal couplings $\klc$ and $\krc$?
It can only be answered by detailed Monte Carlo studies
which are yet to be done.

\section{From the Production of Top Quarks}
 
Here we propose another method to measure 
the couplings $\klc$ and $\krc$
from the production rate of the single-top quark process.

For $m_t=180$ GeV, the sum of the production rates of single-$t$ and 
single-$\bar t$ 
events is about 2\,pb and 14\,pb for $\sqrt{S}= 2$\,TeV and
$\sqrt{S}= 4$\,TeV respectively. The branching ratio of interest is
${\rm Br}=\frac{2}{9}$. The kinematic acceptance 
of this event at $\sqrt{S}= 2$\,TeV is about $0.55$, as shown in Chapter~8.
Assuming the efficiency of $b$-tagging is about 30\%, then
there will be $2\,{\rm pb}\,\times \,10^3\,{\rm pb}^{-1}\,
\times \,{\frac{2}{9}}\,\times \,0.55\, \times \,0.3=75$
events reconstructed for a 1\,$\ifb$ integrated
luminosity.  At $\sqrt{S}= 4$\,TeV, as shown in Chapter~8,  
 the kinematic acceptance of this event is about 
$0.40$ which, from the above calculation, yields
about $3700$ reconstructed events for 10\,$\ifb$ integrated luminosity.
Based on statistical 
error alone, this corresponds to a 12\% and 2\%
measurement on the single-top cross section.
A factor of 10 increase in the luminosity of 
the collider can improve the measurement by a factor of 3 statistically.
Taking into account the theoretical uncertainties, 
as discussed in Chapter~3, we 
examine two scenarios: 20\% and 50\% error on 
the measurement of the cross section for single-top production.
\begin{figure}[p]
\par
\centerline{\hbox{
\psfig{figure=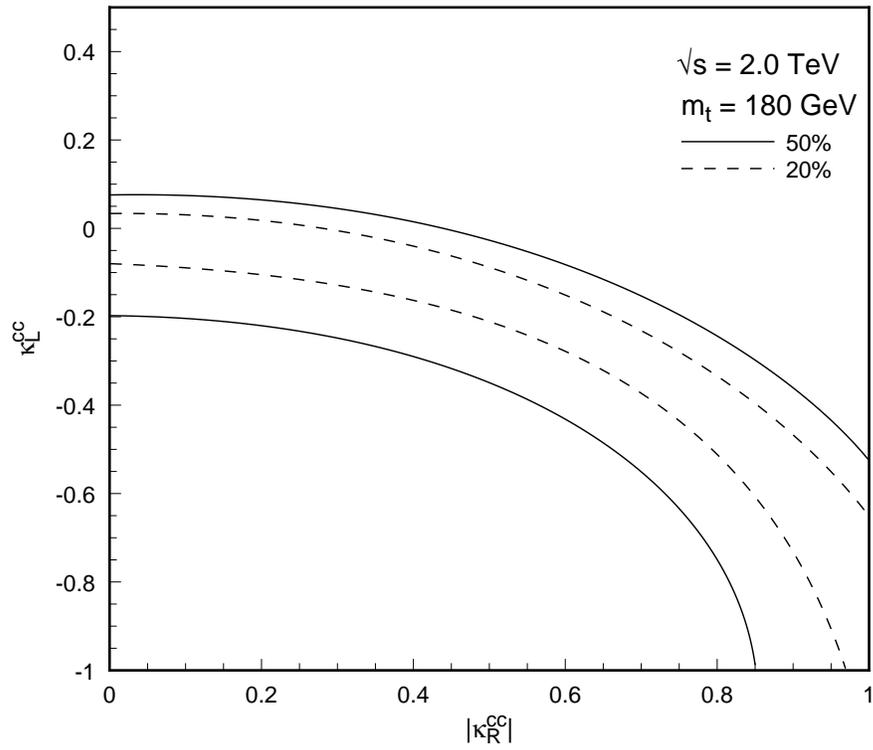,height=4.in}}}
\caption{ Constraint on $|\klc|$ and $\krc$ given $20\%$ and $50\%$ 
error in measurement of Standard Model rate for $W$--gluon fusion. 
Curves are identical for $m_t = 140$ GeV and $m_t = 180$ GeV.
}
\label{klkr}
\par
\end{figure}
The results, which are not sensitive to 
the energies of the colliders considered here (either 2\,TeV or 4\,TeV),
 are shown in Figure~\ref{klkr} for a 180 GeV top quark
at the Tevatron.
We found that $\klc$ and $\krc$ are well constrained inside
the region bounded by two (approximate) ellipses
(cf. Appendix~A).
To further determine the sizes of $\klc$ and $\krc$ one needs to 
study the kinematics of the decay products, such
as the charged lepton $\ell$, of the top quark.
Since the top quark produced from the $W$--gluon fusion process 
is almost one hundred percent left-hand (right-hand) polarized 
for a left-hand (right-hand) $\tbw$ vertex, 
the charged lepton $\ell^+$ from $t$ decay has a harder momentum 
for a right-handed $\tbw$ coupling than for a left-handed coupling.   
(Note that the couplings of 
light-fermions to $W$--boson have been well tested 
from the low energy data to be left-handed as described in the SM.)
As shown in Figures~\ref{left} and~\ref{right}, 
this difference becomes smaller when the top quark is much heavier because
the $W$--boson from the top quark decay tends to be more 
longitudinally polarized.

A right-hand charged current is absent in a 
linearly ${\rm SU(2)}_L$ invariant gauge 
theory with massless bottom quark. 
In this case,  $\krc=0$, 
then $\klc$ can be constrained to within
about $-0.08 < \klc < 0.03$ ($-0.20 < \klc < 0.08$) 
with a 20\% (50\%) measurement on the production rate
of single-top quark at the Tevatron 
\cite{ehab}. (Here we assume the experimental data 
agrees with the SM prediction within 20\% (50\%).) This means that if 
we interpret {\mbox {($1+\klc$)}} as the CKM matrix element $|V_{tb}|$, 
then $|V_{tb}|$ can be bounded as $|V_{tb}| > 0.9$ (or 0.75) for a 20\% 
(or 50\%) measurement on the single-top production rate. 

Before closing this chapter, we remark that
in the Refs.~\cite{ehab} and \cite{fuj} some bounds on the
couplings of $\klc$ and $\krc$ were obtained by studying 
the low energy data with the assumption that
the effects of new physics at low energy can only modify the couplings
of $\klc$ and $\krc$ but not introduce any other light fields
in the effective theory. However, nature might not behave exactly in this 
way. It is possible that some light fields may exist just below the 
TeV scale, then  the bounds
obtained from Refs.~\cite{ehab} and \cite{fuj} may no longer hold.
Thus, it is important to have direct measurements on all 
the form factors listed in Equation~(\ref{eqlag}) from the  production of 
top quarks, in spite of the present bounds on $\kappa$'s derived from 
radiative corrections to low energy data.

\chapter{Probing CP Properties in Top Quarks}

It is known that explicit CP violation requires the presence of both the CP 
non-conserving vertex and the complex structure of the physical amplitude.
Due to the origin of this complex structure, the possible CP-violating
observables can be separated into two categories.
In the first category, this complex structure comes from the absorptive
part of the amplitude due to the final state interactions.
In the second category, this complex structure does not arise from the
absorptive phase but from the correlations in the kinematics of the 
initial and final state particles involved in the physical process.
Hence, it must involve a triple product 
correlation ({\it i.e.}, a Levi-Civita tensor).

To distinguish the symmetry properties between these two cases,
we introduce the transformation $\hatt$, as defined in Reference~\cite{hatt},
which is simply the application of time reversal to all momenta and 
spins without interchanging initial and final states.
The CP-violating observables in the first category are CP-odd and
CP$\hatt$-odd, while those in the second category are
CP-odd and CP$\hatt$-even. Of course, both of them are CPT-even.
 
To illustrate the above two categories, we consider CP-violating
observables for the decay of the top quark.
Consider the partial rate asymmetry
\begin{eqnarray}
{\cal A}_{bW} & \equiv & 
    { \Gamma( t \ra b W^+) - \Gamma({\bar t} \ra {\bar b} W^-)
    \over \Gamma( t \ra b W^+) + \Gamma({\bar t} \ra {\bar b} W^-) }.
\label{abw}
\end{eqnarray}
A non-vanishing ${\cal A}_{bW}$
 clearly violates CP and CP$\hatt$, therefore this observable
belongs to the first category.
We note that 
because of CPT invariance, the total decay width of the top quark
$\Gamma(t)$ has to equal the total decay width of the top anti-quark
$\Gamma(\bar t)$. 
Thus, any non-zero ${\cal A}_{bW}$ implies that 
there exists a state (or perhaps more than one state) $X$ such that
$t$ can decay into $X$ and ${\bar t}$ into ${\bar X}$. 
The absorptive phase 
of $t \ra b W^+$ is therefore generated
by re-scattering through state $X$, {\it i.e.},
$t \ra X \ra b W^+$, where $X \neq b W^+$ because 
the final state interaction should be off-diagonal \cite{wolf}.

Next, let's consider the observable of the second category.
In the decay of $t \ra b W^+ (\ra \ell^+ \nu_{\ell})$, for a polarized $t$
quark, time reversal invariance (T) is violated if 
the expectation value of 
\begin{eqnarray}
\vec{\sigma}_t \times \vec{p}_b \cdot \vec{p}_{\ell^+}
\label{tripro}
\end{eqnarray}
is not zero \cite{toppol}. Assuming CPT invariance, this 
implies CP is violated.
Therefore, this observable is CP-odd but CP$\hatt$-even.
A non-vanishing triple product observable, 
such as that in Equation~(\ref{tripro}), from the decay of the top quark 
violates T. However, it may be entirely due to final state interaction 
effects without involving any CP-violating vertex. 
To construct a truly CP-violating observable, one must combine
information from both the $t$ and $\bar t$ quarks.
For instance, the difference in the expectation values of 
$\vec{\sigma}_t \times \vec{p}_b \cdot \vec{p}_{\ell^+}$ 
and
$\vec{\sigma}_{\bar t} \times \vec{p}_{\bar b} \cdot 
\vec{p}_{\ell^-}$
would be a true measure of an intrinsic CP violation.

There have been many studies
on how to measure the CP-violating effects in the $t \bar t$ system
produced in either electron or hadron collisions.
(For a review, see a recent paper in Reference~\cite{cpcpy}.)
At hadron colliders, the number of 
$t \bar t$ events needed to
measure a CP-violating effect of the order of $10^{-3} - 10^{-2}$
is about $10^7 - 10^8$.
To examine the potential of various current and future
hadron colliders in measuring the CP-violating asymmetries, 
we estimate the total event rates of $t \bar t$ pairs
for a 180 GeV SM top quark produced at these colliders.
At the Tevatron, the Di-TeV and the LHC, an integrated luminosity of
10, 100 and 100 ${\rm fb}^{-1}$ will produce about 
$4.5 \times 10^4$, $2.6 \times 10^6$ and $4.3 \times 10^7$
$t \bar t$ pairs, respectively, as given in Table~\ref{trates}. 
Therefore, the LHC 
would be able to probe the CP asymmetry  of the top
quark at the level of a few percent. A similar 
number of the $t \bar t$ pairs 
is required in electron collisions to probe the CP asymmetry 
at the same level. 
Thus, for a $\sqrt{S}=500\,$GeV $e^-e^+$ collider,
an integrated luminosity of about $10^4 - 10^5$ ${\rm fb}^{-1}$
has to be delivered. This luminosity is 
at least a factor of $100$ higher than the 
planned next linear colliders.
We note that although the initial state in a pp collision 
(such as at the LHC)
is not an eigenstate of a CP transformation,
these CP-odd observables can still be defined as long as 
the production mechanism is dominated by $gg$ fusion. This is 
indeed the case for $t \bar t$ pair production at the LHC. 

In the SM, the top quark  produced via the $W$--gluon fusion
process is about one hundred percent 
left-hand (longitudinally) polarized, see Appendix A.
Given a polarized top quark, one can use 
the triple product correlation, as defined in Equation~(\ref{tripro}), 
to detect CP violation of the top quark.
For  a polarized top quark, one can either use
$\vec{\sigma}_t \times \vec{p}_b$
or $\vec{p}^{\rm Lab}_t \times \vec{p}_b$
to define the decay plane of $t \ra b W (\ra \ell^+ \nu) $.
Obviously, the latter one is easier to implement experimentally.
Define the asymmetry to be
\begin{eqnarray}
{\cal A}_{io} & \equiv & 
      { {   N(\ell^+ \, {\rm out \, of \, the \, decay \, plane}) 
         -  N(\ell^+ \, {\rm into \, the \, decay \, plane }) } \over
           {   N(\ell^+ \, {\rm out \, of \, the \, decay \, plane}) 
           +  N(\ell^+ \, {\rm into \, the \, decay \, plane }) } }~.
\end{eqnarray}
If ${\cal A}_{io}$ is not zero, then the time-reversal T is not conserved,
therefore CP is violated for a CPT invariant theory.
Due to the missing momentum of the neutrino from the decay of the $W$--boson,
it is difficult to reconstruct the azimuthal angle ($\phi_W$)
of the $W$--boson from the decay of the top quark. Once the angle
$\phi_W$ is integrated over, the transverse polarization of the top 
quark averages out and only the longitudinal polarization of the top 
quark contributes to the asymmetry ${\cal A}_{io}$.
Thus, the asymmetry ${\cal A}_{io}$ can be used to study the 
effects of CP violation in the top quark,
which in the SM is about one 
hundred percent left-hand (longitudinally) polarized
as produced from the $W$--gluon fusion process. 
To apply the CP-violating observable 
${\cal A}_{io}$, one needs to reconstruct the directions of both the $t$
and $b$ quarks.
It has been shown in Reference~\cite{onetcp} that it takes about 
$10^7-10^8$ single-top events
to detect CP violation at the order of $\sim 10^{-3} - 10^{-2}$.

For $m_t=180\,$GeV at the Tevatron, the Di-TeV and the LHC, 
an integrated luminosity of
10, 100 and 100 ${\rm fb}^{-1}$ will produce about 
$2 \times 10^4$, $1.4 \times 10^6$ and $2 \times 10^7$
single-$t$ or single-$\bar t$ events, respectively, 
from Table~\ref{trates}. 
At the NLC, the single top quark production rate is 
much smaller. For a $2\,$TeV electron collider,  
the cross sections for 
$e^-e^+ \ra e^- {\bar \nu_e} t {\bar b}$
and
$e^+ \gamma \ra {\bar \nu}_e t {\bar b}$ 
are 8 fb and 60 fb, respectively \cite{eeonetop}.
Hence, it will be extremely difficult
to detect CP violation effects at the order of $\leq 10^{-2}$
in the single-top events produced in electron collisions.

A few comments are in order. First, to extract the {\it genuine} 
CP-violating effects, we need to study the difference in the asymmetry 
${\cal A}_{io}$ measured in the single-$t$ and single-$\bar t$
events because the time-reversal violation in
${\cal A}_{io}$ of the $t$ (or $\bar t$) alone could be 
generated by final state interactions without CP-violating 
interactions.
Second, the detection efficiency for this method is not close
to one, so a good understanding of the kinematics of the decay products and 
how the detector works are needed to make this method useful.

The asymmetry ${\cal A}_{io}$ belongs to the second category of 
CP-violating observables and is CP-odd and CP$\hatt$-even. 
Consider another asymmetry ${\cal A}_t$ which belongs to the first 
category of CP-violating observables and is CP-odd and CP$\hatt$-odd. 
Using ${\cal A}_t$ for detecting CP-violating effects is to make use of 
the fact that $\pbarp$ is a CP eigenstate; therefore, the difference in the
production rates for $\pbarp \ra t X$ and $\pbarp \ra \bar t X$ 
is a signal of CP violation. 
This asymmetry is defined to be 
\begin{eqnarray}
{\cal A}_t & \equiv & 
    { \sigma(\pbarp \ra t X) - \sigma(\pbarp \ra \bar t X) \over
               \sigma(\pbarp \ra t X) + \sigma(\pbarp \ra \bar t X) } ~~.
\end{eqnarray}
As discussed in Chapter~6,
the production rate of $\pbarp \ra t X$ is proportional to 
the decay rate of $t \ra b W^+$ and the rate of 
$\pbarp \ra \bar t X$ is proportional to 
the rate of $\bar t \ra \bar b W^-$. 
This implies that ${\cal A}_t = {\cal A}_{bW}$, cf. Equation~(\ref{abw}).
There have been quite a few models studied in the literature about the
asymmetry in ${\cal A}_{bW}$. For instance, in the Supersymmetric
Standard Model where a CP-violating phase may occur in the
left-handed and right-handed top-squark, ${\cal A}_{bW}$ can 
be as large as a few percent 
depending on the details of the parameters in the model \cite{stopcp}.

Next, let's examine how many top quark events are needed to 
detect a few percent effect in the CP-violating asymmetry ${\cal A}_t$.
Consider $t \ra b W^+ \ra b \ell^+ \nu$, where 
$\ell= e \, {\rm or} \, \mu$. 
Define the branching ratio $B_W$ as the product of 
${\rm Br}(t \ra bW^+)$ and  ${\rm Br}(W^+ \ra \ell^+ \nu)$, where  
${\rm Br}(W^+ \ra \ell^+ \nu)$ is $2/9$.
(${\rm Br}(t \ra bW^+)$ depends on the details of a model and is 
almost 1 in the SM.)
Let us assume that the efficiency of $b$-tagging ($\epsilon_{\rm btag}$)
is about 30\% and the kinematic acceptance 
($\epsilon_k$) of reconstructing the single-top event,
$\pbarp \ra t X \ra b W^+ X \ra b \ell^+ \nu X$,
is about 50\%. (See, a Monte Carlo study in Chapter~8.)
The number of single-$t$ and single-$\bar t$ 
events needed to measure ${\cal A}_t$ is
\beq
{\cal N}_t = { 1 \over B_W \epsilon_{\rm btag} \epsilon_k} 
  \left(1 \over {\cal A}_t \right)^2 ~~.
\enq
Thus, to measure ${\cal A}_t$ of a few percent,
${\cal N}_t$ has to be as large as $\sim 10^6$, which corresponds
to an integrated luminosity of 100\,$\ifb$ at the Di-TeV.

\chapter{ A Monte Carlo Study }

It was shown in Reference~\cite{wgtb} that 
due to the characteristic features of the transverse momentum and
rapidity distributions of the spectator quark which emitted the 
virtual $W$ an almost perfect efficiency 
for ``kinematic $b$ tagging'' can be achieved. 
In addition, the ability of performing $b$-tagging using a 
vertex detector increases the detection efficiency of a heavy top quark
produced via the $W$--gluon fusion process.  
In this chapter we show that this process is useful at 
the Tevatron with the Main Injector.
We also estimate results for $\rtS = 4$ TeV at the Di-TeV
and for $\rtS = 14$ TeV at the LHC in separate subsections.  

To show that a heavy top quark produced from 
the $W$--gluon fusion process can be detected at
the Tevatron, we performed a Monte Carlo study on the $W+\,2\,jets$ mode
of the ($2 \ra 2$) process 
\beq
 q' b \ra q t (\ra b W^+ (\ra {\ell}^+ \nu) )
\label{ethree}
\enq 
with ${\ell}^+=e^+ \,{\rm or}\,\mu^+$.
More specifically, we assume that the $b$-quark jet from the top quark decay 
can be tagged so that the decay mode of interest is 
identified to be  $W+b +\,jet$.
Throughout this study, we assume that the efficiency of 
the $b$-quark tagging is 30\% for $P^b_t > 30\,$GeV
with no misidentifications of a $b$-jet from other QCD jets.
For clarity we only give rates for top quark (not including
top-antiquark) production in this chapter, unless specified otherwise. 
To include the top-antiquark production one can refer to Chapter~3
for its production rate as compared with that of a top quark.

For simplicity we only consider the intrinsic 
backgrounds ({\it i.e.}, those present at the parton level) 
for the $W+b+\,jet$ final state, and will not invoke any detailed
study on effects due to hadronization of the partons or the imperfectness
of the detectors used in experiments. The intrinsic backgrounds
in the SM for the mode $W+b+\,jet$ are the electroweak-QCD process
$q' \bar{q} \ra W + b +\bar b$ and QCD process 
$q{\bar q},\,gg \ra t \bar t \ra W+b+\,jet$. 
\begin{figure}[b]
\hspace{1.in}
\par
\centerline{\hbox{
\psfig{figure=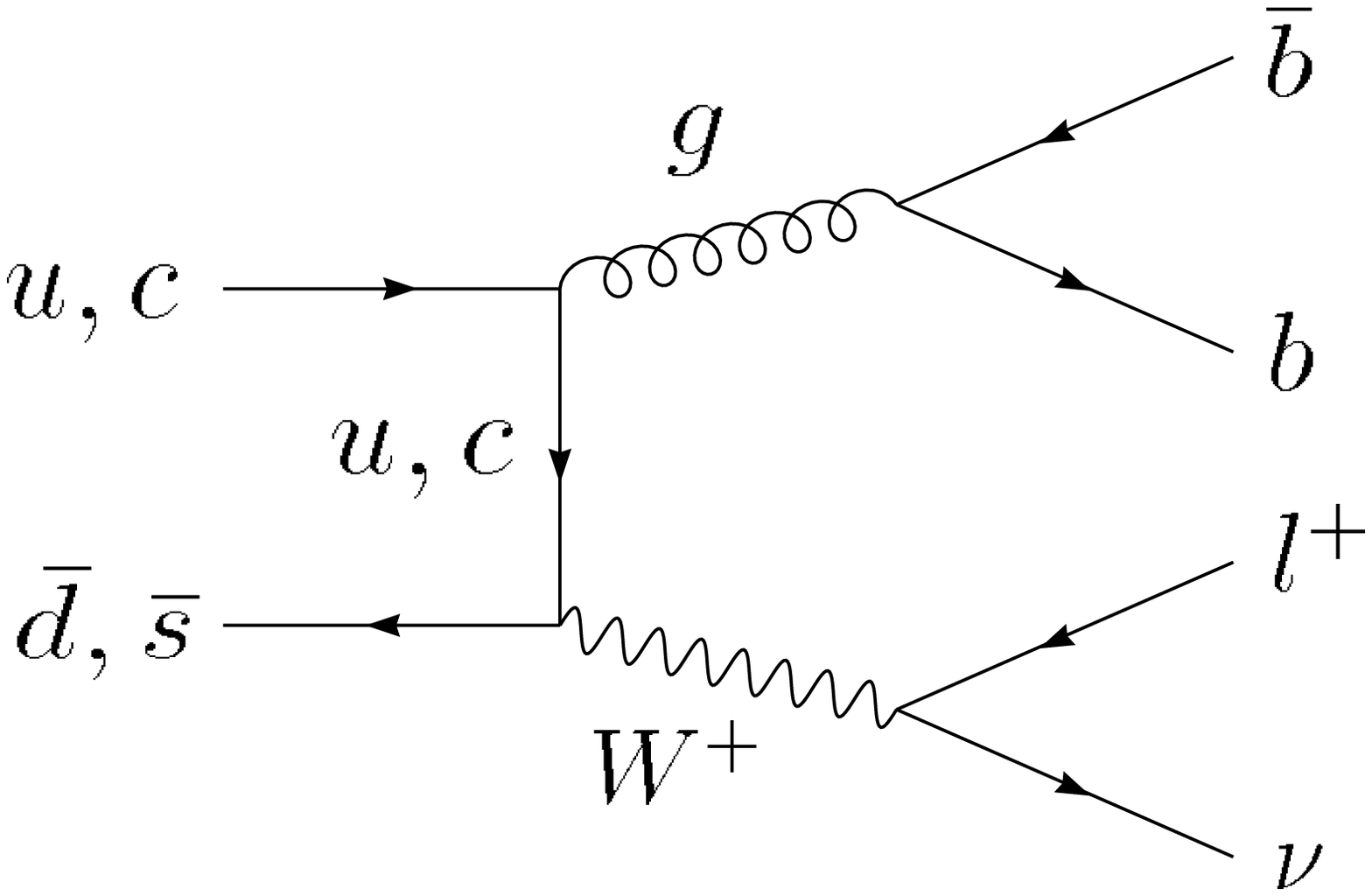,height=1.3in}
\psfig{figure=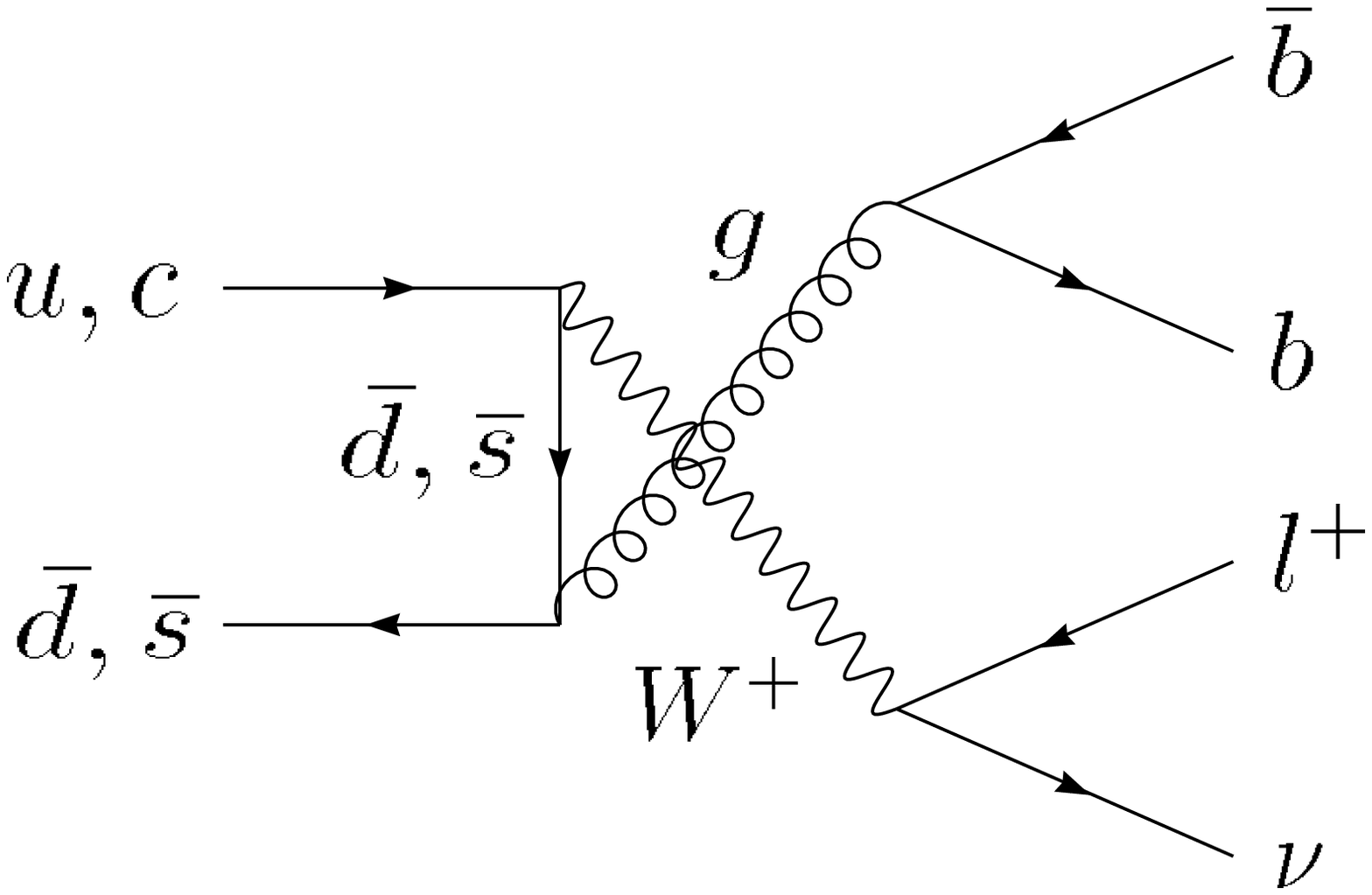,height=1.3in}}}
\caption{ Diagrams for 
$u \bar d, \, c \bar s \ra b \bar b W^+(\ra {\ell}^+ \nu)$. }
\label{udbb}
\par
\hspace{1.in}
\end{figure}
We will show that the dominant backgrounds for the single-top signal 
come from the electroweak-QCD processes (as shown in Figure~\ref{udbb})
\beq                                                            
u \bar d, \, c \bar s \ra b \bar b W^+(\ra {\ell}^+ \nu).
\label{efour}
\enq
The other backgrounds such as $c g \ra b  W^+$ are suppressed due
to the small CKM matrix element 
$|V_{cb}| \simeq 0.03\, {\rm to} \,0.048$ \cite{databook}. 
As done in the previous sections,
we will give our numerical results in this section 
for either a 140\,GeV or a 180\,GeV top quark.

\section{ Tevatron with $\protect\rtS=2$\,TeV }

At the Tevatron 
($\pbarp$, $\sqrt{S}=2\,$TeV) the single-top production rate 
from the $W$--gluon fusion process is 
1(2) pb, as shown in Table~\ref{trates}.
For the final state of Equation~(\ref{ethree}),
the branching ratio of $t \ra b W^+(\ra \ell^+\nu)$ 
for ${\ell}^+=e^+ \,{\rm or}\,\mu^+$ is about $2/9$ in the SM.
As discussed in Chapter~3, we found that
after properly treating the $b$ quark as a parton
inside the proton (or antiproton)
the total rate for the $W$--gluon fusion process 
is about 30\% smaller than that of the ($2 \ra 2$) process.
Hereafter, we shall rescale all the numerical results of our analysis 
for the ($2 \ra 2$) process to the total event rate of the $W$--gluon fusion 
process by multiplying them by a factor of $0.7$. 

To show that a 180 (140) GeV top quark produced from 
this process can be detected at
the Tevatron, we first impose the following kinematic cuts:
\begin{eqnarray}                             
       P_T^q  >  15 \,{\rm GeV},& |\eta^q| < 3.5, \nonumber \\
       P_T^\ell  >  15 \,{\rm GeV},& |\eta^\ell| < 2,   \nonumber \\
       P_T^b  >  35 \,{\rm GeV},& |\eta^b| < 2,   \nonumber \\
 \mynot{E_T}  >  15 \,{\rm GeV},& \Delta R_{qb} > 0.7 ~.
\label{eone}
\end{eqnarray}                             
The efficiency of these cuts for a 180 (140)\,GeV top quark is 
32\%(53\%).
Including the reduction factor from the assumed $b$-tagging
efficiency, 30\%,  
the signal rate is found to be about $0.045 \,(0.063)$ pb.
In Equation~(\ref{eone}), $P_T$ stands for transverse momentum, $\eta$ 
for pseudo-rapidity, $\mynot{E_T}$
for missing transverse momentum, and 
$\Delta R=\sqrt{(\delta \eta)^2 + (\delta \phi)^2}$ with $\phi$ being the
azimuthal angle.
\begin{figure}[p]
\centerline{\hbox{\psfig{figure=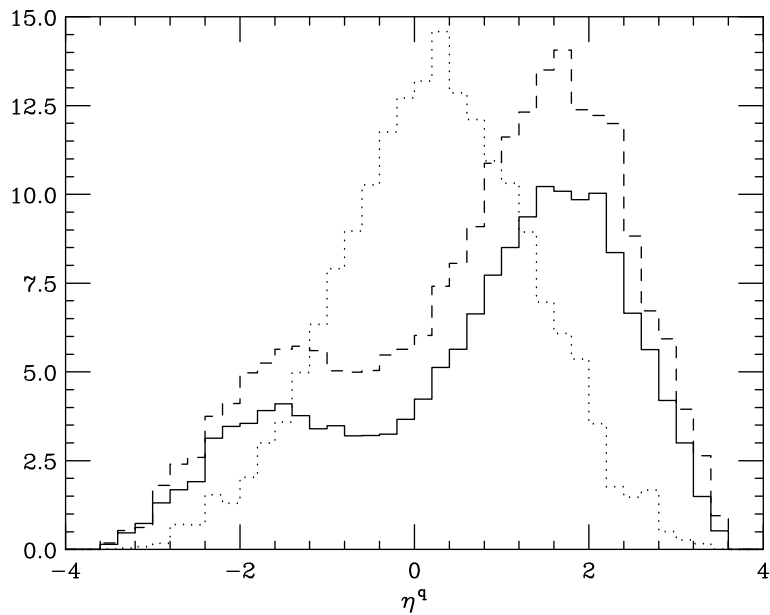,height=4.5in}}}
\caption{ The rapidity distribution
of the spectator quark $q$, after cuts in Equation~(\protect\ref{eone}),
for the signal $q' b \ra q t (\ra b W^+ (\ra \ell^+ \nu) )$, 
and of the spectator quark $\bar b$ for the major background 
$q' \bar{q} \ra \bar b b W^+ (\ra \ell^+ \nu)$ (dots),
for $m_t=180\,$GeV (solid) and 140\,GeV (dash), at the Tevatron. 
(The vertical scale is arbitrary, but the relative size among
these curves are absolute.)
}
\label{etaone}
\end{figure}
It is important to note in Figure~\ref{etaone} that the typical 
rapidity of the spectator jet in the
signal event is about 1.6 although almost all the signal events
have  $|\eta^q| < 3.5$ \cite{wgtb}.
The distribution of $\eta^q$ is asymmetric because the Tevatron is a
$\pbarp$ collider. To produce a heavy top quark,
which decays to a positively charged lepton, the valence quark from 
the proton is most important, implying a large probability for
$\eta^q$ to be positive. (We define the positive z-direction to be the proton
moving direction in the laboratory frame.) 
Similarly, a top-antiquark produced from the $W$--gluon fusion process
would prefer a negative $\eta^q$ due to the large up-antiquark PDF inside 
the antiproton.

In the $W+b \bar b$ background process, the $b \bar b$ pair comes from 
a virtual gluon conversion, therefore its rate is highly suppressed 
if the invariant mass of the $b \bar b$ pair is large.
Since both 
$b$ and $\bar b$ have about the same transverse momentum ($P_T$)
in the background event, the requirement of $P_T^b > 35\,$GeV 
effectively forces a similar $P_T$ cut on $\bar b$. 
This generates a large invariant mass of $b$ and $\bar b$ ({\it i.e.},
the virtuality of the gluon),
strongly suppressing the background rate.
In contrast, in the signal event the 
final parton $q$ (from $q'$, after emitting a virtual $W$) typically has
a smaller $P_T$ than the $b$-quark (from the decay of a heavy top quark). 
Typically, in the signal event, $P_T^b \simeq m_t/3$.
\begin{figure}[p]
\centerline{\hbox{\psfig{figure=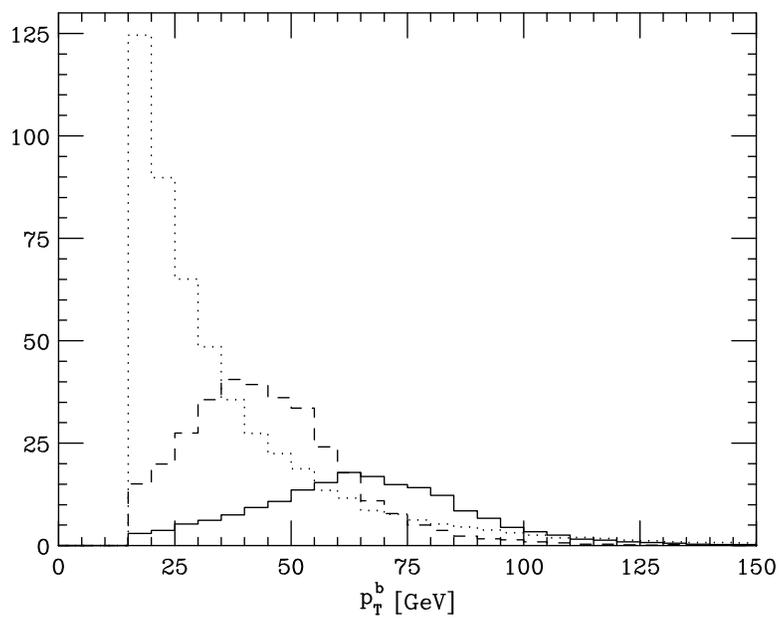,height=4.5in}}}
\caption{ 
$P_T$ distribution of the $b$ quark, after requiring 
$P^b_T > 15\,$GeV along with all the 
other cuts in~(\protect\ref{eone}), for the signal 
$q' b \ra q t (\ra b W^+ (\ra \ell^+ \nu) )$, and the major 
background $q' \bar q \ra \bar b b W^+ (\ra \ell^+ \nu)$,
at the Tevatron.}
\label{ptbone}
\end{figure}
In Figure~\ref{ptbone} we show the $P_T$ distribution of the tagged $b$
from $t$ in both the signal and the background events.
Hence, demanding an asymmetric cut on $P_T$ 
({\it i.e.}, $P_T^q > 15\,$GeV and
$P_T^b > 35\,$GeV) will suppress background effectively and keep
most of the signal events.     
This is why an asymmetric cut on $P_T$ 
was used in our analysis to suppress the major background
process $W+b \bar b$. 
To compare the efficiency of this asymmetric cut in $P_T$, we
note that using $P_T^b > 15$ GeV 
along with all the other cuts in~(\ref{eone}) 
yields a signal-to-background ratio (S/B) of about $1/3$($2/3$).
Requiring $P_T^b > 35$ GeV excludes about $60\%$ of the background
events sacrificing about $10\%$($30\%$) of the signal.

After imposing the kinematic cuts in Equation~(\ref{eone}),
we found that S/B $\simeq 0.9(1.3)$.
However, the signal-to-background ratio can be further 
improved by imposing
\beq
\cos \theta_{{\ell}q} > -0.4.
\label{etwo}
\enq								
Because the top quark produced from the $W$--gluon fusion process
is left-hand polarized, ${\ell}^+$ tends to move against
the moving direction of the top quark in the center-of-mass 
frame of $q$ and $t$, cf. Figure~\ref{left}.    			
\begin{figure}[p]
\centerline{\hbox{\psfig{figure=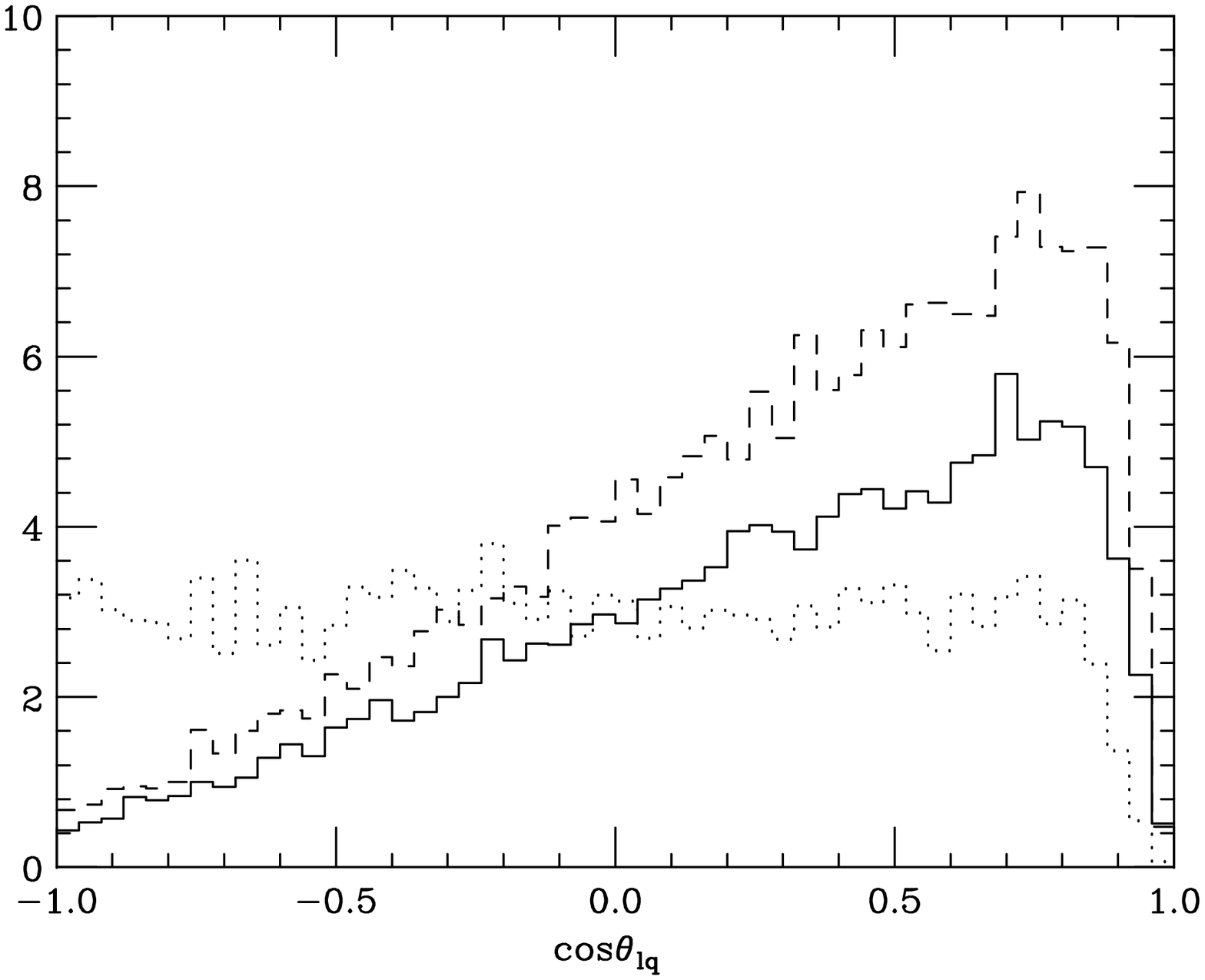,height=4.5in}}}
\caption{ The $\cos \theta_{{\ell}q}$ distribution 
prior to cut Equation~(\protect\ref{etwo}) for the signal 
$q' b \ra q t (\ra b W^+ (\ra \ell^+ \nu) )$ and the major background 
$q' \bar q \ra \bar b b W^+ (\ra \ell^+ \nu)$,
at the Tevatron.}
\label{cthone}
\end{figure}
However, in the background event, the distribution 
of $\cos \theta_{{\ell}q}$,
as shown in Figure~\ref{cthone}, is almost flat after imposing the 
cuts of~(\ref{eone}). 
($\theta_{{\ell}q}=\pi-\theta_\ell,$ where
$\theta_l$ is the the polar angle of ${\ell}^+$ 
in the rest frame of $t$ defined in the center-of-mass frame
of $q$ and $t$.)

To calculate $\cos \theta_{{\ell}q}$, the $P^\nu_{\SST Z}$ information must be
constructed. 
Since both $\ell^+$ and $\nu$ come from a real $W^+$ boson, 
we can use the $W$--boson mass constraint 
\beq
 M^2_{W} = {(p_\ell +p_\nu)}^2
\label{eighteen}
\enq
and the $\mynot{E_T}$ information 
to specify the longitudinal momentum $P^\nu_{\SST Z}$ of the neutrino.  
There are two solutions 
for  $P^\nu_{\SST Z}$ and typically,
both of them are physical solutions for a signal event. 
Therefore, one has to fix a
prescription to choose the one which will most likely give the correct
distribution of the invariant mass of $\ell^+, \nu$ and $b$.
We choose the solution which has the
smaller $\left|P^\nu_{\SST Z}\right|$.  
Here we exploit the fact that the $W$ has finite width.
If a physical solution for $P^\nu_{\SST Z}$ is not found 
with $M_{ W} = 80$ GeV, we generate a resonant mass of the $W$--boson
using a Breit-Wigner distribution.
We use a full half-maximum width
of the $W$--boson, where $\Gamma_W = 2.1\,$GeV,
and solve for $P^\nu_{\SST Z}$, repeating the algorithm for up to three trials
if necessary.
We found that the survival probability for
finding a solution using this algorithm is about $90\% $, and
the difference between this solution and the value actually generated by
the Monte Carlo generator is a Gaussian distribution peaking at 0 with
a width about the order of $\Gamma_W$.
After the additional cut imposed on 
$\cos \theta_{{\ell}q}$, we obtained S/B $\simeq 1.2(1.8)$.
About $55\%$($40\%$) of the total signal event rate remains after applying
the cuts~(\ref{eone}) and~(\ref{etwo}) to the process~(\ref{ethree}).
We conclude that for an integrated luminosity of 1\,${\rm fb}^{-1}$
at a 2\,TeV $\pbarp$ collider, 
there will be about $75(105)$ signal events detected 
with a significance ${\rm S/{\sqrt{B}}}$ of about $10(14)$,  
including both the single-$t$ and single-$\bar t$ events 
as defined in (\ref{ethree}).

\begin{figure}[p]
\centerline{\hbox{\psfig{figure=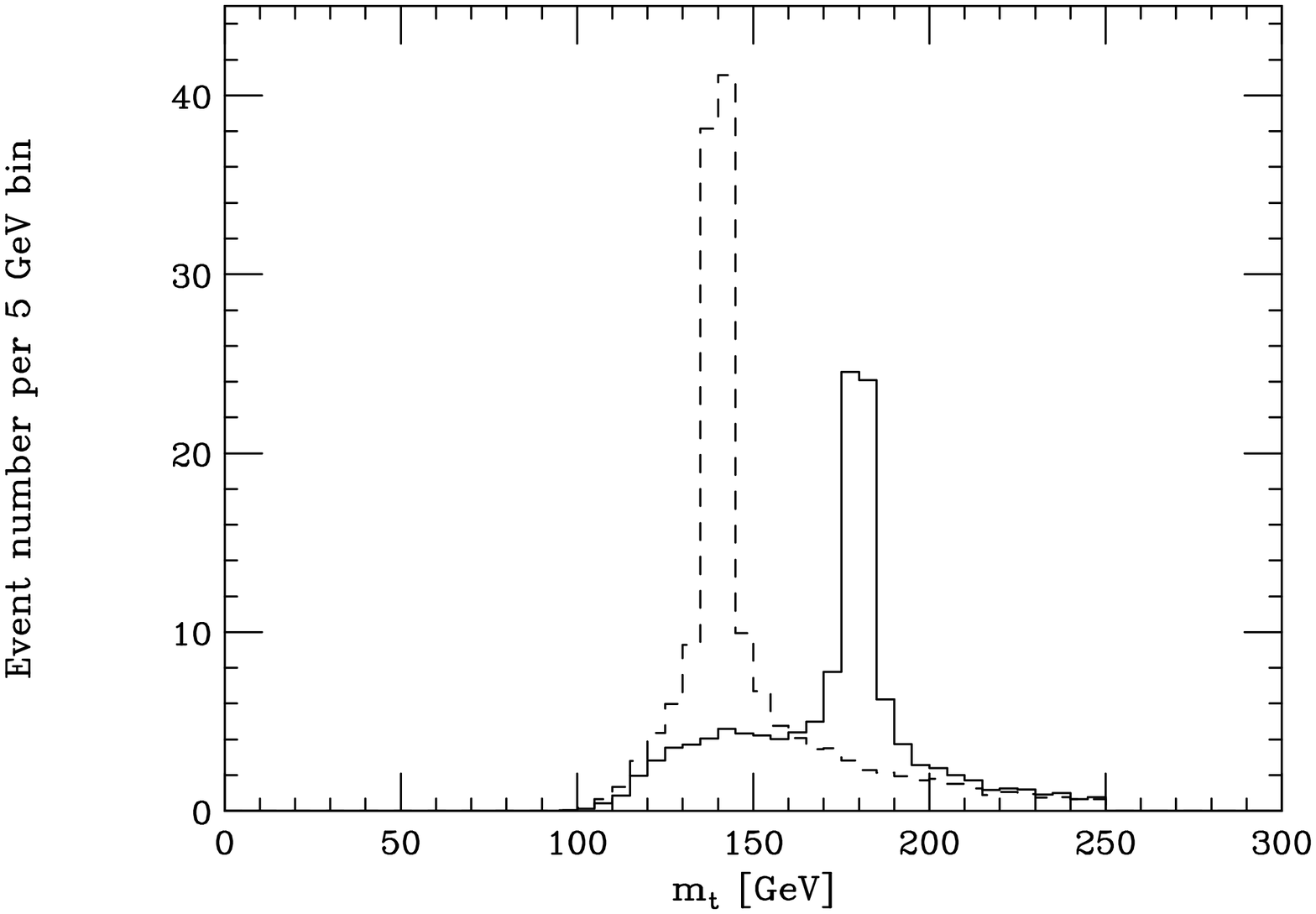,height=4.5in}}}
\caption{ The $m_t$ distribution after the
cuts Equation~(\protect\ref{eone}) and Equation~(\protect\ref{etwo})
for $m_t$ = 180 GeV (solid) and 140 GeV (dash) at the Tevatron 
including both the signal and background events with 
$W^\pm \ra e^\pm \, {\rm or} \, \mu^\pm$. }
\label{mtone}
\end{figure}
To measure the mass of the top quark, we calculate the 
reconstructed invariant mass ($m_t$) of the top quark using
\beq
m^2_t={(p_b + p_\ell + p_\nu)}^2.
\label{nineteen}
\enq
The distribution of $m_t$, including both the signal and the backgrounds,
is shown in  Figure~\ref{mtone}, in which a clear mass peak 
appears unmistakably.
Therefore, we conclude that the top quark can be detected
and studied via this process at the Tevatron.
In Figure~\ref{mtone}, other less important backgrounds, such as 
$t \bar t$ events, were also included.
 
The $t \bar t$ background is not important after
vetoing the events with more than 2 jets \cite{wgtb}.
To support this we did a study for the two decay modes
of $t \bar t$:
\beq                                                         
t \bar t \ra b W^+(\ra \ell^+ \nu) \, \bar b W^-(\ra q' \bar q) 
\enq
and 
\beq
t \bar t \ra b W^+(\ra \ell^+ \nu) \, \bar b W^-(\ra \ell_2^- \bar \nu),
\qquad \ell_{2}=e \, {\rm or} \, \mu.
\enq
For both modes we require one of the jets to be a $b$ or $\bar b$.

Consider first the $\bar t \ra \bar b q' \bar q$ decay mode.  We require at
least one of the $q'$- and $\bar q$-jets and $\bar b$ (if $b$ is tagged) or
$b$ (if $\bar b$ is tagged) to be within rapidity 3.5, otherwise we reject 
the event.  If any one (and only one) of the untagged jets is 
within rapidity 3.5 we call
it the spectator jet and then apply our cuts.
If two of the untagged jets are
within rapidity 3.5, then we require their $\Delta R$ separation
to be less than 0.7 to classify them as
one spectator jet.
When all three untagged jets are within rapidity 3.5 we choose the jet
with the largest $P_T$ and check its $\Delta R$ separation with the other two.
If the lower $P_T$ jets are within $\Delta R = 0.7$ of the
high $P_T$ jet, we call this the 
spectator jet and apply our cuts, otherwise, reject the event.
Recall from Table~\ref{trates} for a 180 GeV top quark, 
the $t \bar t$ rate is about 4.5 pb. 
After applying the cuts (\ref{eone}) and (\ref{etwo}) 
and including the branching ratio for this mode, 
${2 \over 9}{6 \over 9}={4\over 27}$, the event rate, 
of approximately $3\times 10^{-4}$ pb, is very small as compared with
the signal rate of 0.075 pb (including $t$ and $\bar{t}$). 
This is because for most of the $\bar t \ra \bar b q' \bar q$  
decay modes all three jets are within rapidity 3.5 and 
$\Delta R_{bj}$ and $\Delta R_{\bar b j}$ are in general large .

For the $\bar t \ra \bar b \ell_2^- \bar \nu$ decay mode we require that 
$\ell_2$ be undetected.  Specifically, if $\ell_2$ is within rapidity 2 
with $P_T^{\ell_2} > 15 \,$GeV we reject the event.
If $2 < |\eta^{\ell_2}| < 3.5$ we require $P_T^{\ell_2}$ 
to be less than the minimum $P_T$ for detecting leptons,
{\it i.e.}, $< 15\,$GeV in accordance with Equation~(\ref{eone}).
After the cuts in (\ref{eone}), this mode
(with branching ratio ${2 \over 9}{2\over 9}={4\over 81}$)
already suffers, being  about 5\% of the signal rate.
This mode suffers another factor of about two loss 
to the failure of reconstructing $P^\nu_{\SST Z}$ due to 
the presence of two neutrinos in the final state.
After imposing the $\cos \theta_{{\ell}q}$ cut the rate for
$
t \bar t \ra b W^+(\ra \ell^+ \nu) \, \bar b W^-(\ra \ell_2^- \bar \nu)
$
is about $3\times 10^{-3}$ pb which is about a factor of 25
smaller than the signal rate. 
Hence the dominant background (of the same order as the signal rate)
comes from the electroweak-QCD processes
as given in Equation~(\ref{efour}).

As summarized in Figure~\ref{mtone}, even with the very minimum kinematic cuts
of (\ref{eone}) and (\ref{etwo}) the single-top signal can already be
detected, assuming a perfect detector with $b$-tagging efficiency of 30\%.
To incorporate the effects of detector efficiencies, 
we smear the final state
parton momenta using a Gaussian distribution with
\beq
(\Delta E / E)_\ell=15\% / \sqrt{E}, \quad {\rm and} \quad
(\Delta E / E)_{q,b}=50\% / \sqrt{E}.
\label{efive}
\enq
\begin{figure}[p]
\centerline{\hbox{\psfig{figure=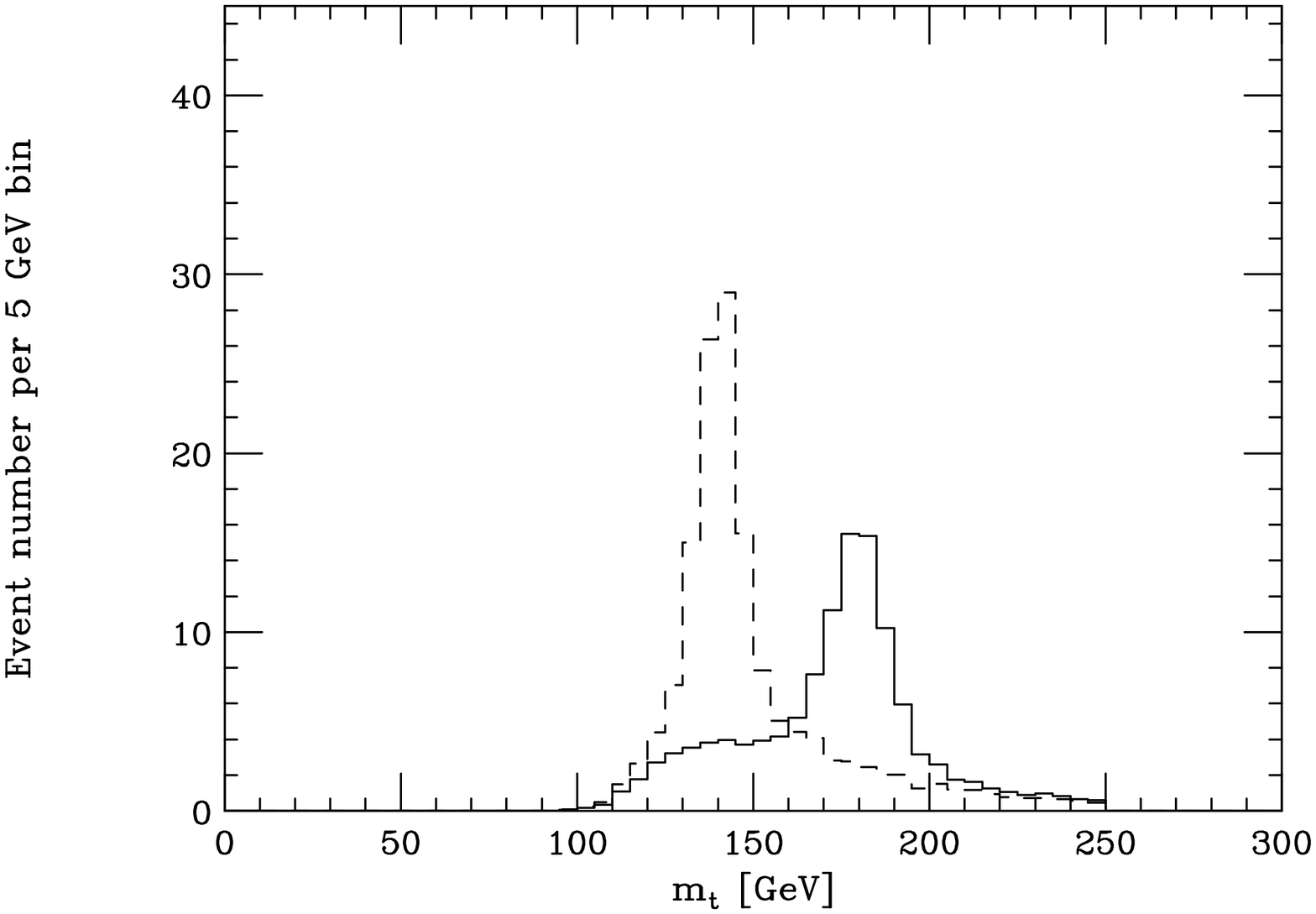,height=4.5in}}}
\caption{ Same as Figure~\protect\ref{mtone}, but with detector 
resolution effects as described in Equation~\protect\ref{efive}. }
\label{mtsone}
\end{figure}
The $m_t$ distribution becomes slightly broader as shown 
in Figure~\ref{mtsone};
however, both the signal and the background rates are almost 
the same as those obtained with a perfect detector.
\begin{figure}
\centerline{\hbox{\psfig{figure=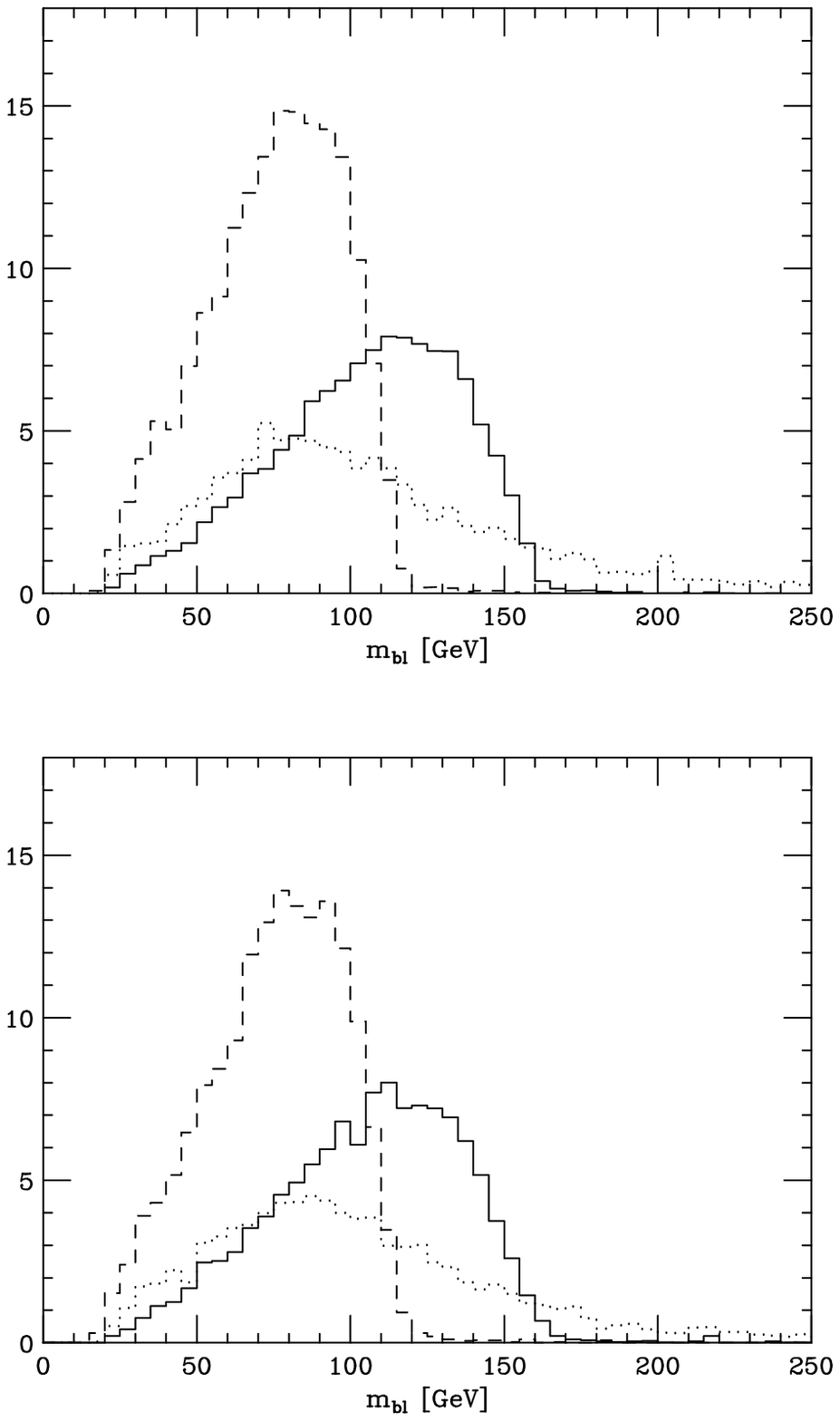,height=8in}}}
\caption{ $m_{b\ell}$ distributions 
without (top) or with (bottom) smearing, after all the above analysis
for the Tevatron. }
\label{mbef}
\end{figure}
As discussed in the previous chapters, the distribution of the invariant
mass $m_{b\ell}$ is extremely useful in either determining the mass of the
top quark or measuring the form factors of $\tbw$. 
For completeness we show in Figure~\ref{mbef} the distribution 
of $m_{b\ell}$ with or without smearing, after all the above analysis.
Because the $b$-jet is required to have large
$P_T$, cf. Equation~(\ref{eone}), so $\Delta E /E$ for the $b$-jet is small,
therefore the two $m_{b\ell}$ distributions do not differ much and
the difference becomes smaller for larger $m_{b\ell}$.
Thus, the position of the bump at large $m_{b \ell}$ 
in the signal events remain a good signature for detecting the 
single-top signals and determining the mass of the top quark 
(discussed in Chapter~4) or the couplings 
of $\tbw$ (discussed in Chapter~6).
 			
We note that the data sample
obtained after all the above analysis
 can be further purified at the cost
of somewhat reducing the signal rates. This can be easily done,
for instance, by noting the distinct differences between 
the signal events and the background events from
the distributions of rapidity of the spectator jet (Figure~\ref{etaone}), 
transverse momentum of the bottom quark (Figure~\ref{ptbone}),
and the angular correlation $\cos \theta_{\ell q}$ (Figure~\ref{cthone})
due to the polarization of the top quark in signal events. 
However, for a more realistic simulation, one should also consider
the possibility of having a charm-jet (or even an ordinary QCD-jet) 
faking a bottom jet in $b$-tagging so that the actual background
rate measured by the detector would be larger than that 
given here. This is outside the scope of our parton level study.

As discussed in Chapter~3, another process which produces a single-top 
is the $W^*$ production
\beq
q' \bar q \ra W^* \ra t \bar b.
\label{ewstar}
\enq
At 2 TeV for a 180 (140) GeV top quark,
the $W^*$ production rate is about $1/5$($1/3$) of the $W$--gluon 
fusion rate.  
Applying the kinematic cuts defined in Equation~(\ref{eone}) we find that 
the $W^*$ process passes with about the same
efficiency as the $W$--gluon fusion process.
However, there are a few obvious differences in the kinematics of 
their final state partons.
First, in the $W^*$ event, there are two $b$-jets (one for $b$ from $t$ decay
and another for $\bar b$ from production), therefore there is 
a 50\% chance of tagging the wrong $b$ and giving the wrong 
reconstructed top quark invariant mass, as defined in 
Equation~(\ref{nineteen}).
To improve the invariant mass distribution of the top quark, one has to 
be able to distinguish a $b$-jet from a $\bar b$-jet by making 
further selections at the cost of 
reducing the single-top rate from this process.
(Some of the techniques have been discussed in Chapter~4.)
\begin{figure}[p]
\centerline{\hbox{\psfig{figure=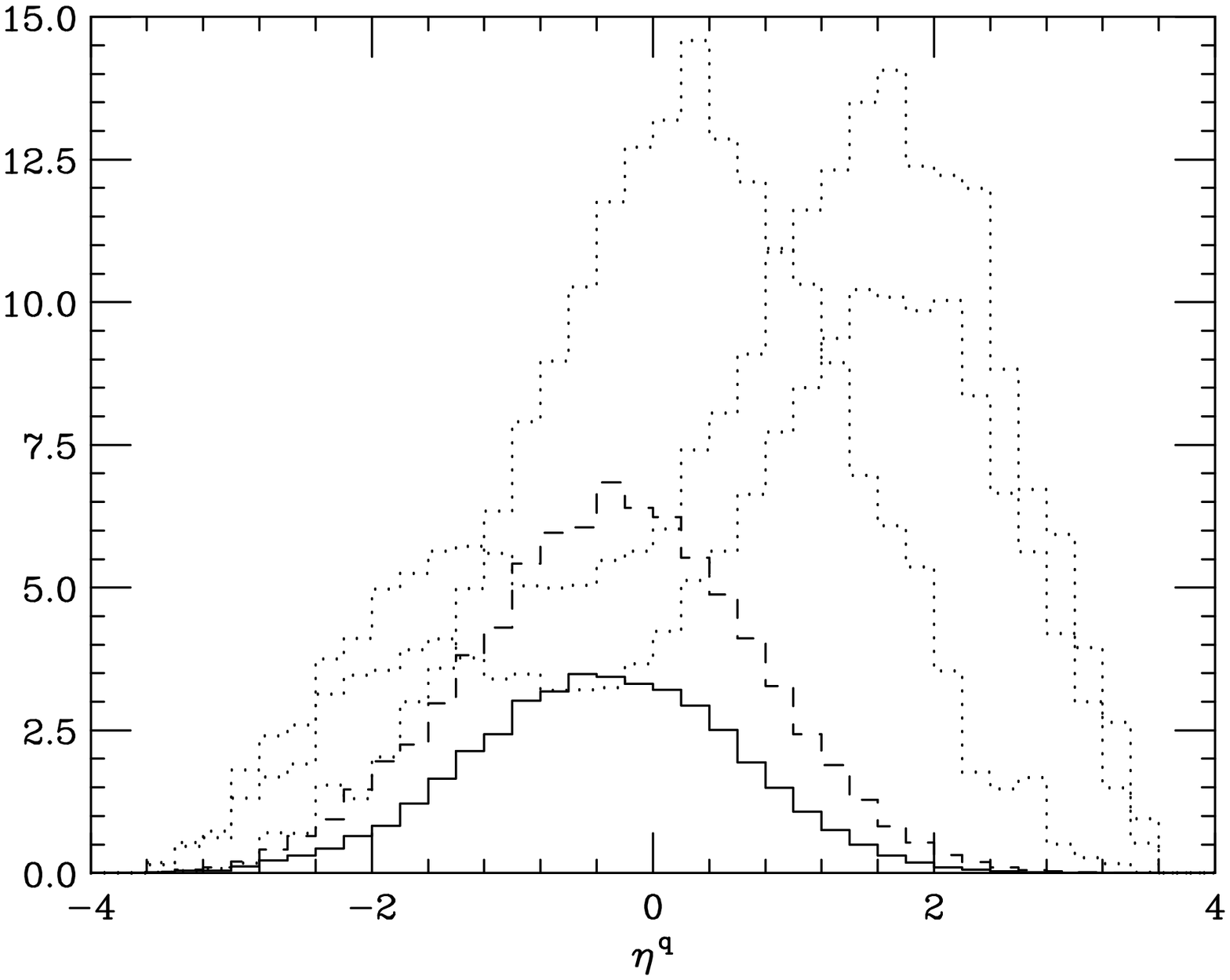,height=4.5in}}}
\caption{ The rapidity distribution, after Equation~(\protect\ref{eone}),
of the spectator jet 
({\it i.e.}, $\bar b$-jet if $b$-jet identified) in the $W^*$ event 
for $m_t$=180 GeV (solid) and 140 GeV (dash)
against Figure~\protect\ref{etaone} (dots), at the Tevatron. }
\label{etaws}
\end{figure}
Second, the rapidity distribution of the spectator jet 
({\it i.e.}, $\bar b$-jet if $b$-jet identified) in 
the $W^*$ event peaks around zero ({\it i.e.}, central, 
as shown in Figure~\ref{etaws})
because $t \bar b$ is produced through 
the s--channel process.\footnote{
We note that the rapidity distribution of the $\bar b$-quark 
in the $W^* \ra t {\bar b}$ event is slightly asymmetric around zero. 
It favors a slightly negative rapidity. (Recall that in the $W$--gluon 
events for producing single-$t$, the rapidity of the spectator 
quark $q$ favors positive values.)   
This is similar to the lepton rapidity asymmetry observed in the 
$\pbarp \ra W^+ \ra \ell^+ \nu$ events due to the ratio of the down-quark
and the up-quark parton distributions inside the proton and the 
anti-proton. 
} This is in contrast to that 
in the $W$--gluon fusion event where the rapidity distribution ($\eta^q$)
of the spectator jet (labeled as $q$-jet)  
is asymmetric and less likely to be around zero.
Third, the polarization of the top quark produced from the
$W^*$ process is not purely left-hand polarized as in the case of
the $q'b \ra qt$ process. 
For a 180 (140)\,GeV top quark, the ratio of the event rates for
producing a left-handed top versus a right-handed top 
in the $W^*$ event is about 3.5(3.4).
\begin{figure}[p]
\centerline{\hbox{\psfig{figure=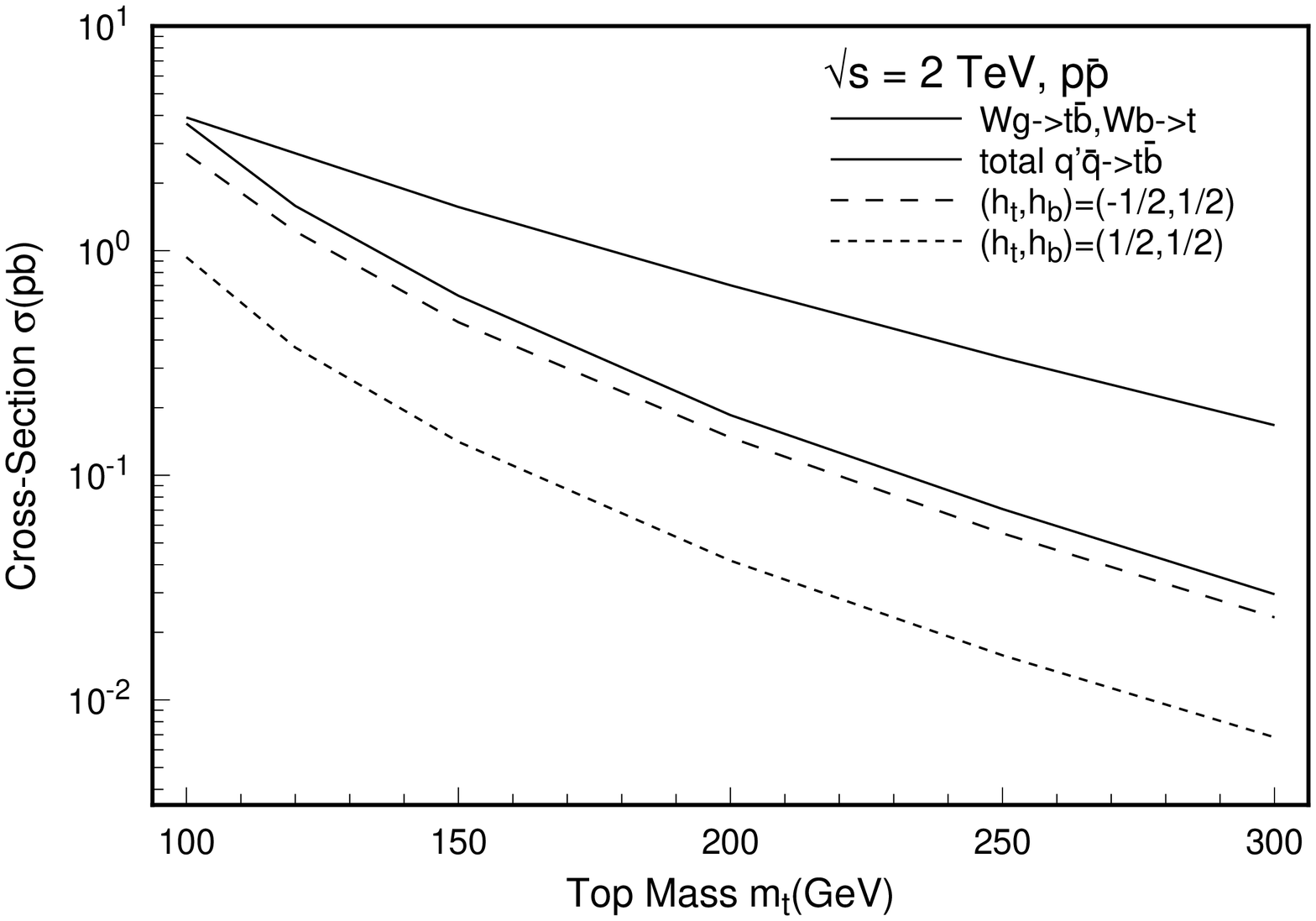,height=3.5in}}}
\caption{ The production rate
for a left-handed (long dash) or a right-handed (short dash) 
top quark from the $W^*$ process. 
The upper solid line is the total rate for the
$W$--gluon fusion process, the
lower solid line for the $W^*$ process.
}
\label{ratews}
\end{figure}
In Figure~\ref{ratews}, we show the production rate
for a left-handed or a right-handed top quark from the $W^*$ process
and compare them with that from the $W$--gluon fusion process
as a function of $m_t$.
Because the top quark is not 100\% polarized in the $W^*$ process, the 
angular correlation of $\ell$ and the spectator jet 
will not be as strong as that in the $W$--gluon fusion process. 
\begin{figure}[p]
\centerline{\hbox{\psfig{figure=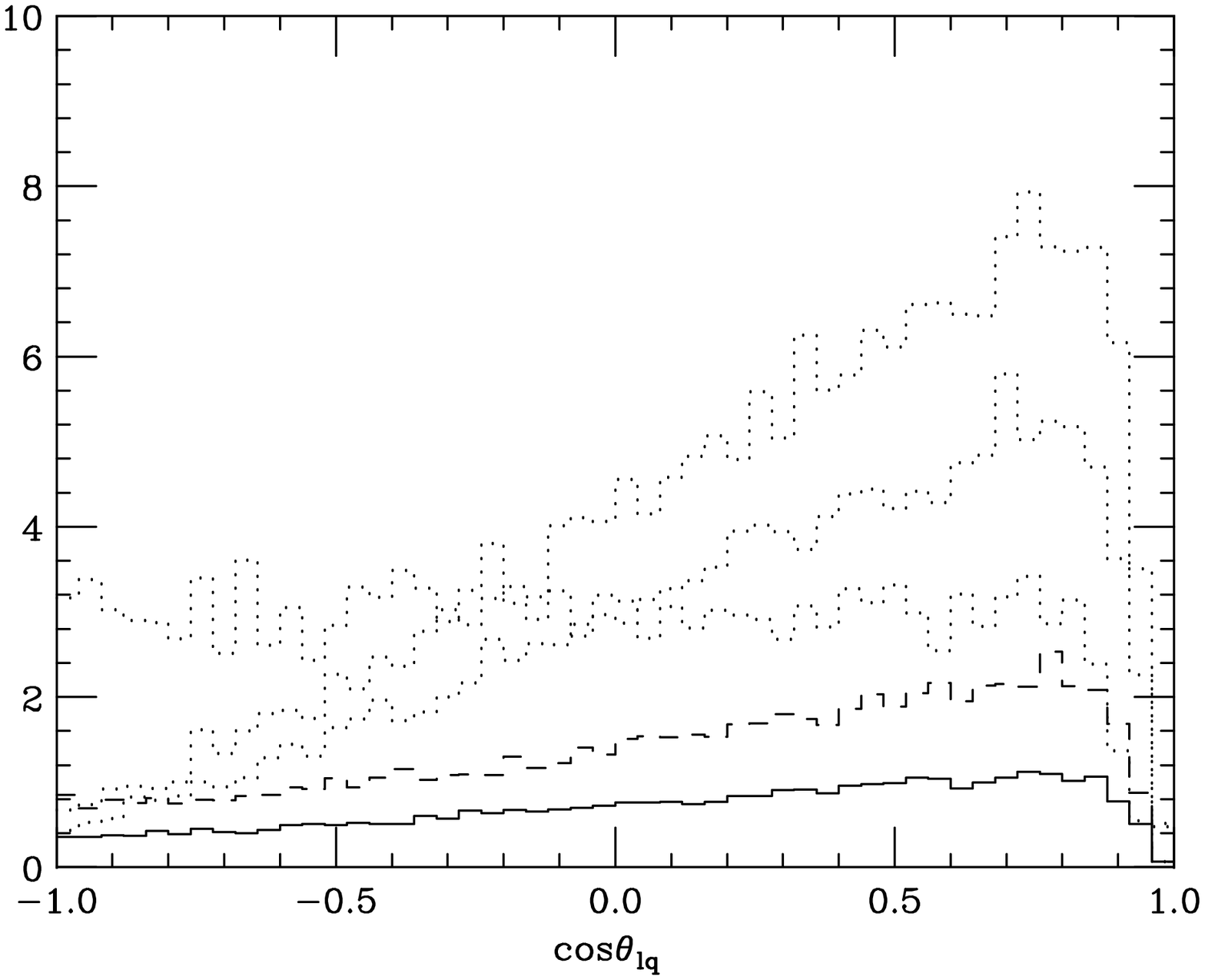,height=4.5in}}}
\caption{ The distribution of
$\cos \theta_{\ell q}$ in $W^*$ event
for $m_t$=180 GeV (solid) and 140 GeV (dash)
against Figure~\protect\ref{cthone} (dots),
at the Tevatron. }
\label{cthws}
\end{figure}
In Figure~\ref{cthws} we show the distribution of
$\cos \theta_{\ell q}$ in $W^*$ events for a 140 and 180 GeV top quark.
(Here, $q$-jet denotes the spectator jet.)
Following through the previous analysis done for the $W$--gluon fusion 
events we found that $W^*$ production compliments the $W$--gluon 
fusion process by increasing the single-top production rate by about
$10\%$.\footnote{
The $W^*$ production rate is about one fifth of the $W$--gluon fusion 
rate for a 180\,GeV top quark, and the kinematic acceptance of the 
$W^*$ event is about half  of that of the $W$--gluon fusion event.}
Therefore, its contribution to our final results of
various distributions is small.

In conclusion we found that at the Tevatron 
($\pbarp$, $\sqrt{S}=2\,$TeV) the single-top production rate 
from the $W$--gluon fusion process after including the branching ratio
for $t \ra b W^+(\ra \ell^+ \nu)$ is about 0.22(0.44) pb for 
a 180 (140) GeV top quark, where $\ell^+=e^+\, {\rm or}\,\mu^+$.
The kinematic acceptance after the kinematic cuts (\ref{eone})
and (\ref{etwo}) is about 55\%(40\%). Assuming a 30\% $b$-tagging efficiency
we concluded that the single-top event rate 
from the $W$--gluon fusion process is about 0.036(0.052) pb.
For an integrated luminosity of 1 $\ifb$, this yields 
36(52) reconstructed single-top events.
(To include top-antiquark production, a factor 2 should be included.)
The dominant background process is the electroweak-QCD process 
$W+b\bar b$ whose rate is about $80\%$($60\%$) of the signal rate in 
the end of the analysis. The $t \bar t$ events are not as important
to our study. The $t \bar t$ rate for a 180 GeV top quark is only 
$0.4\%$ and $4\%$ of the signal rate for its lepton+jet 
and di-lepton mode, respectively.
In both Figs.~\ref{mtone} and~\ref{mtsone} for the distribution of
$m_t$, we have also included another single-top
production process (a single-top produced from $W^*$) which
increases the single-top rate by about 10\%.

Let us make a side remark about the dominant background
$q'\bar q \ra W+b\bar b~$ before we close this section.
In the above analysis we did not include the possibility of having 
an additional QCD jet from either 
the radiation or the conversion of the incoming quark jet ( $q'$ or
$\bar q$). The concern is that this jet may be identified as a forward jet 
which would fake the
single-top signal event. In this case, $b$ and $\bar b$ in 
the  $W+b\bar b+\,jet$ background event
have to both fall into a cone of $\Delta R=0.7$ in order to fake the 
tagged $b$-jet (only one $b$) in the single-$t$ event
(cf. Equation~(\ref{ethree})).
To examine the possibility for this to happen, we have applied the 
eikonal approximation \cite{sterman} to calculate the  
rate of $W+b\bar b+\,jet$ from the square of the 
$W+b\bar b$ amplitude. (The results are shown in Appendix~E.)

After the basic kinematic cuts:
\begin{eqnarray}                             
       P_T^{jet}  >  15 {\rm GeV},& |\eta^{jet}| < 3.5, \nonumber \\
       P_T^\ell  >  15 {\rm GeV},& |\eta^\ell| < 2,   \nonumber \\
       P_T^{b,\bar b}  >  15 {\rm GeV},
       & |\eta^{b,{\bar b}}| < 3.5,   
\label{ewbbj}
\end{eqnarray}                             
the rate for $W+b\bar b+\,jet$ is already about a factor of 5 smaller
than that for $W+b\bar b$.
Naively, one might expect a factor of $\alpha_s(M_W)$ 
($\sim 0.1$) suppression
factor for emitting an additional QCD jet (quark or gluon)
in the hard scattering process. However, as compared to 
the large invariant mass ($M_{Wb\bar b}$) of the $W+b \bar b$ system
a jet $P_T$ of 15\,GeV may be small enough to generate large 
logs, such as $\ln(M_{Wb\bar b}/P_T)$, in the amplitudes.
Hence, because $\alpha_s \ln(M_{Wb\bar b}/P_T)$
is not negligible, the rate for $W+b\bar b+\,jet$ 
is not suppressed  by a factor of 10 relative  
to the rate of $W+b \bar b$, but only a factor of 5.

\begin{figure}[p]
\centerline{\hbox{\psfig{figure=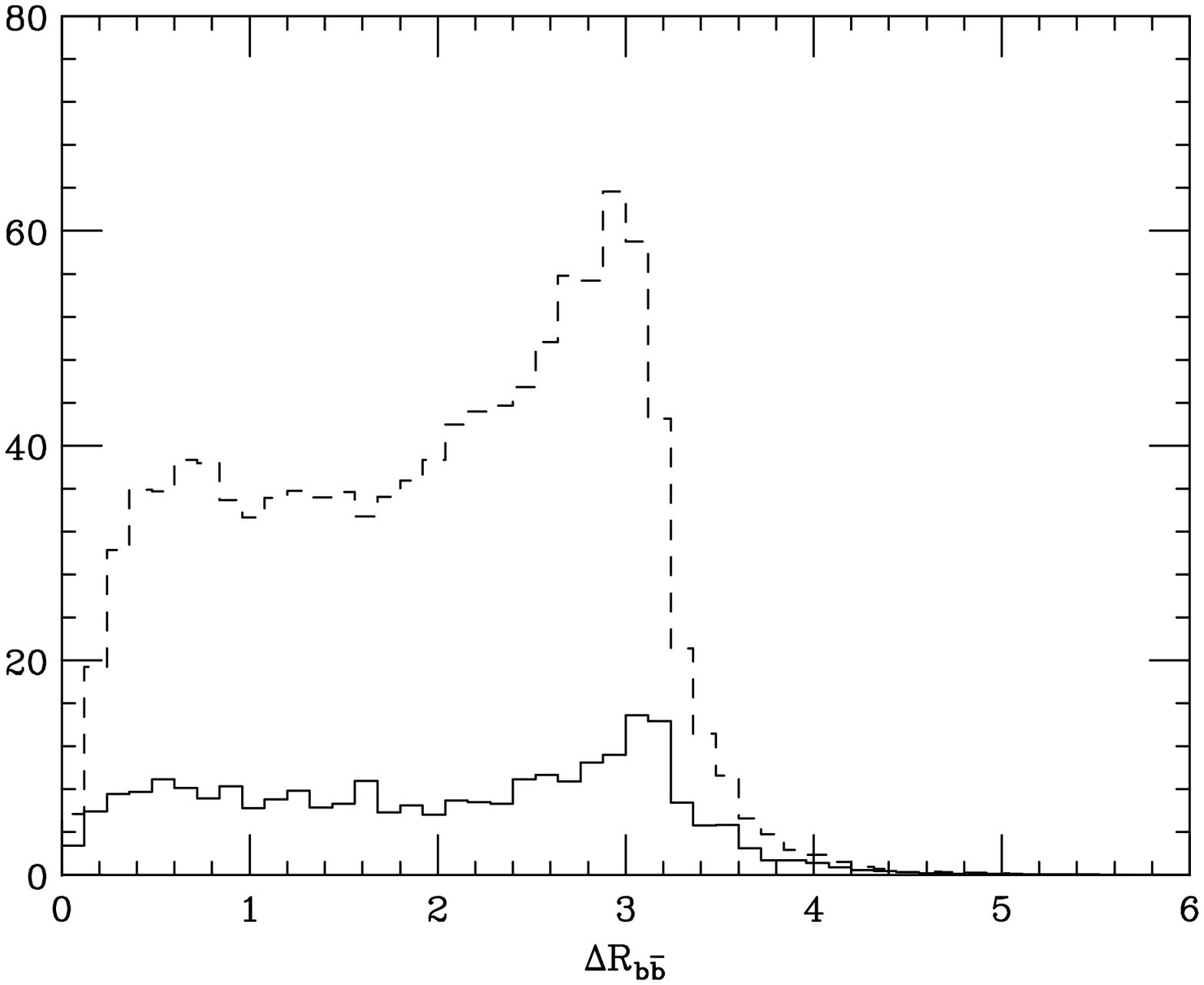,height=4.5in}}}
\caption{ $\Delta R_{b \bar b}$ distributions in 
 $W+b\bar b+\,jet$ (solid) and $W+b\bar b$ (dash) events after applying 
 the cuts listed in (\protect\ref{ewbbj}). }
\label{wbbr}
\end{figure}
\begin{figure}[p]
\centerline{\hbox{\psfig{figure=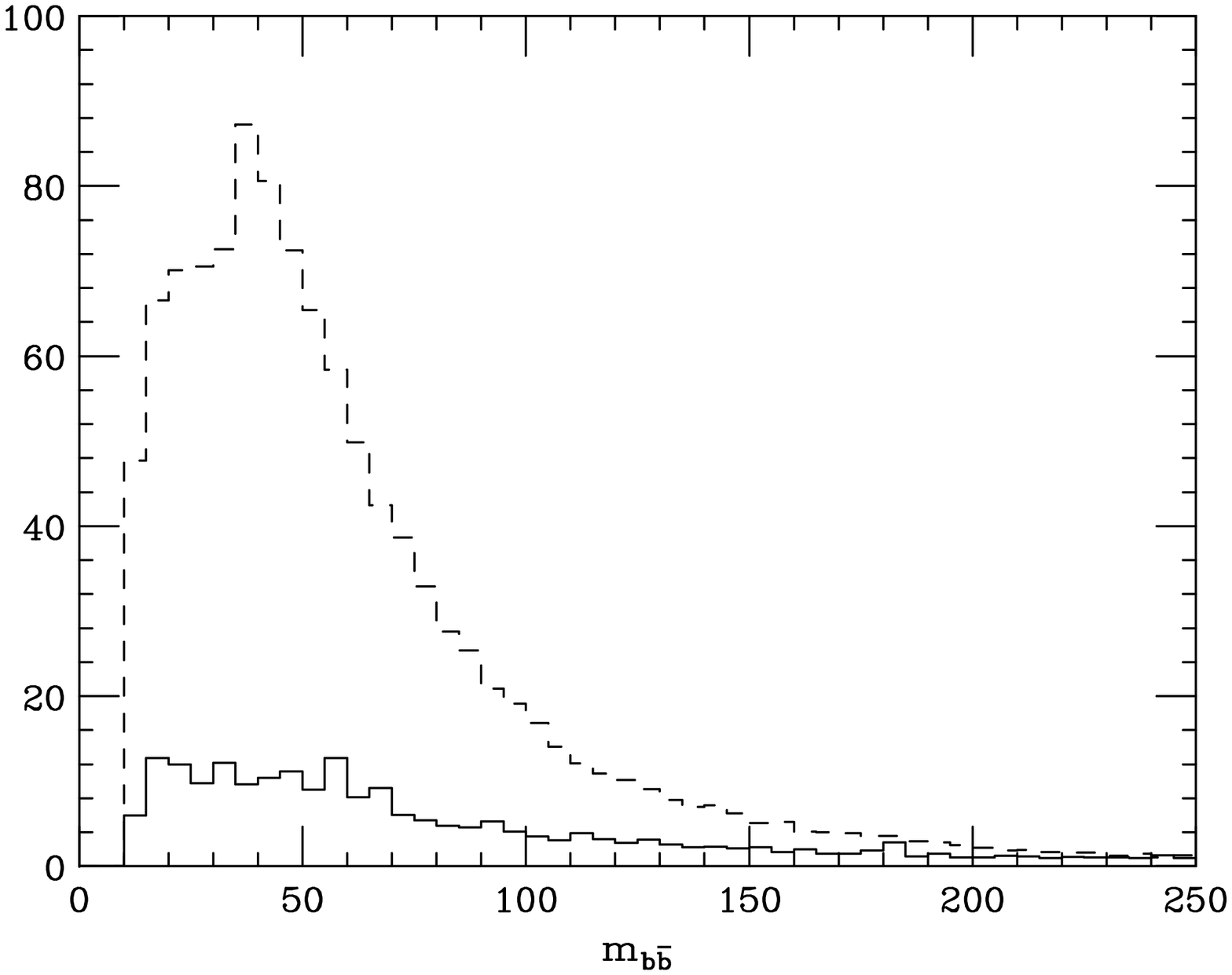,height=4.5in}}}
\caption{ $M_{b \bar b}$ distributions in 
 $W+b\bar b+\,jet$ (solid) and $W+b\bar b$ (dash) events after applying 
 the cuts listed in (\protect\ref{ewbbj}). }
\label{wbbrmbb}
\end{figure}
To see how often $b$ and $\bar b$ will fall into a cone of $\Delta R=0.7$ we
show in Figs.~\ref{wbbr} and \ref{wbbrmbb} 
the $\Delta R_{b \bar b}$ and the $M_{b \bar b}$ distributions 
in $W+b\bar b+\,jet$ events after applying 
the kinematic cuts listed in (\ref{ewbbj}). The same distribution
in $W+b\bar b$ events is also shown for comparison.
The  $\Delta R_{b \bar b}$ distributions look alike, and
the $M_{b \bar b}$ distribution falls slowly as 
$M_{b \bar b}$ increases. Also, the 
$W+b\bar b+\,jet$ event prefers a larger $M_{b \bar b}$
because the $P_T$ of $b$ and $\bar b$ are larger in this process
than that in the $W+b\bar b$ process. 
We find that only about $20\%$ of the $W+b\bar b+\,jet$ events can
possibly fake the single-$t$ event by having $b$ and $\bar b$ 
inside the same jet-cone and thus fake a tagged $b$-jet. 
Hence, the additional background rate from  
$W+b\bar b+\,jet$ events is about a factor 
of ${1 \over 5} \times 20\% =4\%$ of 
the electroweak-QCD background rate.
Although our estimate is not precise, we believe our conclusion 
for this additional background should hold within a factor of 2.
Hence, this additional background is negligible at the Tevatron.
However, it can be important at the LHC. Because the energy of the 
LHC collider is much higher, it is more likely
to have additional radiation in the event and to boost
the $b \bar b$ system to make them closer and thus fall into the same jet-cone.

 \section{ Tevatron with $\protect\rtS=4$\,TeV }

Here we present our results for a possible upgrade of the Tevatron 
with $\rtS=4\,$TeV.  
After the following kinematic cuts:
\begin{eqnarray}                             
       P_T^q > 15 \, {\rm GeV},& |\eta^q| < 3.5, \nonumber \\
       P_T^\ell > 15 \, {\rm GeV},& |\eta^\ell| < 2,   \nonumber \\
       P_T^b > 30 \, {\rm GeV},& |\eta^b| < 2,   \nonumber \\
 \mynot{E_T} > 15 \, {\rm GeV},& \Delta R_{qb} > 0.7 ~, 
\label{eonep}
\end{eqnarray}                             
the signal rate is about $0.28(0.37)$ pb.
(The efficiency of these cuts is 45\%(56\%).)
\begin{figure}[p]
\centerline{\hbox{\psfig{figure=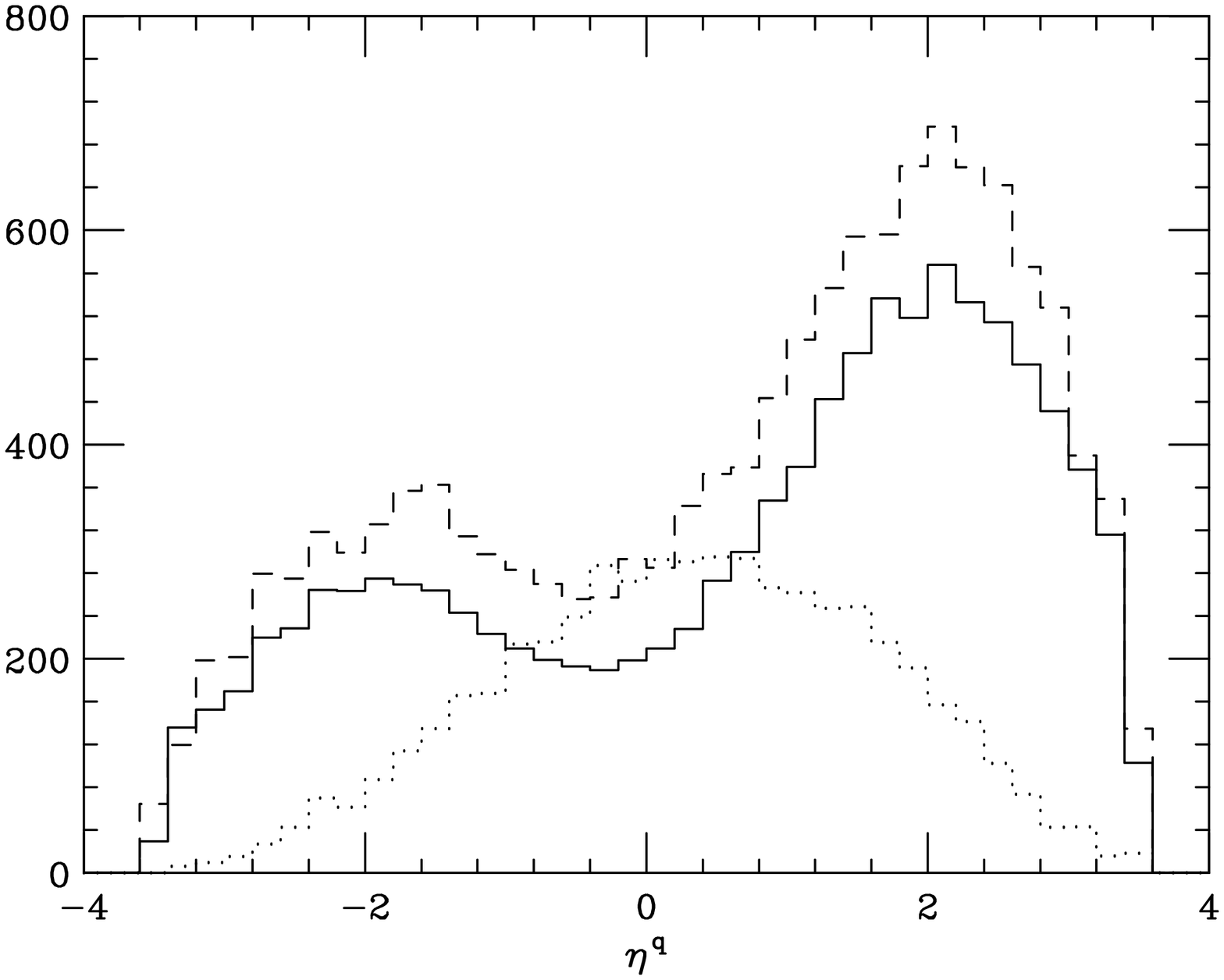,height=4.5in}}}
\caption{ The rapidity distribution
of the spectator quark $q$, after cuts in Equation~(\protect\ref{eonep}),
for the signal $q' b \ra q t (\ra b W^+ (\ra \ell^+ \nu) )$, 
and of the spectator quark $\bar b$ for the major background 
$q' \bar{q} \ra \bar b b W^+ (\ra \ell^+ \nu)$ (dots),
for $m_t=180\,$GeV (solid) and 140\,GeV (dash),
at the \tevs.} 
\label{etatwo}
\end{figure}
In Figure~\ref{etatwo} the typical rapidity of the spectator jet in the
signal event is about 2, and almost all the signal events 
have $|\eta^q| < 3.5$.
An asymmetric cut on $P_T$ was used once again 
to suppress the major background process $W+b+\,jet$. 
Demanding $P_T^b > 15\,$GeV along with the other cuts 
in Equation~(\ref{eonep}),
the signal-to-background ratio (S/B) is about $1.1$($1.7$).
\begin{figure}[p]
\centerline{\hbox{\psfig{figure=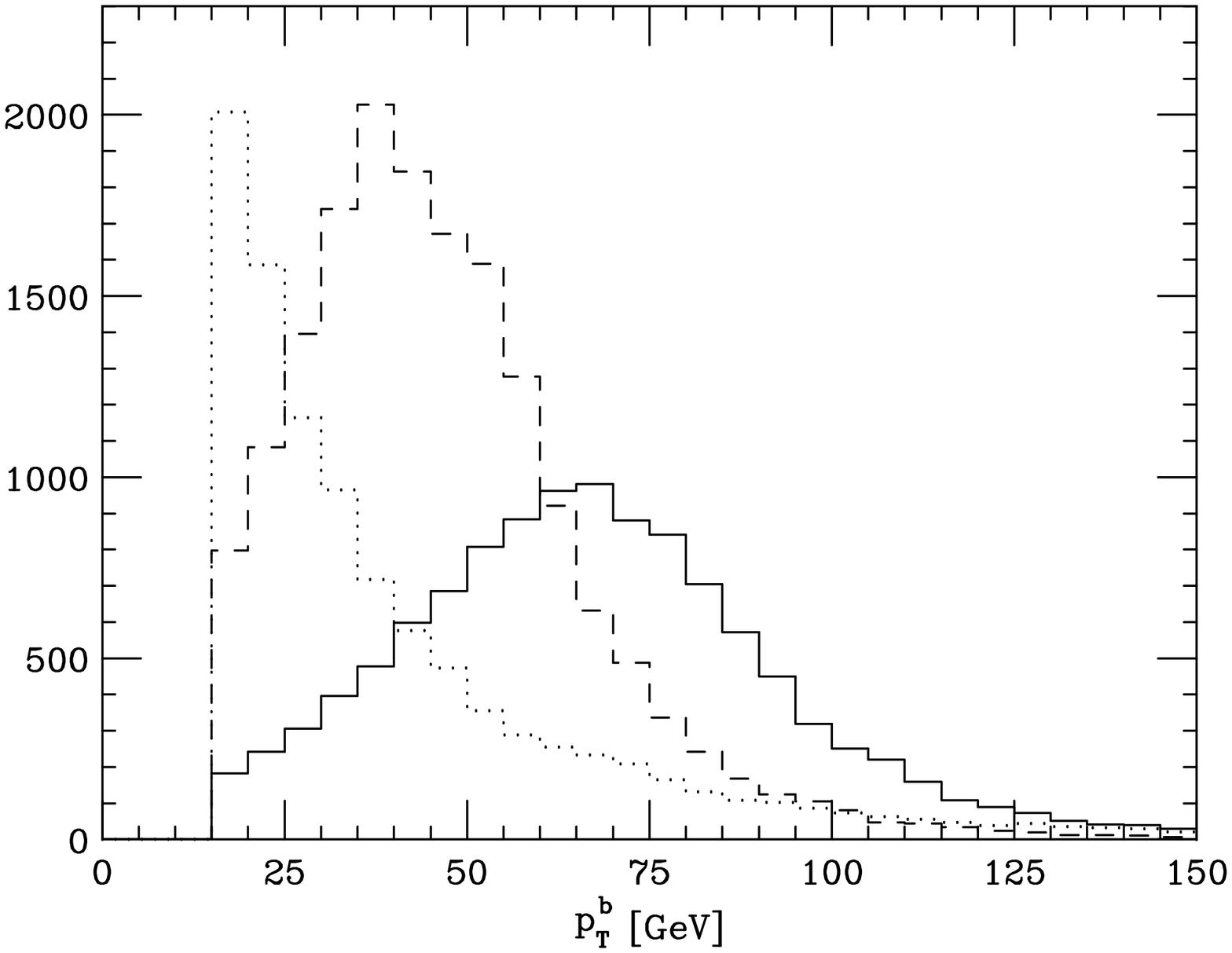,height=4.5in}}}
\caption{
$P_T$ distribution of the $b$ quark, after requiring 
$P^b_T > 15\,$GeV along with all the 
other cuts in~(\protect\ref{eonep}), for the signal 
$q' b \ra q t (\ra b W^+ (\ra \ell^+ \nu) )$, and the major 
background $q' \bar q \ra \bar b b W^+ (\ra \ell^+ \nu)$,
at the \tevs.}
\label{ptbtwo}
\end{figure}
We show in Figure~\ref{ptbtwo} the $P_T$ distribution of the tagged $b$
from $t$.
Requiring $P_T^b > 30$ GeV excludes about half of the background
events sacrificing about $6\%$($20\%$) of the signal.
After all the cuts listed in Equation~(\ref{eonep}), S/B $\simeq 2.3(3.0)$
with the signal rate at 0.28(0.37) pb.

\begin{figure}[p]
\centerline{\hbox{\psfig{figure=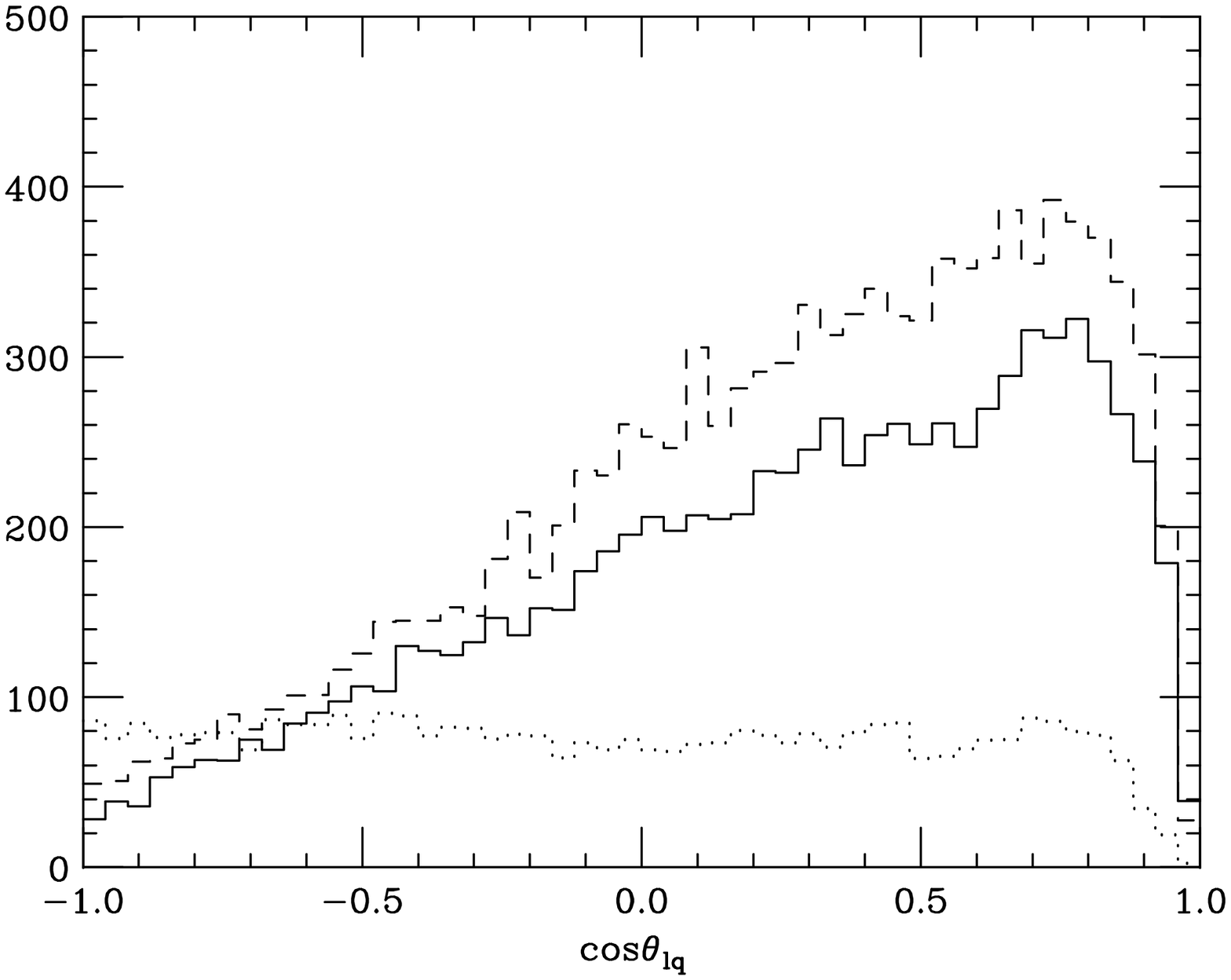,height=4.5in}}}
\caption{ $\cos \theta_{{\ell}q}$ distribution for the signal 
$q' b \ra q t (\ra b W^+ (\ra \ell^+ \nu) )$ and background 
$q' \bar q \ra \bar b b W^+ (\ra \ell^+ \nu)$ 
at the \tevs. }
\label{cthtwo}
\end{figure}

We show in Figure~\ref{cthtwo} the distribution of 
$\cos \theta_{{\ell}q}$.
After applying the cut (\ref{etwo}), the ratio S/B $\simeq 2.9(3.8)$
with the signal rate of 0.22(0.29) pb.
In the end of the analysis there are about $2200(2900)$ single-$t$ events 
for an integrated luminosity of 10 ${\rm fb}^{-1}$
at $\rtS=4$ TeV (a $\pbarp$ collider) with a 
significance ${\rm S/{\sqrt{B}}}$ of about $80(105)$.
The kinematic acceptance of the signal event is about
$43\%$($34\%$). Note that in all the above rates we have included
the reduction factor from a 30\% $b$-tagging efficiency.
\begin{figure}[p]
\centerline{\hbox{\psfig{figure=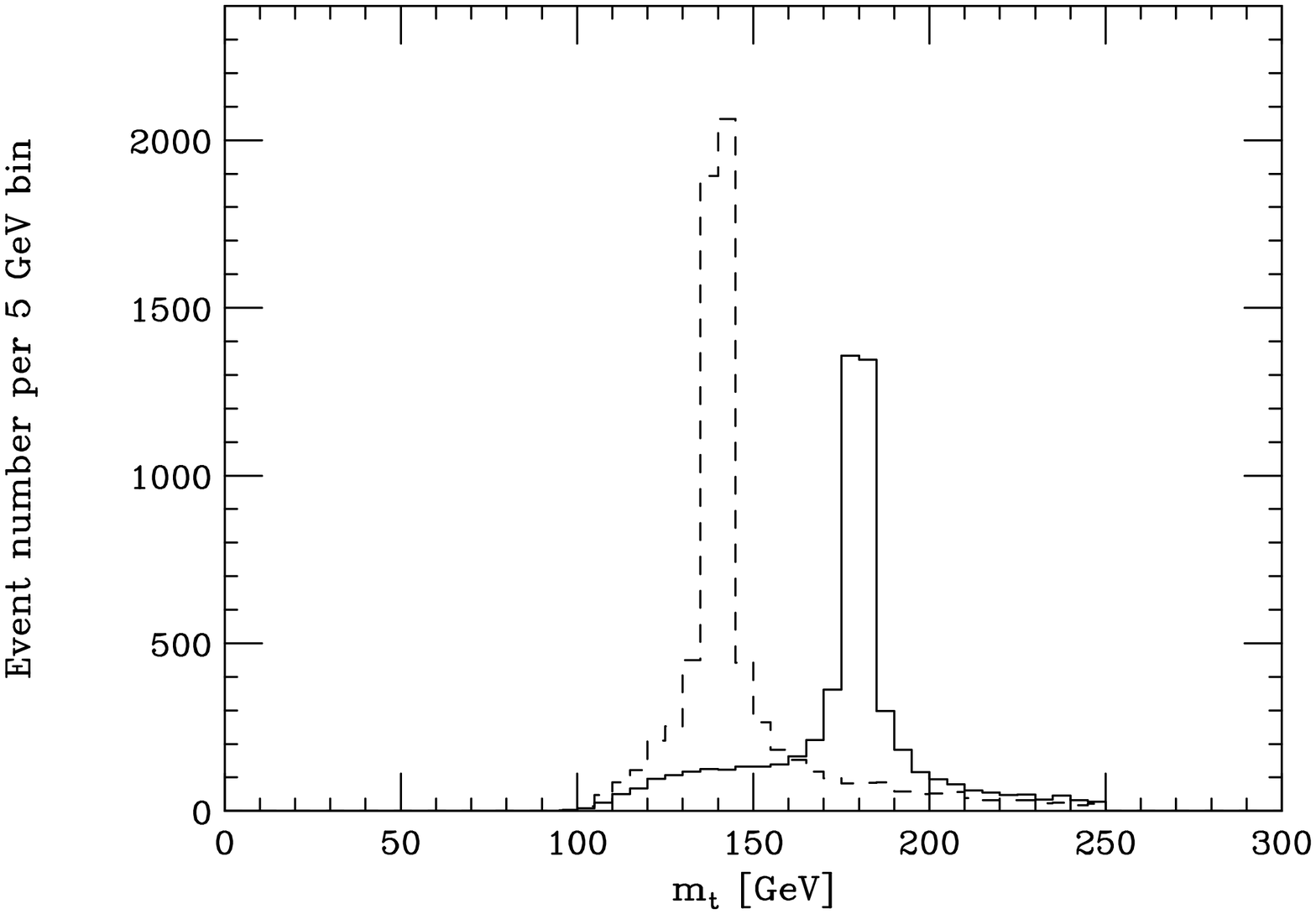,height=4.5in}}}
\caption{ The $m_t$ distribution after the
cuts Equation~(\protect\ref{eonep}) and Equation~(\protect\ref{etwo}) 
for $m_t$ = 180 GeV (solid) and 140 GeV (dash) at the \tevs, 
including both the signal and background events with 
$W^\pm \ra e^\pm \, {\rm or} \, \mu^\pm$. }
\label{mttwo}
\end{figure}
In Figure~\ref{mttwo}, we show the reconstructed 
invariant mass ($m_t$) of the top quark.
\begin{figure}[p]
\centerline{\hbox{\psfig{figure=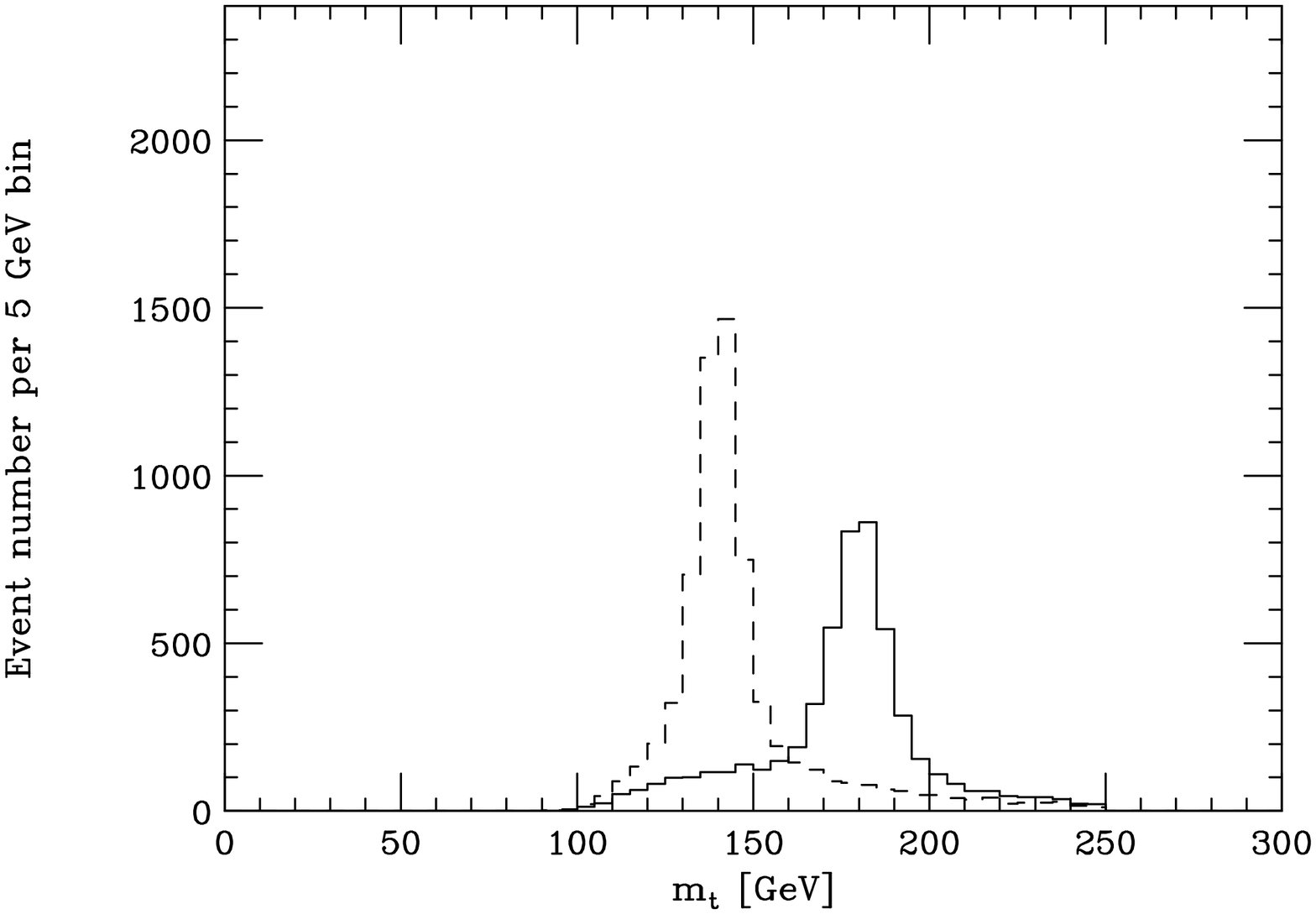,height=4.5in}}}
\caption{ Same as Figure~\protect\ref{mttwo} but with detector 
resolution effects as described in Equation~(\protect\ref{efive}). }
\label{mtstwo}
\end{figure}
Once again, to incorporate the effects of detector efficiencies, 
we smear the final state parton momenta as in~(\ref{efive}).
The $m_t$ distribution becomes slightly broader as 
shown in Figure~\ref{mtstwo};
however, both the signal and the background rates are almost 
the same as those obtained with a perfect detector.

 \section{ LHC with $\protect\rtS=14$\,TeV }

Here we present our results for the LHC with $\rtS=14$\,TeV.
After the following kinematic cuts
\begin{eqnarray}                             
       P_T^q > 40 \, {\rm GeV},&  1 < |\eta^q| < 4, \nonumber \\
       P_T^\ell > 40 \, {\rm GeV},&  |\eta^\ell| < 2,     \nonumber \\
       P_T^b > 40 \, {\rm GeV},&  |\eta^b| < 2,     \nonumber \\
 \mynot{E_T} > 40 \, {\rm GeV},& \Delta R_{qb} > 0.7~, 
\label{eonepp}
\end{eqnarray}                             
the signal rate is about $0.44(0.24)$ pb.
(The efficiency of the cuts is 94\%(98\%). 
We still assume a 30\% efficiency for the $b$-tagging at the LHC.)
Here we did not impose a smaller $P_T^q$ cut 
because a lower $P_T$ jet will be more difficult
to be identified at the LHC. (A typical QCD event at the supercollider
will be engulfed by soft gluon radiation.)
Since the signal event yield is large
at the LHC, we decided to purify our data simply by requiring
a large $P_T^q$ cut.
  Notice that the rate for
$m_t$ = 180 GeV is larger at the LHC than that for $m_t$ = 140 GeV
after our cuts, opposite to the behavior of the rate at 2 TeV and
4 TeV.  This is due in part to less sensitivity to $m_t$ at higher
energies, but mainly because the $b$ from top decay is much harder
for larger top mass and thus is less sensitive to the cut of
$P_T^b > 40$ GeV.  The typical $P_T^b$ for $m_t$ = 180 (140) GeV is
60(40) GeV.  
\begin{figure}[p]
\centerline{\hbox{\psfig{figure=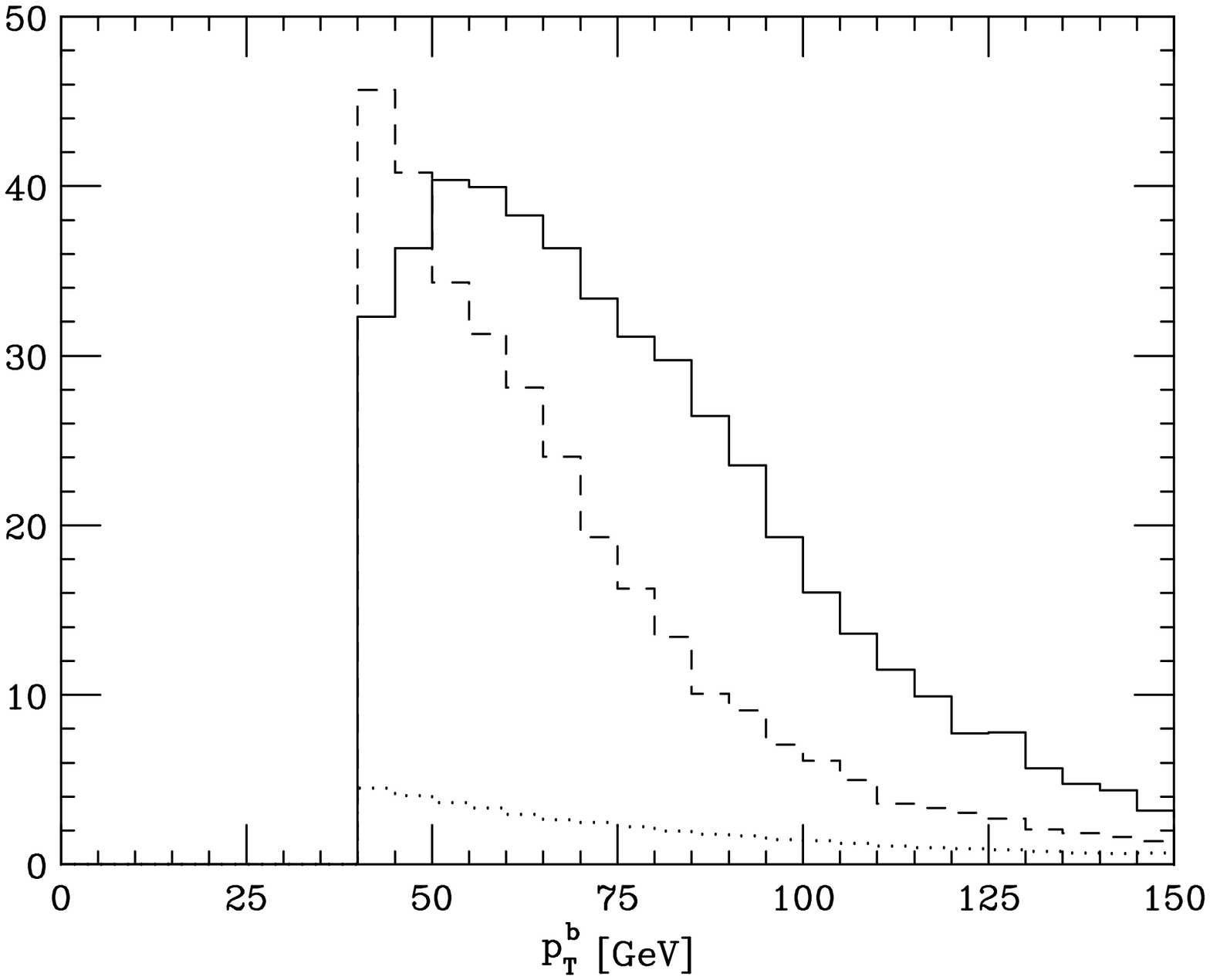,height=4.5in}}}
\caption{ 
$P_T$ distribution of the $b$ quark, after
cuts in~(\protect\ref{eonepp}), for the signal 
$q' b \ra q t (\ra b W^+ (\ra \ell^+ \nu) )$, and the major 
background $q' \bar q \ra \bar b b W^+ (\ra \ell^+ \nu)$,
at the LHC.}
\label{ptbthr}
\end{figure}
We show in Figure~(\ref{ptbthr}) 
the $P_T$ distribution of the tagged $b$ from $t$.

\begin{figure}[p]
\centerline{\hbox{\psfig{figure=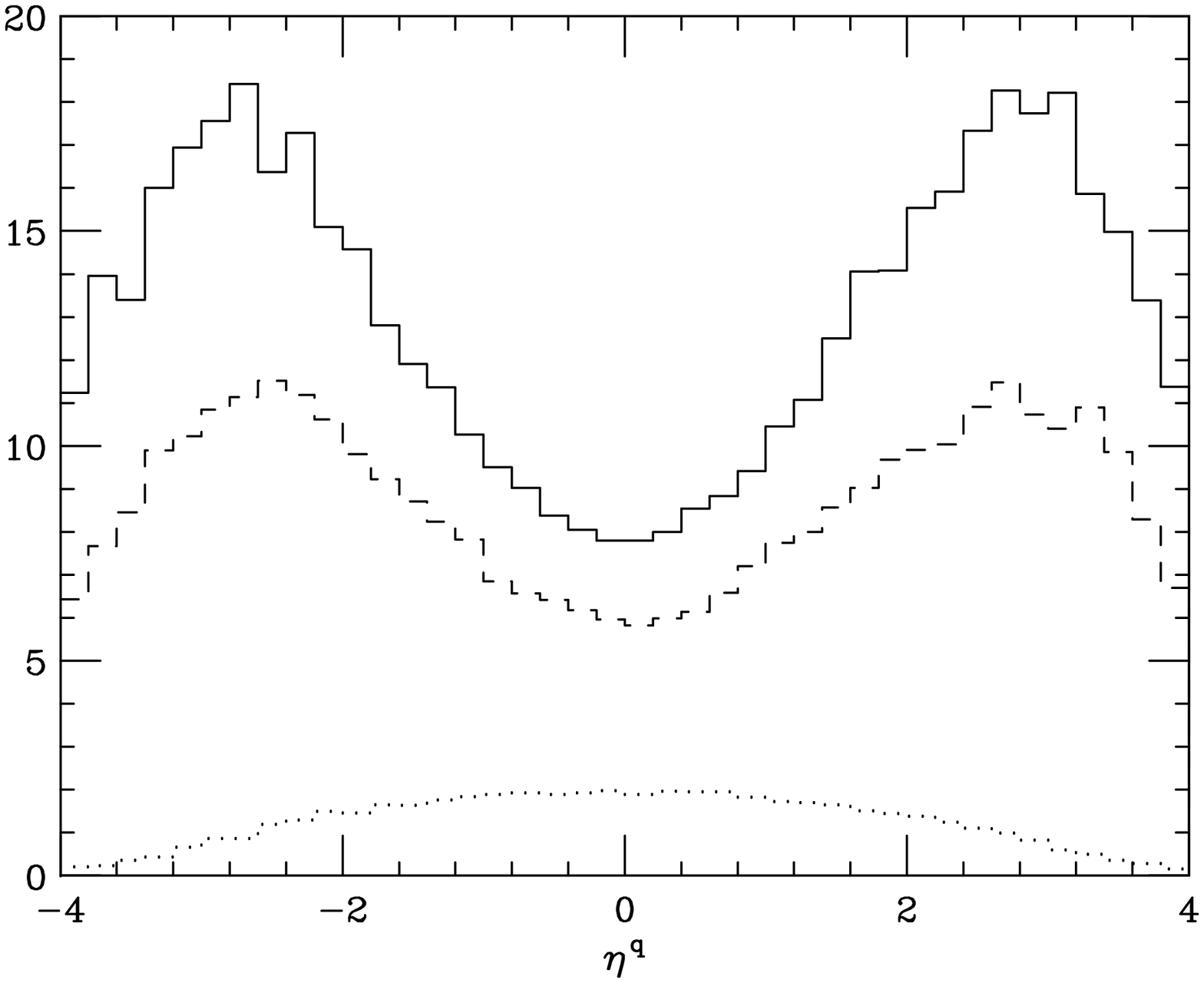,height=4.5in}}}
\caption{ The rapidity distribution
of the spectator quark $q$, after cuts in Equation~(\protect\ref{eonepp}),
for the signal $q' b \ra q t (\ra b W^+ (\ra \ell^+ \nu) )$, 
and of the spectator quark $\bar b$ for the major background 
$q' \bar{q} \ra \bar b b W^+ (\ra \ell^+ \nu)$ (dots),
for $m_t=180\,$GeV (solid) and 140\,GeV (dash),
at the LHC.} 
\label{etathr}
\end{figure}

In Figure~\ref{etathr} the typical rapidity of the spectator jet in the
signal event is about 3, but
a cut on $|\eta^q| < 4$  keeps almost all the signal events.
Excluding the $|\eta^q| > 1$ cut in~(\ref{eonepp}) 
the signal-to-background ratio (S/B) is about 10(7).
Requiring $|\eta^q| > 1$ excludes about $40\%$ of the background
events sacrificing about $15\%$($20\%$) of the signal.
After the kinematic cuts in ~(\ref{eonepp}), the ratio S/B $\simeq 25(14)$.

\begin{figure}[p]
\centerline{\hbox{\psfig{figure=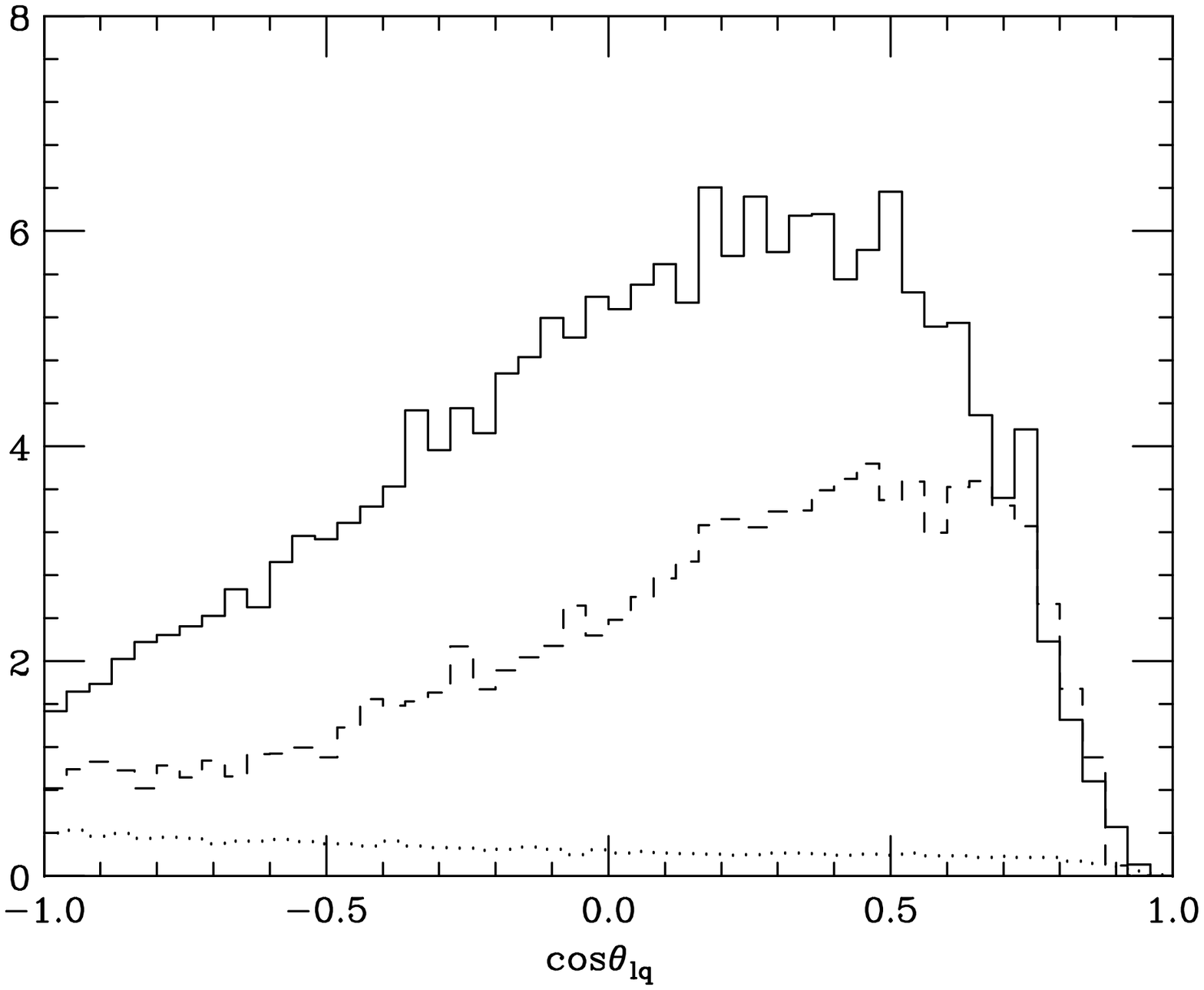,height=4.5in}}}
\caption{ $\cos \theta_{{\ell}q}$ distribution for the signal 
$q' b \ra q t (\ra b W^+ (\ra \ell^+ \nu) )$ and background 
$q' q \ra \bar b b W^+ (\ra \ell^+ \nu)$ 
at the LHC. }
\label{cththr}
\end{figure}

We show in Figure~\ref{cththr} the distribution of $\cos \theta_{{\ell}q}$
at the LHC. After the $\cos \theta_{{\ell}q}$ cut, S/B $\simeq 40(20)$.
In the end of the analysis, there are about $30,000(15,000)$ 
single-top events for an integrated luminosity of 100 ${\rm fb}^{-1}$ 
at $\rtS=14\,$TeV (a $\pp$ collider) with a 
significance ${\rm S/{\sqrt{B}}}$ of about $32(16)$.
Hence, about $4\%$($2\%$) of the total signal event rate remains. 
This is thus the kinematic acceptance 
for the signal process~(\ref{ethree}).
\begin{figure}[p]
\centerline{\hbox{\psfig{figure=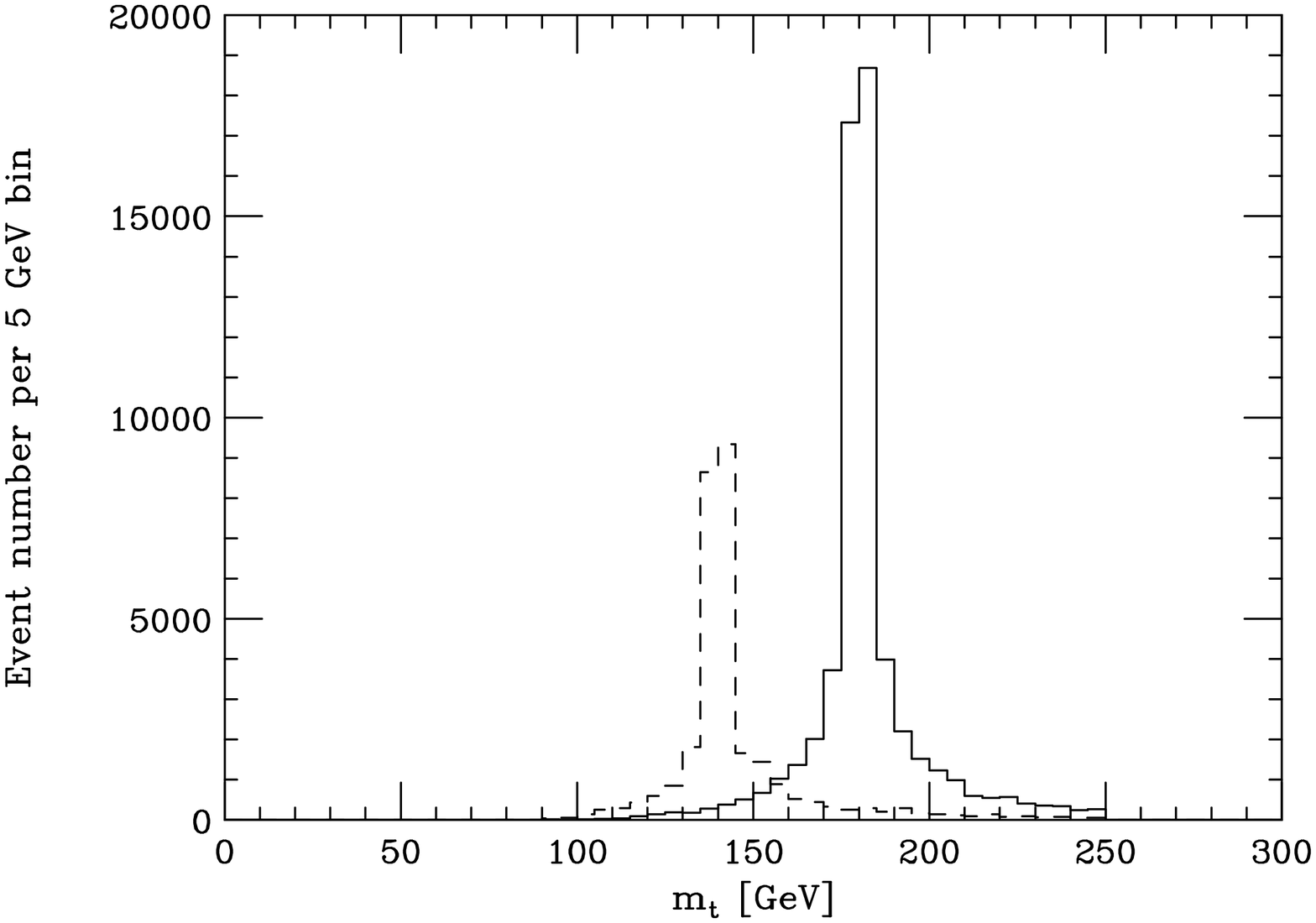,height=4.5in}}}
\caption{ The $m_t$ distribution after the
cuts Equation~(\protect\ref{eonepp}) and Equation~(\protect\ref{etwo})
for $m_t$ = 180\,GeV (solid) and 140\,GeV (dash) at the LHC  
including both the signal and background events with 
$W^\pm \ra e^\pm \, {\rm or} \, \mu^\pm$. }
\label{mtthr}
\end{figure}
In Figure~\ref{mtthr}, we show the reconstructed 
invariant mass ($m_t$) of the top quark for a perfect detector.
Once again, to incorporate the effects of detector efficiencies, 
we smear the final state parton momenta as in~(\ref{efive})
\begin{figure}[p]
\centerline{\hbox{\psfig{figure=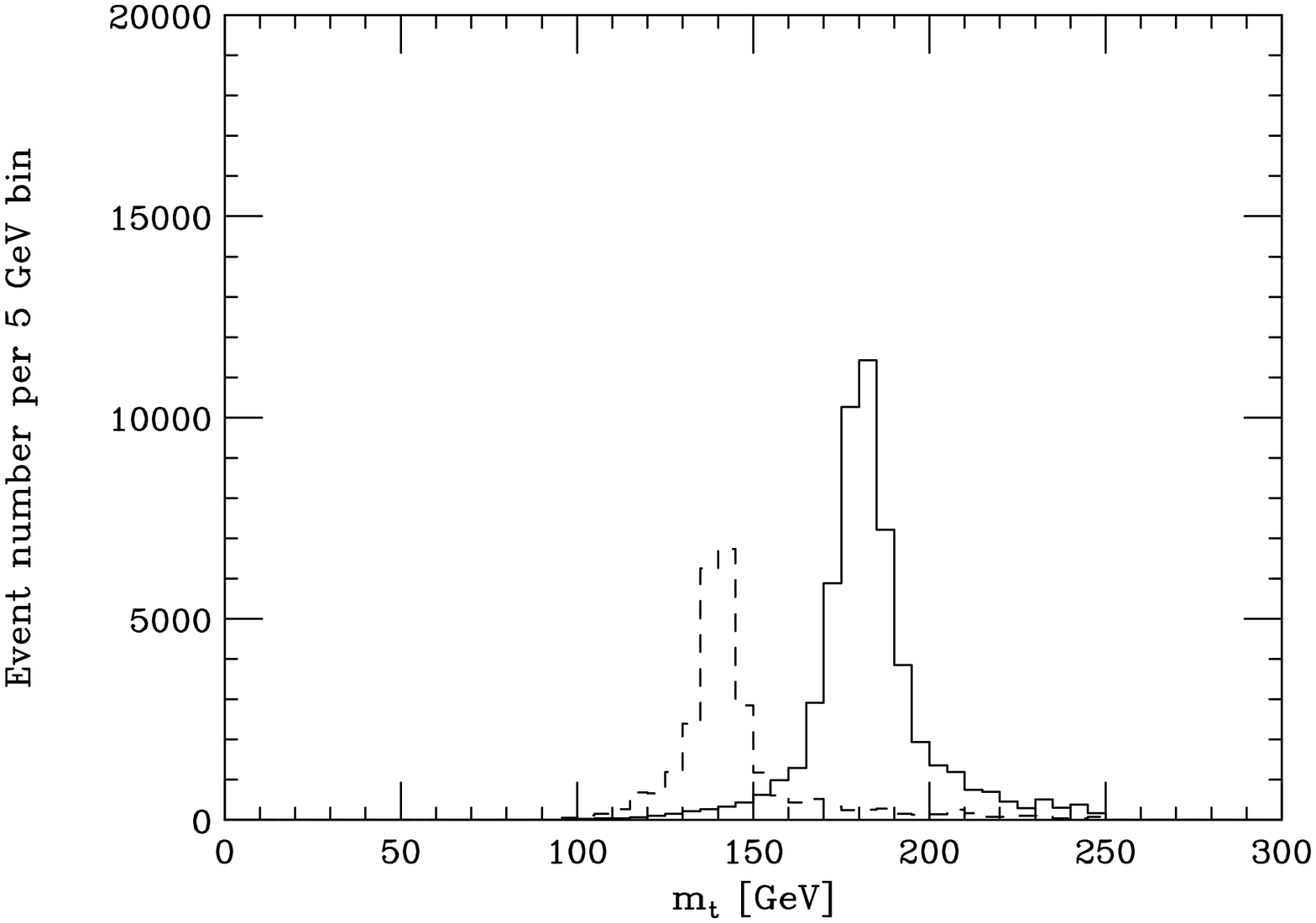,height=4.5in}}}
\caption{ Same as Figure~\protect\ref{mtthr}, but with detector resolution 
effects as described in Equation~(\protect\ref{efive}). }
\label{mtsthr}
\end{figure}
and reconstruct the $m_t$ distribution for the LHC 
is shown in Figure~\ref{mtsthr}.

Notice that the above analysis for the LHC is less reliable because the
energy of the collider is much higher and therefore it is more
likely to have additional soft-jets accompanying
the signal and the background events. 
 As discussed at the end of
section 7.1, it would be more reliable to use a full event generator
such as ISAJET \cite{isajet}, PYTHIA \cite{pythia} or 
HERWIG \cite{herwig} for this study because these
generators contain radiation from either the initial or 
final states.
However, these generators currently do not have the 
correct angular correlations in $\ell$ and jets, as  
discussed in this analysis.
It would therefore be important in the future to improve these generators to 
incorporate the polarization effects of the top quark and the $W$--boson
for studying physics of the top quark in hadron collisions.

\chapter{Discussions and Conclusions}

We discussed the physics of top quark production and decay 
at hadron colliders, such as the Tevatron, the \tevs~and the LHC.
We showed how to use the invariant mass distribution of $m_{b\ell}$
to measure the mass and the width of the 
top quark, produced from either a single-top or a $t \bar t$ pair  
process.
It has been shown in Reference~\cite{steve} that 
the distribution of $m_{b\ell}$ is not sensitive to  
radiative corrections from  QCD interactions. Thus 
it can be reliably used to test the polarization of the 
$W$--boson from $t$ decay (hence,  test the polarization of
the top quark from the production mechanism)
and to measure the mass of the top quark using the observed value
of $\flong$ (the fraction of longitudinal $W$'s from top decays).
We also discussed how well the couplings of $t$-${b}$-$W$ vertex 
can be measured to probe new physics, and how well the 
CP properties of the top quark can be tested in 
electron or hadron colliders.

In Reference~\cite{wgtb} we showed that an almost perfect efficiency 
for ``kinematic $b$ tagging'' can be achieved 
due to the characteristic features of 
$W$--gluon fusion events.
In addition, the ability of performing $b$-tagging using a 
vertex detector increases the detection efficiency 
of a heavy top quark
produced via the $W$--gluon fusion process.  

A detailed Monte Carlo study 
on how to identify the characteristic features  of the signal events
({\it i.e.}, the transverse momentum and
the rapidity distributions of the spectator quark which emitted 
the virtual $W$)
and therefore suppress the background events
was performed in Chapter~8.\footnote{
The fortran code, ONETOP, used for this study is available by request.
In Appendix F we briefly describe the processes included in this 
program.
}
For an integrated luminosity of 1 $\ifb$,
 there will be about 75 (105) 
single-$t$ or single-$\bar t$  events reconstructed 
in the lepton+jet mode 
for $\mt = 180 \,(140)$\,GeV at $\rtS = 2$ TeV.
(The branching ratio of $W \ra e, \, {\rm or} \, \mu$ is included, and
the $b$-quark tagging efficiency is assumed to be 30\% for $P^b_t > 30\,$GeV
with no misidentifications of a $b$-jet from other QCD jets.)
The dominant background process is the electroweak-QCD process 
$W+b\bar b$ whose rate is about $80\%$($60\%$) of the signal rate in 
the end of the analysis. The $t \bar t$ events are not as important
to our study. 
The results for $\rtS = 4$\,TeV at the Di-TeV 
and for $\rtS = 14$\,TeV at the LHC were also discussed. 

Although the $W^* \ra t \bar b$ rate in the SM
is not as large as the $W$--gluon fusion rate for producing 
a heavy top quark, it remains a complementary process for probing
new physics in the single-top quark event. 
The $W^*$ process is particularly useful for detecting new physics
through some possible high mass resonance in the theory.
In that case, its rate will be highly enhanced by the resonance effects.
We however did not study such a possibility in this work because
its rate depends on the details of the models considered.

\appendix
\chapter{Helicity Amplitude Method}


In this appendix we would like to display the rules for doing calculations at
the amplitude level using the Helicity Amplitude Method.  The method
breaks down the algebra of four-dimensional Dirac spinors
and matrices into equivalent two-dimensional ones.  This algebra is
easy to program and more efficient than computing the Dirac algebra
as it stands.  All diagrams are summed and squared numerically.

In what follows we introduce 
the Weyl representation of Dirac spinors and matrices.
We also include several example calculations to illustrate the finer points
of the method.  Throughout this paper
we use the Bjorken-Drell metric
\beq 
g^{\mu\nu} = diag\,(1,-1,-1,-1) \, .
\enq

The four-momenta have the form in spherical coordinates:
\beq 
p^{\mu} = (E, \myabs{p} \; \st \; \cp, \myabs{p} \; \st \; \sp, 
\myabs{p} \; \ct)
\enq
with $E^2 - {\myabs{p}}^2 = m^2$.  We define the right-hand (R), left-hand (L)
and longitudinal (0) polarization vectors for a spin-1 field as
\footnote{ For a massless spin-1 field, only the right-handed and
the left-handed polarizations are physical.}
\bea 
\veps^{\mu}_{(R)} &=& {e^{i\phi}\over\sqrt{2}}
(0, i \; \sp - \cp \; \ct, -i \; \cp - \sp \; \ct, \st) \nonumber \\
\veps^{\mu}_{(L)} &=& {e^{-i\phi}\over\sqrt{2}}
(0, i \; \sp + \cp \; \ct, -i \; \cp + \sp \; \ct, -\st) \\
\veps^{\mu}_{(0)} &=& {1\over m}
(\myabs{p}, E \; \st \; \cp, E \; \st \; \sp, E \; \ct).
\nonumber 
\ena
The above equations satisfy the identities 
$\veps^{\mu}_{(R)} = -\veps^{\mu\ast}_{(L)}$,
$\veps^{\mu}_{(0)} = \veps^{\mu\ast}_{(0)}$,
$p_{\mu} \veps^{\mu}_{(h)} = 0$ \rm{and}
$\veps_{\mu}^{(h)} \veps^{\mu\ast}_{(h')} = 
-\delta_{h h'}$, for $h, h' = R,L \, \rm{or} \, 0$.

In four component form we define the following.
In the Weyl basis Dirac spinors have the form
\beq
\psi=\vect{\psi_{+}}{\psi_{-}}
\enq
where for fermions
\bec
\bea
\psi_{\pm} &=& \left\{\begin{array}{ll}
 u_{\pm}^{(\lambda= 1)} = w_{{\pm}}\*\chi_{ 1/2} \\
 u_{\pm}^{(\lambda=-1)} = w_{{\mp}}\*\chi_{-1/2}
		      \end{array}\right.
\label{fermion}
\ena
\enc
and anti-fermions
\bec
\bea
\psi_{\pm} &=& \left\{\begin{array}{ll}
 v_{\pm}^{(\lambda= 1)} = {\pm} w_{{\mp}}\*\chi_{-1/2} \\
 v_{\pm}^{(\lambda=-1)} = {\mp} w_{{\pm}}\*\chi_{ 1/2}
		      \end{array}\right.
\label{afermion}
\ena
\enc
with $w_{\pm} = \sqrt{E \pm {\myabs{p}}}$.

The $\chi_{\lambda/2}$'s are eigenvectors of the helicity operator 
\beq
h = \hat{p}\cdot\,\vec{\sigma} , \, \hat{p} = {\vec{p} / {\myabs{p}}}
\enq
with eigenvalue $\lambda$ where $\lambda = +1$ is for ``spin-up'' 
and $\lambda = -1$ is for ``spin-down''.
\beq                                                                       
\chi_{1/2} = 
     \vect{\cos {\theta /2}}{e^{i\phi} \sin {\theta /2}},\quad 
\chi_{-1/2} = 
     \vect{-e^{-i\phi} \sin {\theta /2}}{\cos {\theta /2}} \, .
\label{heigen}
\enq
Later it proves useful to represent $\chi_{\lambda/2}$'s using bra-ket 
notation where 
\beq
\ket{p}{+}\ \equiv \chi_{1/2}, \quad \ket{p}{-}\ \equiv \chi_{-1/2} \, .
\label{braket}
\enq

Gamma matrices in the Weyl basis have the form
\beq
  \gamma^0 = \gamz, \quad
  \gamma^j = \gamj, \quad
  \gamma^5 = \gamma_5 = \gam5,
\enq
where $\sigma_j$ are the Pauli $2\times 2$ spin matrices
\beq
  \sigma_1 = \sone, \quad
  \sigma_2 = \stwo, \quad
  \sigma_3 = \sthr.
\enq
The chirality projection operators are defined by
\beq
P_{\pm} = \Ppm \, .
\enq
Notice that $P_{+}$ ($P_{-}$) projects out the 
``right-handed'' (``left-handed'') component of the Weyl spinor 
effectively reducing
the algebra from one involving four component spinors and matrices
to one involving two component spinors and matrices.
\vspace{-.3in}
\bec
\bea
 P_{-} {\psi} & =&  {\Pm} \vect{\psi_{+}}{\psi_{-}} = 
\vect{0}{\psi_{-}} \nonumber \\
 {\bar{\psi}} P_{+} &= &
(\psi_{+}^{\dagger} \, \psi_{-}^{\dagger}) \gamz \Pp =
(\psi_{-}^{\dagger} \,\, 0) 
\ena
\enc

In the Weyl basis $\slash p$ has the form
\beq
{\slash p} \equiv p_{\mu}\*\gamma^{\mu} = 
\mat{0}{p_0 + \vec{\sigma} \cdot \vec{p}}
    {p_0 - \vec{\sigma} \cdot \vec{p}}{0} \equiv
\mat{0}{\slash p_{+}}{\slash p_{-}}{0} \equiv
p_{\mu} {\mat{0}{\gamma_{+}^{\mu}}{\gamma_{-}^{\mu}}{0}}
\enq
where
\beq
\gamma_{\pm}^{\mu} = (1,\mp \vec{\sigma}) \, .
\enq
Products of these $\gamma_{\pm}^{\mu}$'s have the following useful property
when Lorentz indices are contracted:
\beq
(\gamma^{\mu}_{+})_{ij} (\gamma_{\mu +})_{kl} = 
(\gamma^{\mu}_{-})_{ij} (\gamma_{\mu -})_{kl} =
2[ \delta_{ij} \delta_{kl} - \delta_{il}\delta_{kj} ]
\label{F1}
\enq
and
\beq
(\gamma^{\mu}_{+})_{ij} (\gamma_{\mu -})_{kl} = 
(\gamma^{\mu}_{-})_{ij} (\gamma_{\mu +})_{kl} =
2 \delta_{il}\delta_{kj},
\label{F2}
\enq
where the Roman indices are not vector indices in the usual sense, but are
labels identifying bras and kets.
For instance, for arbitrary kets $\kt{i}, \kt{j}, \kt{k}$ and $\kt{l}$ we 
have
\beq
(\gamma^{\mu}_{+})_{ij} (\gamma_{\mu -})_{kl} =
\br{i} \gamma^{\mu}_{+} \kt{j} \br{k} \gamma_{\mu -} \kt{l} =
2 \, \bk{i}{l} \bk{k}{j} =
2 \delta_{il}\delta_{kj} \, .
\enq
Equations~(\ref{F1}) and (\ref{F2}) are simply the two-dimensional 
version of the well known Fiertz identities.

\begin{figure}
\hspace{1.in}
\par
\centerline{\hbox{
\psfig{figure=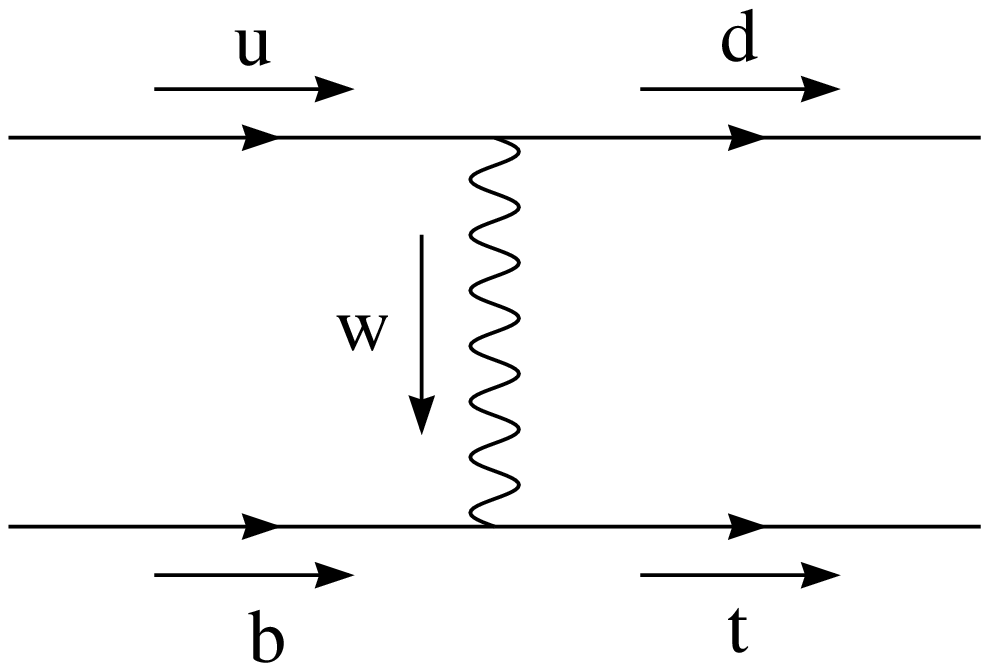,height=1.3in}}}
\caption{ Diagram for the ($2 \ra 2$) process 
$ u \, b \ra d \, t $.}\label{ubdt0}
\par
\hspace{1.in}
\end{figure}

\section{ Helicity Amplitudes for $u b \ra d t$ }
To illustrate the use of helicity amplitudes we calculate the matrix 
element for the ($2 \ra 2$) process $u b \ra d t$ which contributes to the
total rate for $W$--gluon fusion. Figure~\ref{ubdt0} shows the Feynman
diagram for this process with the $t$ decay included.
In this example we use the generalized $\tbW$ coupling
\beq
\Gamma^{\mu} = (1 + \klc)\,\gamma^{\mu} P_{-} + 
\krc \gamma^{\mu} P_{+}
\enq
where $\klc$ and $\krc$ parameterize deviations from the Standard Model
in which $\klc = 0$ and $\krc = 0$.
We calculate the matrix element in the 
't~Hooft-Feynman gauge ignoring for simplicity the factors
due to vertices and propagators.  We obtain
\beq
\M = [\bar u(t) \Gamma^{\mu} u(b)] 
[\bar u(d) \gamma_{\mu} P_{-} u(u)]
\enq
where u, b, d and t are the momenta of the external legs and we retain only
the mass of the top quark.
We use the algebraic properties of the projection operators 
$P_{\pm}^2 = P_{\pm}, P_{\pm}P_{\mp} = 0$ \rm{and}
$P_{\pm} \gamma^{\mu} = \gamma^{\mu} P_{\mp}$
to project out the chirality states.  In this example the 
amplitude contains both the
left-handed and right-handed currents.  For pure vector or axial vector
currents one must first insert 
$1 = P_{+} + P_{-}$ or $\gamma_5 = P_{+} - P_{-}$ respectively.
Therefore $\M$ takes the form
\bea
\lefteqn{ \M = (1 + \klc) [u^{\dagger}_{-}(t) \gamma^{\mu}_{+} u_{-}(b)] 
                    [u^{\dagger}_{-}(d) \gamma_{\mu +} u_{-}(u)] }
 \nonumber\\ & & 
                + \, \krc [u^{\dagger}_{+}(t) \gamma^{\mu}_{-} u_{+}(b)]
                    [u^{\dagger}_{-}(d) \gamma_{\mu +} u_{-}(u)] 
\ena
According to Equations~(\ref{fermion}) and (\ref{braket}) we see that 
\vspace{-.3in}
\bec
\bea
u_{-}(u) &=& \sqrt{2 E_u} \, \ket{u}{-} \nonumber \\
u_{-}(d) &=& \sqrt{2 E_d} \, \ket{d}{-} \nonumber \\
u_{-}(b) &=& \sqrt{2 E_b} \, \ket{b}{-} \nonumber \\
u_{+}(b) &=& \sqrt{2 E_b} \, \ket{b}{+} \nonumber \\
u_{-}(t) &=& \left\{\begin{array}{ll}
                       \sqrt{E_t-{\myabs{t}}} \, \ket{t}{+} \nonumber \\
                       \sqrt{E_t+{\myabs{t}}} \, \ket{t}{-}
		  \end{array}
	   \right. \\
u_{+}(t) &=& \left\{\begin{array}{ll}
                       \sqrt{E_t+{\myabs{t}}} \, \ket{t}{+} \\
                       \sqrt{E_t-{\myabs{t}}} \, \ket{t}{-}
		  \end{array}
	   \right.
\ena
\enc

Therefore,
\bea
\lefteqn{ {\M}(+) = (1 + \klc) \sqrt{E_t-{\myabs{t}}}
            \, \bra{t}{+} \gamma^{\mu}_{+} \ket{b}{-}
     	       \bra{d}{-} \gamma_{\mu +} \ket{u}{-} } \nonumber \\
        & &     + \, \krc \sqrt{E_t+{\myabs{t}}}
            \, \bra{t}{+} \gamma^{\mu}_{-} \ket{b}{+}
     	       \bra{d}{-} \gamma_{\mu +} \ket{u}{-} \nonumber \\
\lefteqn{ {\M}(-) = (1 + \klc) \sqrt{E_t+{\myabs{t}}}
            \, \bra{t}{-} \gamma^{\mu}_{+} \ket{b}{-}
     	       \bra{d}{-} \gamma_{\mu +} \ket{u}{-} } \nonumber \\
        & &     + \, \krc \sqrt{E_t-{\myabs{t}}}
            \, \bra{t}{-} \gamma^{\mu}_{-} \ket{b}{+}
     	       \bra{d}{-} \gamma_{\mu +} \ket{u}{-} 
\label{Mpm}
\ena
where ${\M}(\pm)$ denotes the amplitude with $t$ helicity 
$\lambda_{t} = {\pm 1}$ and we ignore for now the common factor of
$\sqrt{2 E_u} \sqrt{2 E_b} \sqrt{2 E_d}$.

One is now tempted to move on ahead and contract the Lorentz indices
as in Equation~(\ref{F1}).  However, contracting $\gamma_{\mu}$'s with the
same chirality introduces extra terms into the matrix element.  
It would be more useful with processes containing many
branchings from decay, such as in Supersymmetry, if there were a 
way to utilize Equation~(\ref{F2}) instead.  To this end, we digress a moment.

Consider some current of the form
\beq
{\psi_{f\pm}^{\dagger}}[\gamma_{\mu_1\mp}\cdots\gamma_{\mu_n\mp}]
{\psi_{i\pm}}
\label{cur}
\enq
where $n$ is odd.  We note in passing,
recalling the properties of the projection operators, consecutive matrices
of the same sign would give zero current.
Since Equation~(\ref{cur}) is just a number, it is identical to taking its
transpose as
\beq
{\psi_{i\pm}^{\top}}[\gamma_{\mu_n\mp}^{\top}\cdots\gamma_{\mu_1\mp}^{\top}]
{\psi_{f\pm}^{*}} \, .
\label{trans}
\enq
We now utilize the following 
algebraic properties of the Pauli matrices and in particular $\sigma_2$:
\beq
\sigma_2 \sigma_2 = 1, \quad \sigma_2^{\top} = -\sigma_2, \quad
\sigma_2{\gamma^{\top}_{\mu\mp}}\sigma_2 = \gamma_{\mu\pm}
\enq
and define
\beq
\widetilde{\psi_{\pm}} \equiv i\sigma_2\psi^{*}_{\pm} \, .
\enq
By inserting pairs of $\sigma_2$ between each pair of objects in
Equation~(\ref{trans}) we therefore obtain
\beq
{\psi_{f\pm}^{\dagger}}[\gamma_{\mu_1\mp}\cdots\gamma_{\mu_n\mp}]
{\psi_{i\pm}} =
\widetilde{\psi_{i\pm}}^{\dagger}[\gamma_{\mu_n\pm}\cdots\gamma_{\mu_1\pm}]
\widetilde{\psi_{f\pm}} \, .
\enq
It is easy to show that $(i\sigma_2\psi_{i\pm})^{\top} = 
\widetilde{\psi_{i\pm}}^{\dagger}$\,.  
In addition, for an even number $n$ of gamma matrices we have
\beq
{\psi_{f\pm}^{\dagger}}[\gamma_{\mu_1\mp}\cdots\gamma_{\mu_n\pm}]
{\psi_{i\mp}} =
\widetilde{\psi_{i\mp}}^{\dagger}[\gamma_{\mu_n\mp}\cdots\gamma_{\mu_1\pm}]
\widetilde{\psi_{f\pm}} \, .
\enq

An important result has occurred, which allows us to 
take advantage of Equation~(\ref{F2}) avoiding
the number of terms that would otherwise occur.
\footnote{ $i\sigma_2$ acts as a kind of charge
conjugation operator on the chirality states of the Weyl spinors. }
For a fermion
or anti-fermion with momentum $\vec{p}$ 
and helicity $\lambda$, $\psi_{\pm}$ is 
proportional to either $\ket{p}{+}$ or $\ket{p}{-}$.  It is easy to show that
\bec
\bea
\widetilde{\ket{p}{+}} &=& -\ket{p}{-}, \nonumber \\
\widetilde{\ket{p}{-}} &=& +\ket{p}{+}, \nonumber \\
\widetilde{\bra{p}{+}} &=& -\bra{p}{-}, \nonumber \\
\widetilde{\bra{p}{-}} &=& +\bra{p}{+} \,.
\ena
\enc

Finally, recalling Equation~(\ref{Mpm}) the term
\beq
\bra{d}{-}\gamma_{\mu +}\ket{u}{-} = 
\bra{u}{+} \gamma_{\mu -} \ket{d}{+}
\enq
giving, via Equation~(\ref{F2}),
\bea
\lefteqn{ {\M}(+) = 2 (1 + \klc) \sqrt{E_t-{\myabs{t}}} \,
		\braket{t}{+}{d}{+} \braket{u}{+}{b}{-} } \nonumber \\
        & &    + \, 2 \krc \sqrt{E_t+{\myabs{t}}} \,
		\braket{t}{+}{u}{-} \braket{d}{-}{b}{+} \nonumber \\
\lefteqn{ {\M}(-) = 2 (1 + \klc) \sqrt{E_t+{\myabs{t}}} \,
		\braket{t}{-}{d}{+} \braket{u}{+}{b}{-} } \nonumber \\
        & &    + \, 2 \krc \sqrt{E_t-{\myabs{t}}} \,
		\braket{t}{-}{u}{-} \braket{d}{-}{b}{+}. 
\ena
We now have the matrix element in the form we require for our Monte
Carlo package ONETOP\footnote{ A FORTRAN code.},
remembering to include coupling constants, propagators, color factors
and $\sqrt{2 E_u} \sqrt{2 E_b} \sqrt{2 E_d}$.

\section{ Helicity Amplitudes for $u b \ra d t$ in the CMS.}
To illustrate our claim that in the SM (\ie \, $\klc = 0$ and $\krc = 0$) 
only the left-handed top quark is produced from the $u b \ra d t$
process in the $d$-$t$ center of mass frame (CMS), 
we evaluate the matrix element in terms of CMS variables.
Define the four-momenta:
\bec 
\bea
u^{\mu}& =& (\rsh/2,0,0,-\rsh/2) \nonumber \\
b^{\mu}& =& (\rsh/2,0,0, \rsh/2) \nonumber \\
d^{\mu}& =& (t, -t \; \st, 0, -t \; \ct) \nonumber \\
t^{\mu}& =& (E_t, t \; \st, 0, t \; \ct)
\ena
\enc
where
$E_t = (\hat{s}+m_t^2)/2\rsh$,
$t = (\hat{s}-m_t^2)/2\rsh$
and we have chosen $\phi = 0$ to be the scattering plane.

Using these, we obtain from Equations~(\ref{fermion}), (\ref{heigen}) 
and (\ref{braket}) 
\bec
\bea
\ket{u}{+} &=& \vect{0}{-1} \nonumber \\
\ket{u}{-} &=& \vect{1}{ 0} \nonumber \\
\ket{b}{+} &=& \vect{1}{ 0} \nonumber \\
\ket{b}{-} &=& \vect{0}{ 1} \nonumber \\
\ket{d}{+} &=& \vect{ \sin{\theta /2}}{-\cos{\theta /2}} \nonumber \\
\ket{d}{-} &=& \vect{ \cos{\theta /2}}{ \sin{\theta /2}} \nonumber \\
\ket{t}{+} &=& \vect{ \cos{\theta /2}}{ \sin{\theta /2}} \nonumber \\
\ket{t}{-} &=& \vect{-\sin{\theta /2}}{ \cos{\theta /2}}
\ena
\enc
Therefore, 
\vspace{-.3in}
\bec
\bea
\braket{u}{+}{b}{-} &=& -1 		 \nonumber \\
\braket{d}{-}{b}{+} &=&  \cos{\theta /2} \nonumber \\
\braket{t}{+}{d}{+} &=&  0 		 \nonumber \\
\braket{t}{+}{u}{-} &=&  \cos{\theta /2} \nonumber \\
\braket{t}{-}{d}{+} &=& -1 		 \nonumber \\
\braket{t}{-}{u}{-} &=& -\sin{\theta /2}.
\ena
\enc
Including the common factor
\beq
\sqrt{2 E_u} \sqrt{2 E_b} \sqrt{2 E_d} = 
              \sqrt{\sqrt{\hat{s}} (\hat{s} - m_t^2)}
\enq
and 
\bea
\sqrt{E_t+{\myabs{t}}} &=& \sqrt{\sqrt{\hat{s}}} \nonumber \\
\sqrt{E_t-{\myabs{t}}} &=& \sqrt{m_t^2 \over \sqrt{\hat{s}}} \nonumber
\ena
we see that
\bea
\M(+) & = & 2 (\krc) \sqrt{\hat{s}\,(\hat{s}-m_t^2)} \cos^2{\theta /2}, \\
\M(-) & = & 2 (1 + \klc) \sqrt{\hat{s} (\hat{s}-m_t^2)} - 2 (\krc) \sqrt{m_t^2 (\hat{s}-m_t^2)}\,\sin{\theta /2} \cos{\theta /2} \, . \nonumber 
\ena
Notice in the SM, the top
quark is $100\%$ left-hand polarized in the CMS.

Having outlined the general procedure for calculating amplitudes using
the helicity amplitude method, we list the matrix elements contributing
to single top production and the major background $W b \bar b$
in the Standard Model. We 
include the decay of $t \ra b W^+$ and $W^+ \ra \ell^+ \nu_{\ell}$ in the 
final form. 

\section{ Helicity Amplitudes for
$u \, g \ra d \, t (\ra b \, W^+ (\ra \ell^+ \, \nu_{\ell}))\, \bar b$ }

In this and the following sections, we give the diagrams for the process 
listed, indicating the momentum flow and particle momentum labels:  
w's are for $W^+$ bosons, b's for $b$ or $\bar b$ quarks,
e for $e^+$, n for $\nu_e$ and u, d, t and g are for 
$u, d, t$ quarks and gluon, respectively.

\begin{figure}
\hspace{1.in}
\par
\centerline{\hbox{
\psfig{figure=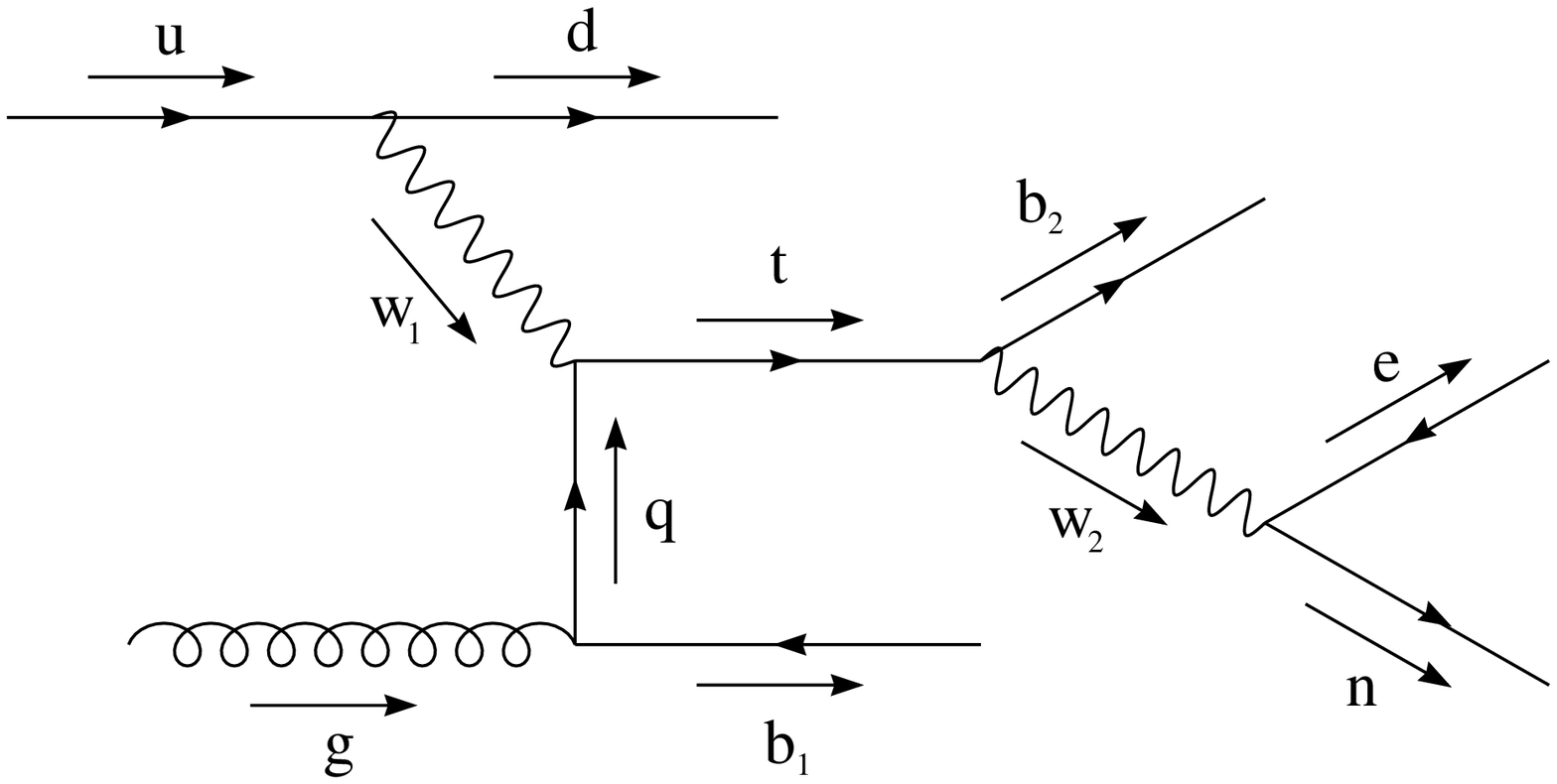,height=1.3in}
\psfig{figure=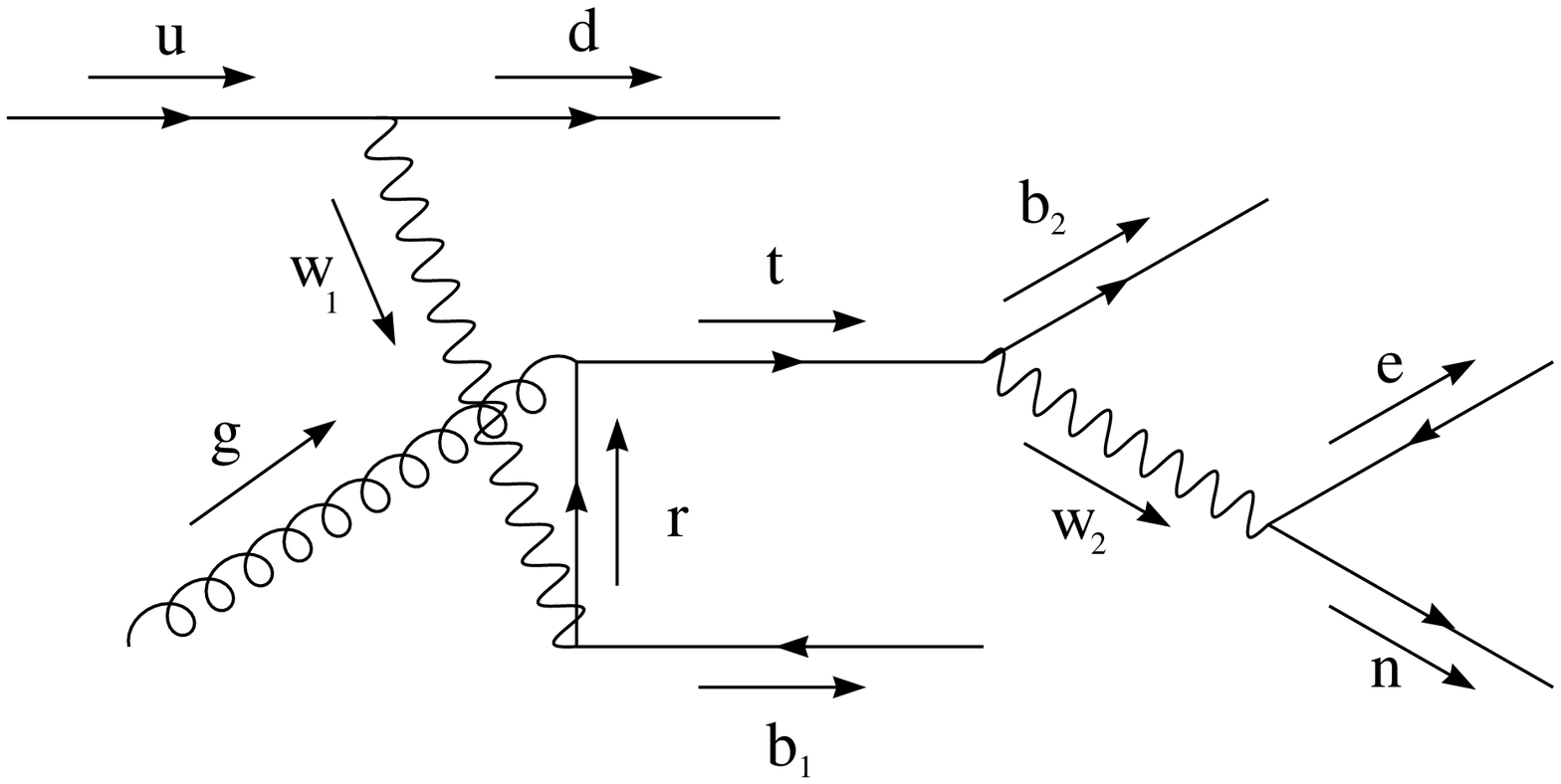,height=1.3in}}}
\caption{ Diagrams for $
u \, g \ra d \, t (\ra b \, W^+ (\ra \ell^+ \, \nu_{\ell}))\, \bar b\,$. }
\label{wgtb12}
\par
\hspace{1.in}
\end{figure}

In Figure~\ref{wgtb12}, we show the diagrams for 
$u \, g \ra d \, t (\ra b \, W^+ (\ra \ell^+ \, \nu_{\ell}))\, \bar b$.
${\M}_i(h_g,\lambda_{b_1})$ will represent the $i$th diagram ($i = 1,2$ 
from left to right in Figure~\ref{wgtb12})
where $h_g$ represents the two transverse
gluon polarizations and $\lambda_{b_1}$ represents the two helicity states
of $\bar b$. 
The matrix element in the helicity amplitude formalism for this process is
\bea
\lefteqn{ {\M}_1(h_g,\mp) = 
4 \braket{b_2}{-}{n}{+} \bsk{e}{+}{t_{-}}{d}{+} \times} \nonumber \\ & &
{ { \pm \sqrt{E_{b_1} \! \mp \myabs{b_1}} \, 
\bssk{u}{+}{q_-}{\veps(h_g)_+}{b_1}{\pm} 
\mp m_b \sqrt{E_{b_1} \! \pm \myabs{b_1}} \, 
\bsk{u}{+}{\veps(h_g)_-}{b_1}{\pm}  } \over (q^2-m_b^2)} \nonumber \\
\lefteqn{ {\M}_2(h_g,\mp) = \pm 4 \sqrt{E_{b_1} \! \mp \myabs{b_1}} \,
\braket{b_2}{-}{n}{+} \braket{u}{+}{b_1}{\pm} \times} \nonumber \\ & &
{ {  \bsssk{e}{+}{t_-}{\veps(h_g)_+}{r_-}{d}{+} 
+ m_t^2 \bsk{e}{+}{\veps(h_g)_-}{d}{+}  } \over (r^2-m_t^2)}
\ena
where we have indicated the four different helicity states involved in this 
process.  We keep the mass of the $\bar b$ parton to avoid
the case where the $b$ propagator goes on shell.
For simplicity we have omitted a common factor of
\beq
{g^2_S} {\left({g_W \over \sqrt{2}}\right)^4}
{\sqrt{(2 E_u)(2 E_d)(2 E_{b_2})(2 E_e)(2 E_n)}
\over {(t^2-m_t^2)(w_1^2-M_W^2)(w_2^2-M_W^2)}} 
\enq
and color matrices.\footnote{ The color factor for the amplitude squared
is $3\times 4 \times {1/3} \times {1/8}$ for this process.}  
We note that the polarization vectors for spin-1 gauge bosons may be expressed
in terms of spin-$1\over 2$ bras and kets.
We define
\beq
\kt{+}\ = \vect{1}{0} \quad \kt{-}\ = \vect{0}{1}, \, 
\enq
then for the transverse polarizations
\bea
\slash{\veps^{(R)}_{\pm}} &=& \mp \sqrt{2} \ket{g}{+} \bra{g}{-} \nonumber \\
\slash{\veps^{(L)}_{\pm}} &=& \pm \sqrt{2} \ket{g}{-} \bra{g}{+} 
\ena
and for massive spin-1 gauge bosons 
\bea
\slash{\veps^{(0)}_{+}} &=& \left({\myabs{g}\over m}-{E\over m} \right)
\left( \kt{+} \br{+} + \kt{-} \br{-} \right) +
2 {E\over m}\ket{g}{+} \bra{g}{+} \nonumber \\
\slash{\veps^{(0)}_{-}} &=& \left({\myabs{g}\over m}-{E\over m} \right)
\left( \kt{+} \br{+} + \kt{-} \br{-} \right) +
2 {E\over m}\ket{g}{-} \bra{g}{-}.
\ena


\section{ Helicity Amplitudes for
$u \, b \ra d \, t (\ra b \, W^+ (\ra \ell^+ \, \nu_{\ell}))$ }

\begin{figure}
\hspace{1.in}
\par
\centerline{\hbox{
\psfig{figure=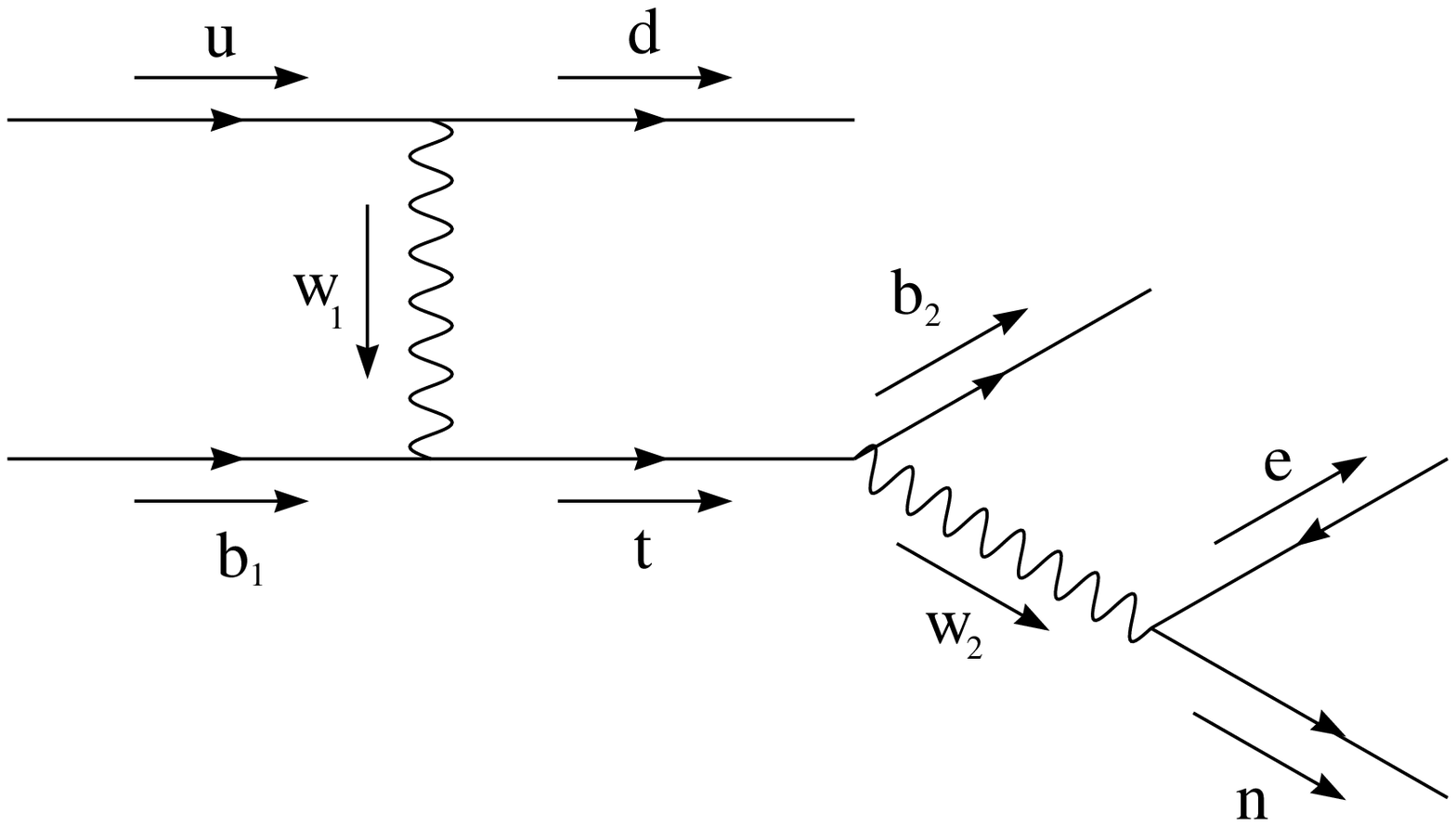,height=1.3in}}}
\caption{ Diagram for $
u \, b \ra d \, t (\ra b \, W^+ (\ra \ell^+ \, \nu_{\ell}))$.}\label{ubdt}
\par
\hspace{1.in}
\end{figure}

In Figure~\ref{ubdt}, we show the diagram for 
$u \, b \ra d \, t (\ra b \, W^+ (\ra \ell^+ \, \nu_{\ell}))$.
The matrix element in the helicity amplitude formalism for this process is
\beq
{\M} = 4 \braket{u}{+}{b_1}{-} \bsk{e}{+}{t_{-}}{d}{+} \braket{b_2}{-}{n}{+}
\enq
where we ignore the $b$ parton mass.
We have omitted a common factor of
\beq
{\left({g_W \over \sqrt{2}}\right)^4}
{\sqrt{(2 E_u)(2 E_d)(2 E_{b_1})(2 E_{b_2})(2 E_e)(2 E_n)}
\over {(t^2-m_t^2)(w_1^2-M_W^2)(w_2^2-M_W^2)}} 
\enq
and color matrices.\footnote{ The color factor for the amplitude squared
is $3\times 3 \times {1/3} \times {1/3}$ for this process.}

\section{ Helicity Amplitudes for
$u \, \bar d \ra W^{*} \ra \bar b \, t (\ra b \, W^+ (\ra \ell^+ \, \nu_{\ell}))$ }
\begin{figure}
\hspace{1.in}
\par
\centerline{\hbox{
\psfig{figure=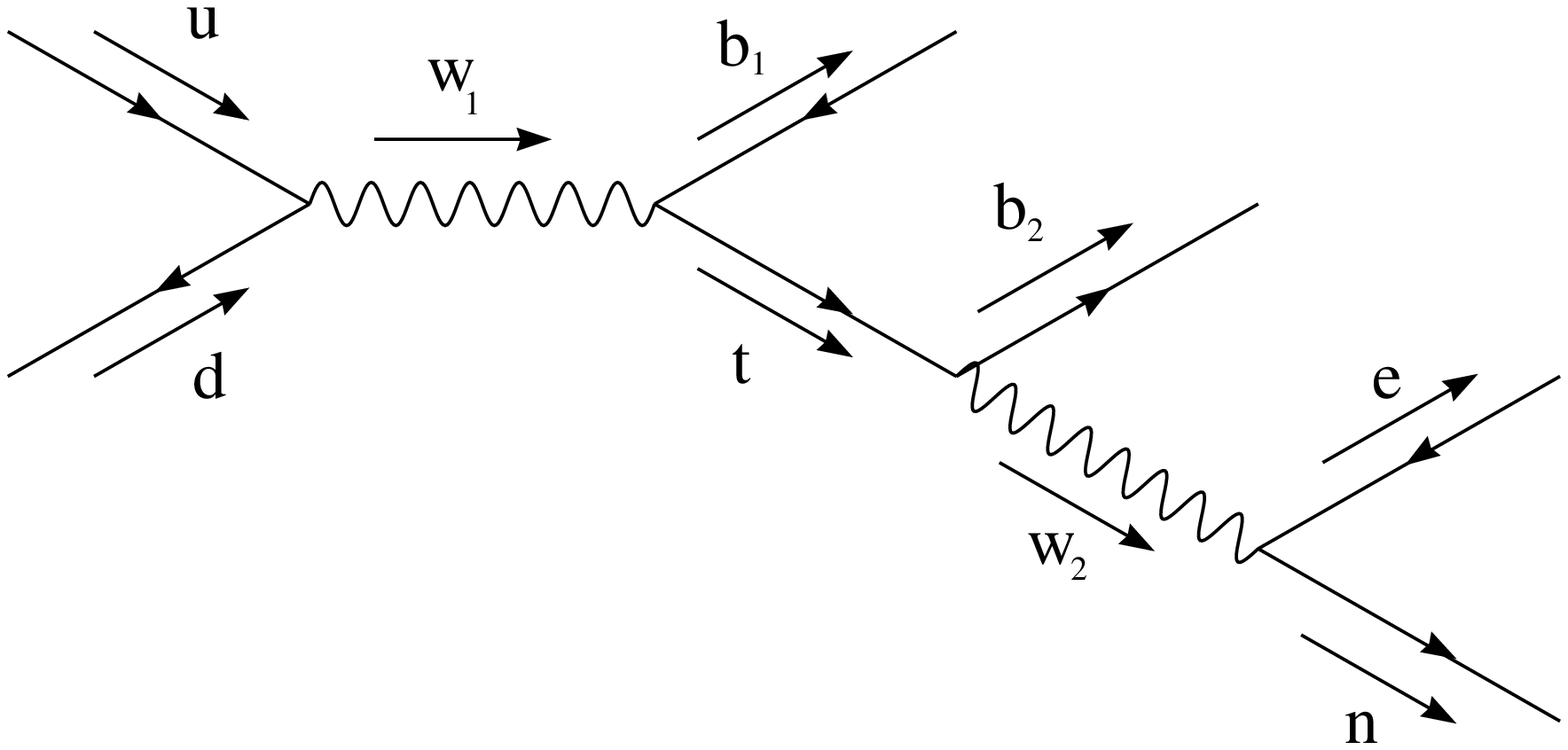,height=1.3in}}}
\caption{ Diagram for $
u \, \bar d \ra \bar b \, t (\ra b \, W^+ (\ra \ell^+ \, \nu_{\ell}))$.}
\label{udbt}
\par
\hspace{1.in}
\end{figure}

In Figure~\ref{udbt}, we show the diagram for 
$u \, \bar d \ra W^{*} \ra
\bar b \, t (\ra b \, W^+ (\ra \ell^+ \, \nu_{\ell}))$.
Aside from a possible phase, the matrix element in the helicity amplitude 
formalism for this process ($W^*$ production) is identical to that of 
Figure~\ref{ubdt}.
This is because one process is the cross diagram of the other and therefore
only the momentum assignments are different.
For clarity, it is 
\beq
{\M} = 4 \braket{u}{+}{b_1}{-} \bsk{e}{+}{t_{-}}{d}{+} \braket{b_2}{-}{n}{+}.
\enq
Again we ignore the $b$ parton mass and omit the common factor of
\beq
{\left({g_W \over \sqrt{2}}\right)^4}
{\sqrt{(2 E_u)(2 E_d)(2 E_{b_1})(2 E_{b_2})(2 E_e)(2 E_n)}
\over {(t^2-m_t^2)(w_1^2-M_W^2)(w_2^2-M_W^2)}}
\enq
and color matrices.\footnote{ The color factor for the amplitude squared
is $3\times 3 \times {1/3} \times {1/3}$ for this process.}

\section{ Helicity Amplitudes for
$u \, \bar d \ra \bar b \, b \, W^+ (\ra \ell^+ \, \nu_{\ell})$ }

\begin{figure}
\hspace{1.in}
\par
\centerline{\hbox{
\psfig{figure=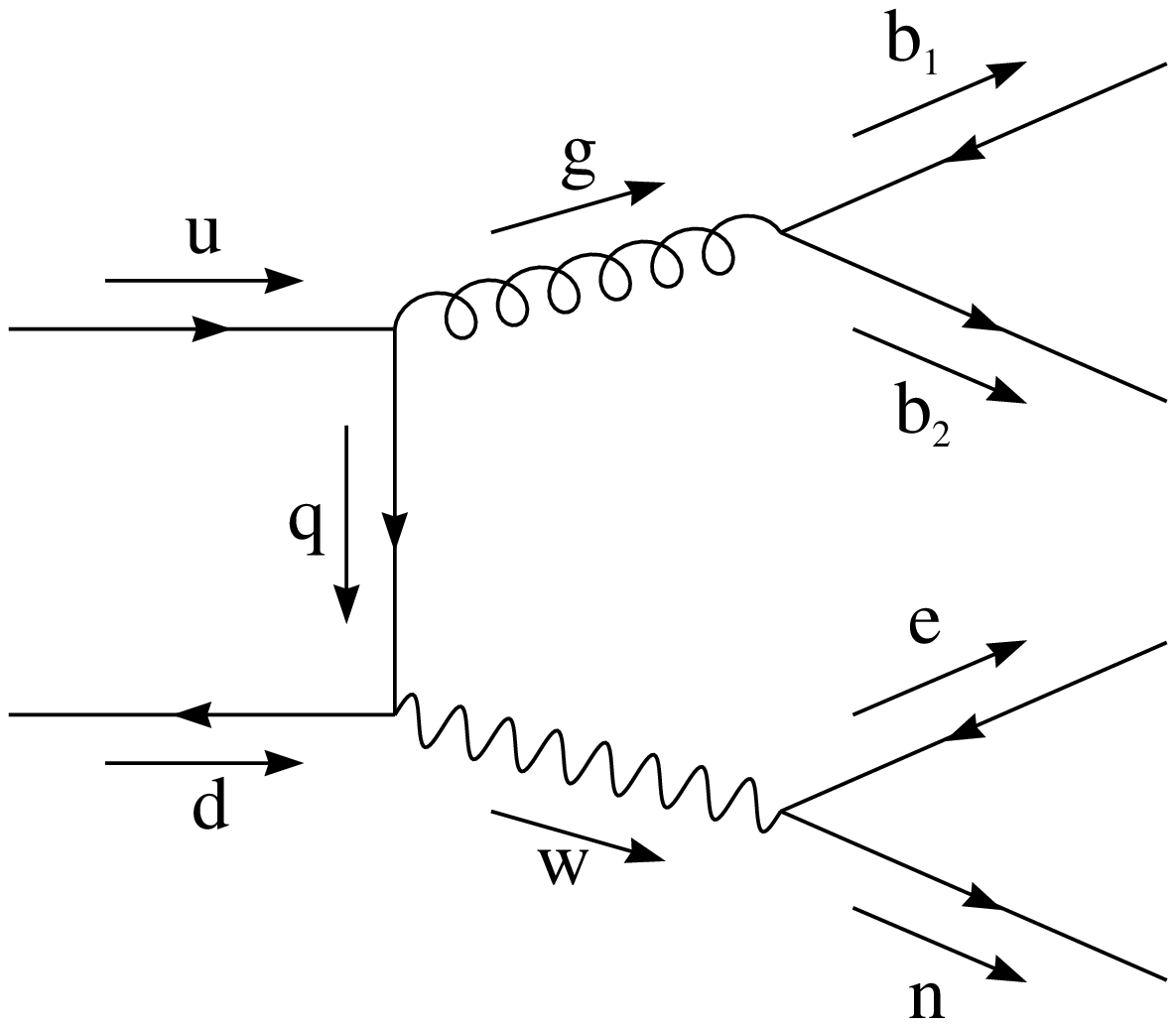,height=1.3in}
\psfig{figure=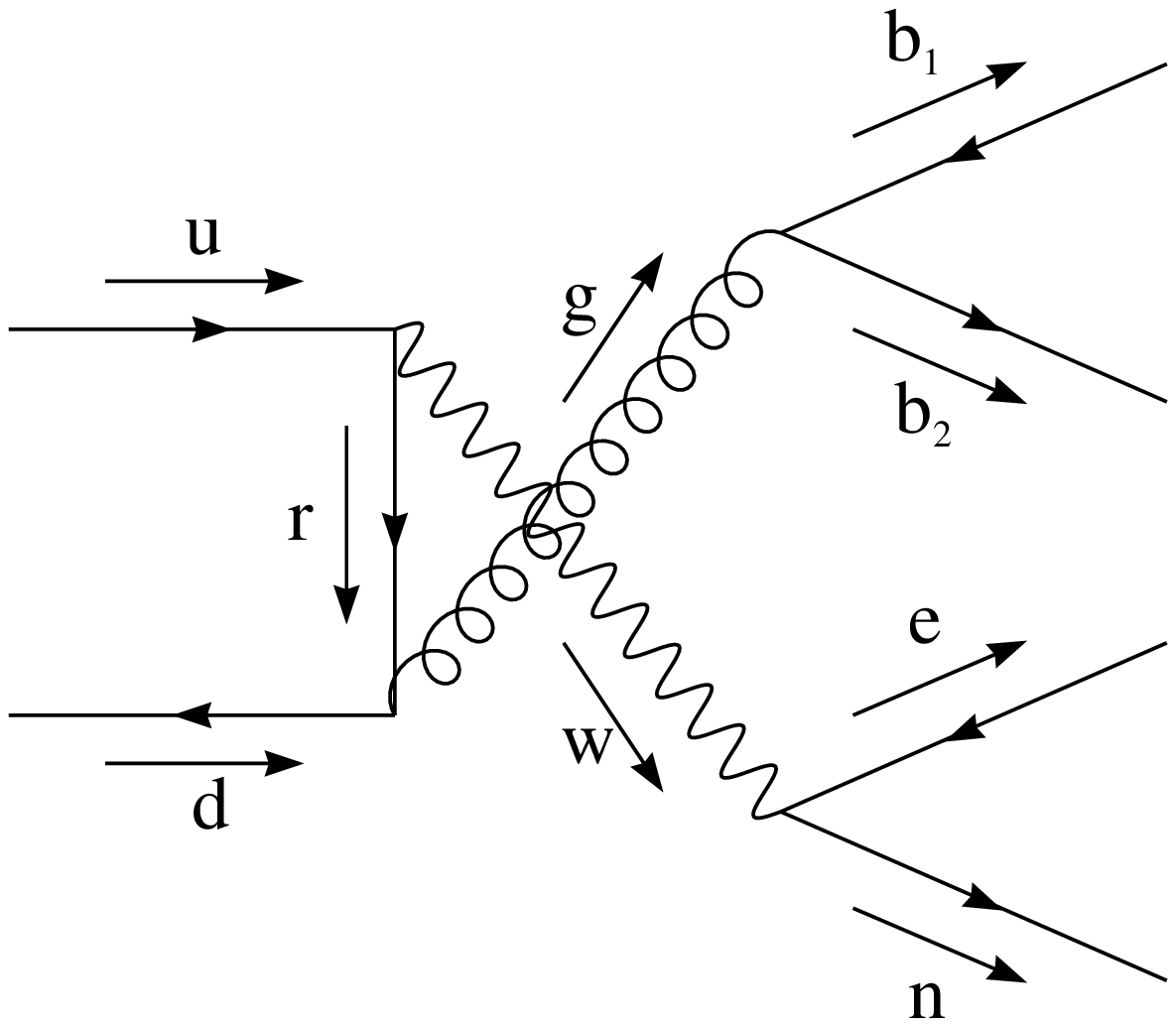,height=1.3in}}}
\caption{ Diagrams for $
u \, \bar d \ra \bar b \, b \, W^+ (\ra \ell^+ \, \nu_{\ell})$. }\label{udbb12}
\par
\hspace{1.in}
\end{figure}

In Figure~\ref{udbb12} we show the diagrams for 
$u \, \bar d \ra \bar b \, b \, W^+ (\ra \ell^+ \, \nu_{\ell})$, the major
$W + 2$ jets background to $W$--gluon fusion including $b$-tagging. The matrix 
element in the helicity amplitude formalism is
\bea
\lefteqn{ {\M(+,-)} = 4 \braket{d}{-}{n}{+} \braket{b_1}{+}{u}{-} 
\bsk{e}{+}{q_{-}}{b_2}{+}/q^2} \nonumber \\ 
& & {+ \, 4 \braket{d}{-}{b_2}{+} \braket{e}{+}{u}{-} 
\bsk{b_1}{+}{r_{-}}{n}{+}/r^2} \nonumber \\
\lefteqn{ {\M(-,+)} = 4 \braket{d}{-}{n}{+} \braket{b_2}{+}{u}{-} 
\bsk{e}{+}{q_{-}}{b_1}{+}/q^2} \nonumber \\ 
& & {+ \, 4 \braket{d}{-}{b_1}{+} \braket{e}{+}{u}{-} 
\bsk{b_2}{+}{r_{-}}{n}{+}/r^2}
\ena
where we have indicated the helicity states of $\bar b$ and $b$ as 
${\M(\lambda_{b_1},\lambda_{b_2})}$.
We have once again left out the factor of 
\beq
{g_S^2}{\left({g_W \over \sqrt{2}}\right)^2}
{\sqrt{(2 E_u)(2 E_d)(2 E_{b_1})(2 E_{b_2})(2 E_e)(2 E_n)}
\over {(g^2)(w^2-M_W^2)}} 
\enq
and color matrices.
\footnote{ The color factor for the amplitude squared
is $2\times {1/3} \times {1/3}$ for this process.}

\chapter{Event Rate of the ($2\ra 3$) Process $u g\ra d t \bar b$}

Monte Carlo integration is an indispensable tool in phenomenology. However,
when performing a calculation one often encounters singularities which
make it impossible to obtain meaningful results from a Monte Carlo program. 
In the case of delta functions one is forced to integrate by hand.  Other
singularities may occur when propagators go on mass--shell.
These types of divergences may be regularized by applying suitable cuts on
the external particles in the process. However, when one is interested
in obtaining a total rate, part of the calculation must be performed by
hand if there is any singularity present.

\begin{figure}[b]
\hspace{1.in}
\par
\centerline{\hbox{
\psfig{figure=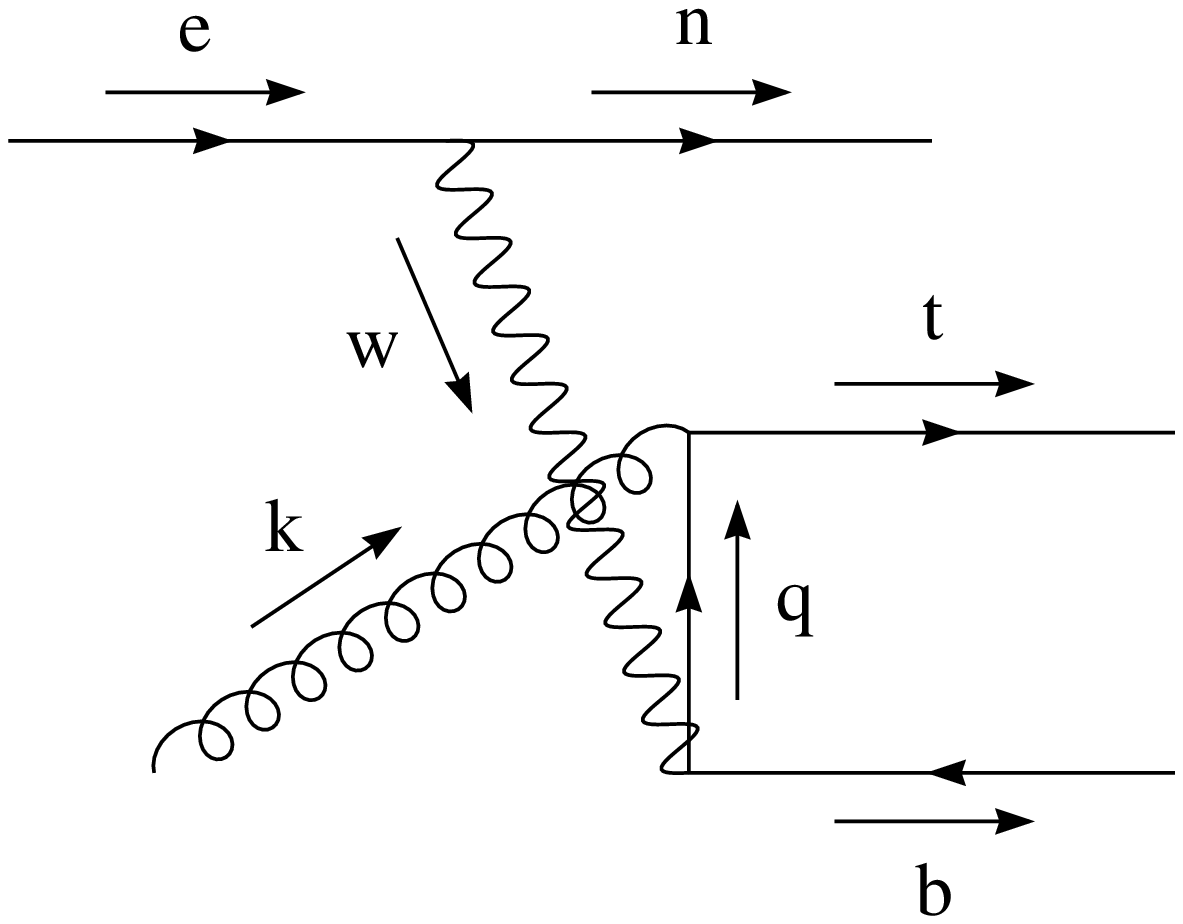,height=1.3in}
\psfig{figure=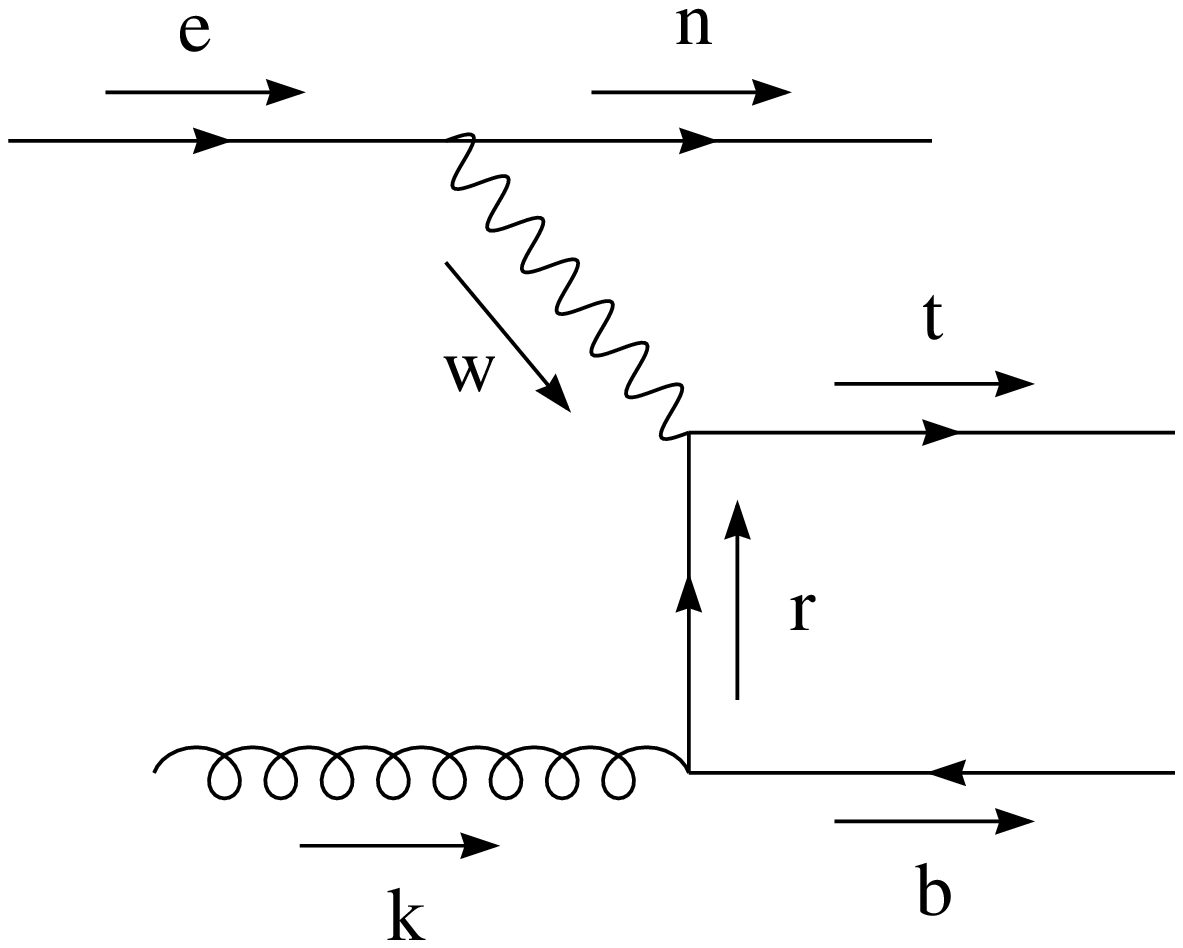,height=1.3in}}}
\caption{ Diagrams for $u \, g \ra d \, t \, \bar{b}\,$. }
\label{wgtb34}
\par
\hspace{1.in}
\end{figure}

To obtain the total rate for the ($2\ra 3$) process $u g\ra d t \bar b$ 
via Monte Carlo integration,
the idea is to integrate out the $W^+ g \ra t b$ sub-cross section 
in Figure~(\ref{wgtb34}) by hand. 
The alternative is to use Monte Carlo for the full 3-body phase space, but
the singularity due to the small mass of the $b$ quark 
in the right diagram would require too
much computer time even for the powerful technique applied in VEGAS,
a fortran code for calculating multiple dimensional 
integrations \cite{vegasdoc}. 
An added benefit of performing the sub-cross section integration
in the way shown in this appendix is an understanding of the validity of the
effective--$W$ approximation\cite{equiv}.

The differential cross section for this 
process is 
\beq
\sigma = {1 \over 2s} {\int\!{d^3n \over {(2\pi)^3 (2E_n)}}} \sigma_{sub}
\enq
where
\beq
\sigma_{sub} \equiv {\int\!{d^3t \over {(2\pi)^3 (2E_t)}}
                      {d^3b \over {(2\pi)^3 (2E_b)}}\,
		      {\overline{|\M|^2}}\,
		      {(2\pi)^4}{\delta^4(e + k - n - t - b)}
	            }.
\enq
Since $\sigma_{sub}$ is a Lorentz invariant, we choose to evaluate it 
and the matrix element in the 
$t \bar b$ center of mass (CMS) frame.  We express the 4-vector components of
$t, b, e, n$ and $k$ explicitly as
\bea
t &=& (E_t, p \; \st \; \cp, p \; \st \; \sp, p \; \ct) \nonumber\\
b &=& (E_b, -p \; \st \; \cp, -p \; \st \; \sp, -p \; \ct) \nonumber\\
e &=& E_e (1, \st_e, 0, \ct_e) \nonumber\\
n &=& E_{n} (1, \st_{n}, 0, \ct_{n}) \nonumber\\
k &=& E_k (1, 0, 0, 1) .
\ena
In the $t \bar b$ rest frame $\sh \equiv (t + b)^2$, so
\bea
E_t &=& {\sh + m_t^2 - m_b^2} \over 2\rsh \nonumber\\
E_b &=& {\sh - m_t^2 + m_b^2} \over 2\rsh \nonumber\\
  p &=& {1 \over 2\rsh} \sqrt{[\sh - (m_t + m_b)^2][\sh - (m_t - m_b)^2]} .
\ena

It is easy to see in the $t \bar b$ CMS, that $\vec{e}$, $\vec{k}$, $\vec{n}$ 
all lie in the same plane.  Momentum conservation 
($e + k = n + t + b$) and $\vec{t} + \vec{b} = \vec{0}$\
imply $\vec{e} + \vec{k} = \vec{n}$.  It only takes two vectors to 
describe a plane and any vector which is a linear combination of those
two vectors lies in that plane.
Therefore, we choose the vectors $\vec{e}$, $\vec{k}$, $\vec{n}$ to define
the $x$--$z$ plane with the momentum of the gluon ($\vec{k}$) along
the $z$--axis.

For a 3-body final state, there are only $5$ ($=3 \times 3 - 4$) 
independent variables.  (The minus 4 is for energy-momentum conservation.)
We shall choose the 5 independent variables to be
$\theta$, $\phi$, $E_{n}$, $\theta_{n}$ and $\phi_{n}$ where 
$\phi_{n}$ can be trivially integrated as $\int\!d\phi_{n} = 2\pi$.
This is a consequence of the arbitrariness of the choice of scattering 
planes.
Hence, we should be able to express all vectors in terms of the 4 variables
$\theta$, $\phi$, $E_{n}$ and $\theta_{n}$.

What we have done so far is express all 4-vectors in the $t \bar b$ CMS.
It is now a fairly trivial exersize to evaluate $\sigma_{sub}$.\footnote{
This integration is done analytically to avoid bad convergence in the 
numerical integration method.  All the singular terms are of the nature of 
$\ln(m_b^2)$. Because the analytic form for this result is long, it will not
be explicitly given here.}
We found
\beq
{\sigma_{sub} = {1 \over {4(2\pi)^2}}\,{p \over{\rsh}}\,
                {\int\!{d\cos\theta}\,{d\phi}}\,{\overline{|\M|^2}}}.
\enq
Having done so, we now have a Lorentz invariant expression for 
$${\sigma_{sub} = \sigma_{sub}(\sh,E_{n},\theta_{n})}.$$
To compare our result with the effective--$W$ approximation, we express
$E_{n}$ and $\theta_{n}$ in terms of $e k$ CMS quantities with the
aid of some projection operators defined below.

To extract out the information of $e$, $k$ and $n$ in the $t \bar b$ 
CMS frame 
we use the fact that $\vec{e} + \vec{k} - \vec{n} = 0$ and
$(e + k -n)^2 = \sh$.  We then define the projection operators
${\bf P}_E$ and ${\bf P}_{\parallel}$ as follows:
\bea
{\bf P}_E V &\equiv& {(e + k - n) \cdot V \over \rsh} \\
{\bf P}_{\parallel} V &\equiv& {{-k \cdot V + E_k E_V} \over E_k}
\ena
where $V = e, n$ or $k$.  ${\bf P}_E$ and ${\bf P}_{\parallel}$ 
project out the energy of $V$ and component of $V$ parallel to $k$
(\ie, the $z$--axis) respectively.  We express the 4-vector components of
$e, n$ and $k$ explicitly as
\bea
k &=& {\rts \over 2} (1, 0, 0, 1) \nonumber\\
e &=& {\rts \over 2} (1, 0, 0, -1) \nonumber\\
n &=& E_{n}^{'} (1, -\st_{n}^{'}, 0, -\ct_{n}^{'})
\ena
where $s = (e + k)^2$.
From now on we will distinguish $e k$ CMS components from $t \bar{b}$
CMS components by a prime.  We assume 
$$E_{n}^{'} = E_{e}^{'}{(1-x)} = {\rts \over 2}{(1-x)}$$
and use the following results
\bea
\sh &=& (e + k - n)^2 = x s \nonumber\\
W^2 &\equiv& (e-n)^2 = - {s \over 2}{(1-x)}{(1-\ct_{n}^{'})}.
\ena
Using the projection operators it is easy to show 
\bea
E_k &=& {1 \over {2 \sqrt{x s}}}{(x s - W^2)} \nonumber\\
E_e &=& {1 \over {2 \sqrt{x s}}}{(s + W^2)} \nonumber\\
E_n &=& {1 \over {2 \sqrt{x s}}}{s (1-x)} \nonumber\\
k_{\parallel} &=& E_k \nonumber\\
e_{\parallel} &\equiv& E_e \ct_{e} =
{-1 \over {2 E_k}}{(W^2 + s(1-x))} + E_n \nonumber\\
n_{\parallel} &\equiv& E_n \ct_{n} = {-s \over {2 E_k}} + E_e 
\ena
and therefore $${\sigma_{sub} = \sigma_{sub}(s,x,W^2)}.$$
Also, the Lorentz invariant phase space integral
in terms of $e k$ CMS components becomes
\beq
{\int\!{d^3n^{'} \over {(2\pi)^3 (2E_{n}^{'})}}} = 
{1 \over (4\pi)^2}{\int_{0}^{1}\!{dx}}{\int_{-s(1-x)}^{0}\!{dW^2}}.
\enq
Finally, the following cross section 
\beq
\sigma =
{1 \over 2s(4\pi)^2}{\int_{0}^{1}\!{dx}}{\int_{-s(1-x)}^{0}\!{dW^2}}
{\sigma_{sub}(s,x,W^2)}
\enq
may be folded in with parton
distributions and safely evaluated using a Monte Carlo program.

We note that $\sqrt{W^2}$ is the virtuality of the $W$--boson line emitted
from the $u$--quark line (with momentum $e$).  ($1-x$) is the fraction of the 
incoming $u$--quark energy carried away by the outgoing $d$--quark line in the
$ek$ CMS. If desired, one can approximate the above equation as the result 
of the effective--$W$ approximation. However, we shall not pursue it further 
here.

\chapter{ Helicity Amplitudes
of ${t\ra W^+ b}$ and ${\bar t \ra W^- \bar b}$ }
 
In Equations~(\ref{eqlag}) and~(\ref{eqlag2}) we have listed
the most general form factors for the decay processes
$t\rightarrow W^+\ +\ b$ and
$\bar{t}\rightarrow W^-\ +\ \bar{b}$.
Here we use those equations to calculate the helicity amplitudes
for an on--shell $W$--boson. (We take the limit of $m_b\ra 0$ in the
following for $m_t \gg m_b$.)
 
For the decay process $t\rightarrow W^+b$,
the top quark is taken to decay in its rest frame
where the top quark momentum is $p_t=(m_t,0,0,0)$.
Spherical coordinates are used to describe the outgoing particles; $\theta$ is
taken from the positive $Z$--axis and $\phi$ is taken
from the positive $X$--axis in the
$X-Y$ plane.  The bottom quark and the $W$--boson are taken on their mass shells
with the four--momenta for the bottom quark ($p_b$) and the $W$--boson
($p_W$) taken as
\bec
\bea
p_b &= & (E_b,-E_b\sin\theta\cos\phi,-E_b\sin\theta\sin\phi,-E_b\cos\theta),
\nonumber \\
p_W &= & (E_W,E_b\sin\theta\cos\phi,E_b\sin\theta\sin\phi,E_b\cos\theta),
\ena
\enc
where we have neglected the bottom quark mass, and
\beq
E_b={m_t^2-M_W^2 \over 2 m_t}.
\enq
The angles $\theta$ and $\phi$ refer to the direction of the $W$--boson.
 
Denote the helicity amplitudes as $(h_t,\lambda_W,h_b)$ with
$\lambda_W=-, +, 0$ being a left-handed, right-handed,
and longitudinal $W$--boson.
After suppressing the common factor
\beq
{- g \over \sqrt{2}} \sqrt{2 E_b m_t},
\enq
there are 8 nonvanishing helicity amplitudes
in the rest frame of the top quark for $m_b=0$:
\bec
\bea
(-\ 0\ -) &= & \left(\tow\fol+\cpyftr\right)\stt,  \nonumber \\
(---) &= & \rt\left(\fol+\tow\cpyftr\right)\ctt\phip,  \nonumber \\  
(+\ 0\ -) &= & \left(\tow\fol+\cpyftr\right)\ctt\phip, \nonumber \\
(+--) &= & -\rt\left(\fol+\tow\cpyftr\right)\stt e^{2i\phi}, \nonumber \\
(-\ 0\ +) &= & -\left(\tow\for+\cpyftl\right)\ctt\phim,      \nonumber \\
(-++) &= & -\rt\left(\for+\tow\cpyftl\right)\stt e^{-2i\phi}, \nonumber \\
(+\ 0\ +) &= & \left(\tow\for+\cpyftl\right)\stt,  \nonumber \\
(+++) &= & -\rt\left(\for+\tow\cpyftl\right)\ctt\phim.
\label{helamps}     
\ena
\enc
To obtain the averaged amplitude squared, a spin factor
$ {1\over 2}$ should be included.
We note that there is no right-handed $W$--boson produced with
a massless left-handed $b$ from a top quark decay. Similarly, from helicity
conservation, it is not possible to have a left-handed $W$--boson
produced with a
massless right-handed $b$ from $t$ decay. 

For an unpolarized top quark decay, 
after summing over the helicities of the bottom quark, 
the amplitudes squared for
various $W$ polarizations are, apart from a common factor $(g^2E_bm_t)$,
\bec
\bea
\overline{|M(\lambda_W=-)|^2} &=& {\left|\fol+\tow {f_2^R}\right|}^2 
\, , \nonumber \\
\overline{|M(\lambda_W=+)|^2} &=& {\left|\for+\tow {f_2^L}\right|}^2
\, , \nonumber \\
\overline{|M(\lambda_W=0)|^2} &=& {1\over 2}{\left|\tow\fol+{f_2^R}\right|}^2
+ {1\over 2}{\left|\tow\for+{f_2^L}\right|}^2 \, .
\ena
\enc
The fraction ($\flong$) of longitudinally polarized $W$-boson produced
in the rest frame of the top quark
is defined as the ratio of the number of longitudinally
polarized $W$--bosons produced with respect to the total number of
$W$--bosons produced in top quark decays:
\bec
\bea
\flong &=& {\Gamma(\lambda_W=0)
\over{\Gamma(\lambda_W=0)+\Gamma(\lambda_W=-)+\Gamma(\lambda_W=+)}}
\nonumber \\
 \qquad &=& { \overline{|M(\lambda_W=0)|^2}
\over \overline{|M(\lambda_W=0)|^2} + \overline{|M(\lambda_W=-)|^2}
+ \overline{|M(\lambda_W=+)|^2} }
\, ,
\ena
\enc
where we use $\Gamma(\lambda_W)$ to refer to the decay rate for a
top quark to decay into a $W$--boson with polarization $\lambda_W$.

Using a parallel definition for the process
$\bar t \ra W^- \bar b$, we
obtain the helicity amplitudes $(h_{\bar t},\lambda_{W},h_{\bar b})$,
similar to the ones listed in Equation~(\ref{helamps})
for the $t \ra W^+ b$ process, provided we replace
$f_1^L$ by ${f_1^R}^*$, $f_1^R$ by ${f_1^L}^*$,
$f_2^L$ by ${f_2^R}^*$, and $f_2^R$ by ${f_2^L}^*$. (Here the 
superscript $*$ means complex conjugate.)
 
The helicity amplitudes of the process $W^+ \ra e^+ \nu_e$
are well known. After suppressing the common factor ($ g M_W$),
they are
\bec
\bea
(\lambda_W=-) &= & -e^{-i\phi^*_e}
              \left({1-\cos\theta^*_e \over 2}\right), \nonumber \\
(\lambda_W=0) &= & -{\sin\theta^*_e \over \sqrt{2}}, \nonumber \\
(\lambda_W=+) &= & -e^{i\phi^*_e}
             \left({1+\cos\theta^*_e \over 2}\right), 
\label{wenu}
\ena
\enc
where $\theta^*_e$ and $\phi^*_e$ refer to $e^+$ in the rest frame of $W^+$.
 
The helicity amplitudes $(\lambda_W)$
for the decay process $W^- \ra e^- \bar{\nu_e}$
can be obtained from Equation~(\ref{wenu}) by replacing $\theta^*_e$ by
$\pi-\theta^*_e$ and $\phi^*_e$ by $\pi +\phi^*_e$.
In this case, $\theta^*_e$ and $\phi^*_e$ refer to $e^-$ in
the rest frame of $W^-$.

\chapter{The Total Rate for $W$--gluon Fusion}

As discussed in Section 2, the total rate for the $W$--gluon fusion
process is obtained by  
$$Total = (2 \ra 2)\, + \, (2 \ra 3) \, - \, ({\rm splitting\,piece})$$
and the rates of
\bea
\lefteqn{ (2 \ra 2) = \int\!{d\xi_1\,d\xi_2}\,{f_{q'/A}(\xi_1, \mu)}\,
                      {f_{b/B}(\xi_2, \mu)}\,{\hat{\sigma}(\ubdt)}} \nonumber \\
	        & & + \int\!{d\xi_1\,d\xi_2}\,{f_{b/A}(\xi_1, \mu)}\,
		      {f_{q'/B}(\xi_2, \mu)}\,{\hat{\sigma}(\budt)} \\
\lefteqn{ (2 \ra 3) = \int\!{d\xi_1\,d\xi_2}\,{f_{q'/A}(\xi_1, \mu)}\,
                      {f_{g/B}(\xi_2, \mu)}\,{\hat{\sigma}(\ugtb)}} \nonumber \\
	        & & + \int\!{d\xi_1\,d\xi_2}\,{f_{g/A}(\xi_1, \mu)}\,
		      {f_{q'/B}(\xi_2, \mu)}\,{\hat{\sigma}(\gutb)} \\
\lefteqn{ ({\rm splitting\,piece}) = 
	       \int\!{d\xi_1\,d\xi_2}\,{f_{q'/A}(\xi_1, \mu)}\,
     {\widetilde{f_{b/B}}(\xi_2, \mu)}\,{\hat{\sigma}(\ubdt)}} \nonumber \\
	& & + \int\!{d\xi_1\,d\xi_2}\,{\widetilde{f_{b/A}}(\xi_1, \mu)}\,
		      {f_{q'/B}(\xi_2, \mu)}\,{\hat{\sigma}(\budt)}\label{dumb} 
\ena
where, for instance,
$f_{b/A}(\xi_1, \mu)$ denotes the parton distribution function (PDF)
of the $b$ quark inside hadron $A$, carrying the fraction $\xi_1$ of the 
hadron momentum, and $\mu$ is the energy scale at which the PDF is evaluated.
\begin{figure}[b]
\centerline{\hbox{\psfig{figure=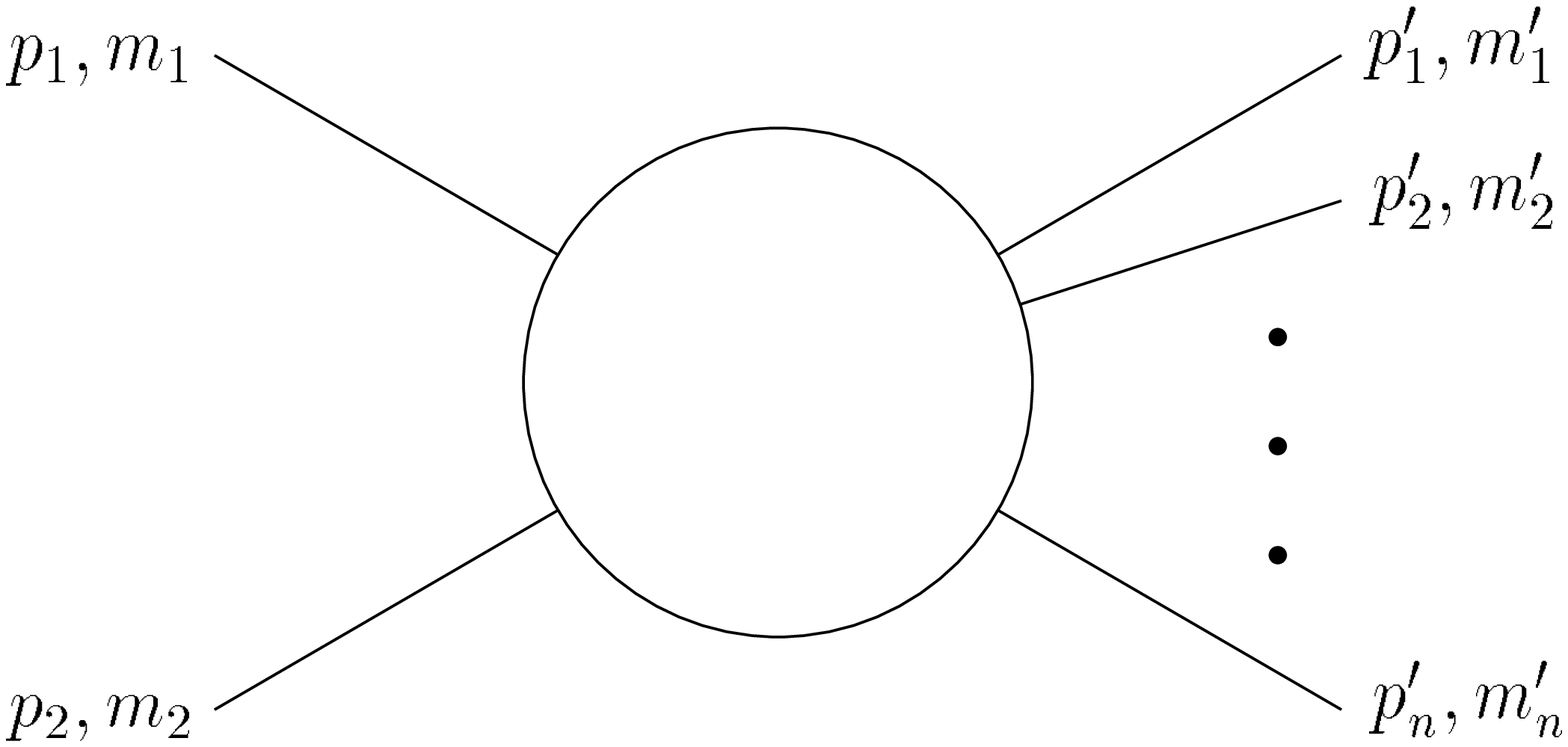,height=1.5in}}}
\caption{$n$--body scattering}
\label{nbdy}
\end{figure}
The constituent cross section $\hat{\sigma}$ is given by the differential cross
section for
\beq
p_1(m_1) + p_2(m_2) \ra p'_1(m'_1) + p'_2(m'_2) + \cdots + p'_n(m'_n),
\enq
as shown in Figure~\ref{nbdy}.
The differential cross section is
\beq
d\hat{\sigma} = {{\overline{| {\M} \, |^2} \, {{d\Phi_n}
     (p_1 + p_2; p'_1, \cdots ,p'_n)}} \over
     {4 \sqrt{(p_1 \cdot p_2)^2 - m_1^2 m_2^2}}}
\enq
with
\beq
{{d\Phi_n}(p_1 + p_2; p'_1, \cdots ,p'_n)} = 
  (2 \pi)^4 \delta^4(p_1 + p_2 - \sum_{i=1}^{n} p'_i) 
  \prod_{i=1}^{n} {{{d^3}{p'_i}} \over {(2 \pi)^3 dE'_i}}
\enq
and $\overline{| {\M} \, |^2}$ is the square of the amplitude 
after summing over
the final state polarization and colors and averaging over the initial state
polarization and colors.  Notice that the differential cross section for
$d\hat{\sigma}(\budt)$ can be obtained from $d\hat{\sigma}(\ubdt)$ by
interchanging the 4--momenta $p_1$ and $p_2$ in the scattering amplitudes.
In terms of the polar angle $\theta^{*}$ and azimuthal angle $\phi^{*}$
defined in the center-of-mass frame of the partons, this means replace 
$\theta^{*}$ by $\pi - \theta^{*}$ and $\phi^{*}$ by $\pi + \phi^{*}$.
In Equation~(\ref{dumb}), the effective parton density
\beq
\widetilde{f_{b/A}}(\xi, \mu) = {{{\alpha_S}(\mu) \over {2 \pi}}
    \ln \left({\mu^2 \over m_b^2} \right) \int\!{{dz \over z} 
    \left[{{z^2 + (1-z)^2} \over 2} \right]}
    {f_{g/A} \left({\xi \over z}, \mu \right)}}
\enq
in the $\overline{\rm MS}$ scheme.
The coupling constant 
\beq
{\alpha_S}(\mu) = {{12 \pi} \over {(33 - 2 n_f) \ln({\mu^2 \over \Lambda^2})}}.
\enq
Here, $n_f$ is the number of quarks with mass less than the energy scale $\mu$.
The QCD parameter $\Lambda \equiv \Lambda^{(n_f)}_{\overline{\rm MS}}$ 
is an experimentally determined parameter. Perturbative QCD is presumed
to be applicable for $\mu \gg \Lambda$.
For CTEQ2L PDF, $\Lambda^{(4)}_{\overline{\rm MS}} = 190$ MeV.

\chapter{ The Eikonal Approximation for 
$\sigma(AB(u\bar{d}) \ra \bar{b}bW^+ + jet)$ }

Applying the Eikonal approximation, we can write the cross section of
$\sigma(AB(u\bar{d}) \ra \bar{b}bW^+ + jet)$ in terms of the amplitude
square of the $u\bar{d} \ra \bar{b}bW^+$ process as follows.
\bea
\lefteqn{ \sigma(AB(u\bar{d}) \ra \bar{b}bW^+ + jet) =
\int\!dQ^2\int\!dy\int\!dq^2_T\int\!d\Phi_3 \, {1\over{2 S}} \,
\left({{\alpha_S(Q)}\over{2\pi q^2_T}}\right) \times } \nonumber\\ & &
\left\{ f_{u/A}(x_A,Q) \,
\left[ \sum_j P^{(1)}_{\bar{d}\leftarrow j} \circ f_{j/B} +
P^{(1)}_{\bar{d}\leftarrow g} \circ f_{g/B} \right]\!(x_B,Q) \: + \right.
\\ & &
\left. f_{u/A}(x_A,Q) \, f_{\bar{d}/A}(x_B,Q) \,
{\left[C_F \ln\left({Q^2\over q^2_T}\right) - {3\over 2} C_F \right] }
\right\}
\, \overline{{| \M(u\bar{d} \ra \bar{b}bW^+) \, |}^2} \: + \nonumber\\ & &
(A\leftrightarrow B) , \nonumber
\ena
where $d\Phi_3$ is the usual 3-dimensional phase space volume
as defined in Appendix~B.

$Q, y$ and $q_T$ are the invariant mass, rapidity and transverse momentum of 
the ($W + \bar{b} + b$) system. 
$\overline{{| \M(u\bar{d} \ra \bar{b}bW^+) \, |}^2}$ is the
amplitude square of $u\bar{d} \ra \bar{b}bW^+$ after summing over the spin
and color factors in the final state and averaging over the spin
and color factors in the initial state.
For a given $Q, y$ and $q_T$
\beq
x_A = {Q\over \sqrt{S}} e^y, \quad x_B = {Q\over \sqrt{S}} e^{-y},
\enq
where $\sqrt{S}$ is the center-of-mass energy of the hadrons $A$ and $B$.

The splitting functions are
\bea
P^{(1)}_{k\leftarrow j}\!(z) &=& C_F 
{\left({{1+z^2}\over {1-z}}\right)}_{\hspace{-.15cm}+} \delta_{kj}, 
\\ \nonumber
P^{(2)}_{k\leftarrow g}\!(z) &=& {1\over 2}{\left(z^2 + (1-z)^2\right)} \\
\ena
and
\bea
(P^{(1)}\circ f)(x,Q) &=& \int^1_x\!{d\xi\over \xi} \, 
P^{(1)}\!\left({x\over \xi}\right) \, f(x,Q), \\
\ena
where the indices $j$ and $k$ denote the flavor of quark or antiquark, 
$\delta_{kj}$ is equal to 1 for $k = j$ and zero otherwise.
In QCD, $C_F = {4\over 3}$ for three colors.
The ``$+$'' prescription is defined by
\bea
\int^1_x\;dz {\left({{1+z^2}\over {1-z}}\right)}_{\hspace{-.15cm}+} f(z) = 
\int^1_0\;dz{{(1+z^2) f(z) \theta(z-x) - (1+z^2) f(1)}\over{1-z}},
\ena
where 
\beq
\theta(z-x) = \left\{ \begin{array}{ll}
			1 & \mbox{for $z>x$} \\
			0 & \mbox{otherwise}
		      \end{array} ~~.
	      \right.
\enq

The above result holds in the soft-gluon approximation.
We have also assumed that the initial state QCD radiation dominates
the soft-gluon radiation from $u\bar{d} \ra \bar{b}bW^+$.
This should be a good approximation because the $b$ quark is 
massive and is less likely to radiate gluons as compared to the
initial state quark or gluon.

\chapter{ The Computer Program ONETOP }

Our analysis is based on our Monte Carlo program ONETOP,
created by modifying 
PAPAGENO (version 3.07), written by Ian Hinchliffe.  ONETOP contains
code for parton level analyses of
single top-quark production at hadron colliders as 
well as the major background. The top quark decays on-shell
to $b W^+$ with branching ratio $\rm{Br} = 1$.
All final state $W$'s decay on-shell
to $e \nu$ with branching ratio $\rm{Br} ={1 \over 9}$.
In addition, we implemented QCD $t \bar{t}$ 
production with the top quark decaying on-shell according to the 
Effective Lagrangian of Equation(\ref{eqlag}), which includes
the most general $\tbW$ couplings. 

Only the CTEQ2 leading order parton distribution is implemented\cite{pdf}.
ONETOP accepts matrix 
elements calculated using the helicity amplitude method described in 
Appendix A. Squaring of the matrix elements 
and sums over spin and color are performed numerically.
We include a simple histogramming package which allows the 
plotting of one- and two-dimensional differential cross-sections
to aid in analyzing event topologies.

We list below the processes included in ONETOP. The structure of these
processes are fully discussed in Section 7. 
\begin{itemize}
  \item $q' b \ra q \,t (\ra b W^+ (\ra \ell^+ \nu))$,
  \item $q' g \ra q \,t (\ra b W^+ (\ra \ell^+ \nu)) \,\bar b$,
  \item $q' \bar{q} \ra W^{*} \ra \bar{b} \,t (\ra b W^+ (\ra \ell^+ \nu))$,
  \item $q' \bar{q} \ra \bar{b} b W^+ (\ra \ell^+ \nu)$,
  \item $q \bar q, \, g g \ra t (\ra b W^+ (\ra \ell^+ \nu))
                     \, \bar{t} (\ra \bar{b} W^- (\ra \ell^- \bar{\nu}))$.

\end{itemize} 

\clearpage
\singlespacing


\begin{thebibliography}{99}
\bibitem{cdfd0}
The specific methods used by the CDF and the \D0 groups can be found 
in Refs.~\cite{CDFft} and \cite{D0ft}, respectively.
\bibitem{lepjet} 
This is the kind of analysis that has currently been performed at LEP for 
separating a quark jet from a gluon jet.
\end{thebibliography}
\end{document}